\documentclass[dvipsnames,twocolumn]{aastex631}

\usepackage{placeins}
\usepackage{graphicx}
\usepackage{amsmath}
\usepackage{txfonts}
\usepackage{comment}
\usepackage{color}
\usepackage{soul}
\graphicspath{{Figures/}}

\newcommand{\mdisk}[0]{\ensuremath{M_{\rm disk}}}
\newcommand{\mgas}[0]{\ensuremath{M_{\rm gas}}}

\newcommand{\rgas}[0]{\ensuremath{R_{\rm CO,\ 90\%}}}
\newcommand{\rdust}[0]{\ensuremath{R_{\rm dust,\ 90\%}}}
\newcommand{\rc}[0]{\ensuremath{R_{\rm c}}}
\newcommand{\macc}[0]{\ensuremath{\dot{M}_{\rm acc}}}
\newcommand{\mstar}[0]{\ensuremath{M_{*}}}
\newcommand{\msun}[0]{\ensuremath{\mathrm{M}_{\odot}}}
\newcommand{\nhp}[0]{\ensuremath{\mathrm{N}_2\mathrm{H}^+}}
\newcommand{\co}[0]{\ensuremath{^{12}\mathrm{CO}}}
\newcommand{\xco}[0]{\ensuremath{^{13}\mathrm{CO}}}
\newcommand{\cyo}[0]{\ensuremath{\mathrm{C}^{18}\mathrm{O}}}
\newcommand{\zet}[0]{\ensuremath{\zeta_{\rm CR}}}

\newcommand{\abu}[0]{\ensuremath{x_{\rm CO}}}

\begin{document}

\title{The ALMA Survey of Gas Evolution of PROtoplanetary Disks (AGE-PRO): V.  Protoplanetary gas disk masses}

\correspondingauthor{Leon Trapman}
\email{ltrapman@wisc.edu}
\author[0000-0002-8623-9703]{Leon Trapman}
\affiliation{Department of Astronomy, University of Wisconsin-Madison, 
475 N Charter St, Madison, WI 53706, USA}

\author[0000-0002-0661-7517]{Ke Zhang}
\affiliation{Department of Astronomy, University of Wisconsin-Madison, 
475 N Charter St, Madison, WI 53706, USA}

\author[0000-0003-4853-5736]{Giovanni P. Rosotti}
\affiliation{Dipartimento di Fisica, Università degli Studi di Milano, Via Celoria 16, I-20133 Milano, Italy}

\author[0000-0001-8764-1780]{Paola Pinilla}
\affiliation{Mullard Space Science Laboratory, University College London, 
Holmbury St Mary, Dorking, Surrey RH5 6NT, UK}

\author[0000-0002-1103-3225]{Beno\^it Tabone }
\affiliation{Universit\'e Paris-Saclay, CNRS, Institut d'Astrophysique Spatiale, Orsay, France}

\author[0000-0001-7962-1683]{Ilaria Pascucci}
\affiliation{Lunar and Planetary Laboratory, the University of Arizona, Tucson, AZ 85721, USA}

\author[0000-0002-7238-2306]{Carolina Agurto-Gangas}
\affiliation{Departamento de Astronom\'ia, Universidad de Chile, Camino El Observatorio 1515, Las Condes, Santiago, Chile}

\author[0009-0004-8091-5055]{Rossella Anania}
\affiliation{Dipartimento di Fisica, Università degli Studi di Milano, Via Celoria 16, I-20133 Milano, Italy}

\author[0000-0003-2251-0602]{John Carpenter}
\affiliation{Joint ALMA Observatory, Avenida Alonso de C\'ordova 3107, Vitacura, Santiago, Chile}

\author[0000-0002-2828-1153]{Lucas A. Cieza}
\affiliation{Instituto de Estudios Astrofísicos, Universidad Diego Portales, Av. Ejercito 441, Santiago, Chile}

\author[0000-0003-0777-7392]{Dingshan Deng}
\affiliation{Lunar and Planetary Laboratory, the University of Arizona, Tucson, AZ 85721, USA}

\author[0000-0003-4907-189X]{Camilo Gonz\'alez-Ruilova}
\affiliation{Instituto de Estudios Astrofísicos, Universidad Diego Portales, Av. Ejercito 441, Santiago, Chile}
\affiliation{Millennium Nucleus on Young Exoplanets and their Moons (YEMS), Chile} 
\affiliation{Center for Interdisciplinary Research in Astrophysics and Space Exploration (CIRAS), Universidad de Santiago, Chile}

\author[0000-0001-5217-537X]{Michiel R. Hogerheijde}
\affiliation{Leiden Observatory, Leiden University, PO Box 9513, 2300 RA Leiden, the Netherlands}
\affiliation{Anton Pannekoek Institute for Astronomy, University of Amsterdam, the Netherlands}

\author[0000-0002-2358-4796]{Nicol\'as T. Kurtovic}
\affiliation{Max Planck Institute for Extraterrestrial Physics, Giessenbachstrasse 1, D-85748 Garching, Germany}
\affiliation{Max-Planck-Institut fur Astronomie (MPIA), Konigstuhl 17, 69117 Heidelberg, Germany}

\author[0000-0002-6946-6787]{Aleksandra Kuznetsova}
\affiliation{Center for Computational Astrophysics, Flatiron Institute, 162 Fifth Ave., New York, New York, 10025}

\author[0000-0002-1575-680X]{James Miley}
\affiliation{Departamento de Física, Universidad de Santiago de Chile, Av. Victor Jara 3659, Santiago, Chile}
\affiliation{Millennium Nucleus on Young Exoplanets and their Moons (YEMS), Chile} 
\affiliation{Center for Interdisciplinary Research in Astrophysics and Space Exploration (CIRAS), Universidad de Santiago, Chile}

\author[0000-0002-1199-9564]{Laura M. P\'erez}
\affiliation{Departamento de Astronom\'ia, Universidad de Chile, Camino El Observatorio 1515, Las Condes, Santiago, Chile}

\author[0000-0003-3573-8163]{Dary A. Ru\'iz-Rodr\'iguez}
\affiliation{National Radio Astronomy Observatory, 520 Edgemont Rd., Charlottesville, VA 22903, USA}

\author[0000-0002-6429-9457]{Kamber Schwarz}
\affiliation{Max-Planck-Institut fur Astronomie (MPIA), Konigstuhl 17, 69117 Heidelberg, Germany}

\author[0000-0002-5991-8073]{Anibal Sierra}
\affiliation{Departamento de Astronom\'ia, Universidad de Chile, Camino El Observatorio 1515, Las Condes, Santiago, Chile}
\affiliation{Mullard Space Science Laboratory, University College London, 
Holmbury St Mary, Dorking, Surrey RH5 6NT, UK}

\author[0000-0001-9961-8203]{Estephani TorresVillanueva}
\affiliation{Department of Astronomy, University of Wisconsin-Madison, 
475 N Charter St, Madison, WI 53706, USA}

\author[0000-0002-4147-3846]{Miguel Vioque}
\affiliation{European Southern Observatory, Karl-Schwarzschild-Str. 2, 85748 Garching bei München, Germany}
\affiliation{Joint ALMA Observatory, Alonso de Córdova 3107, Vitacura, Santiago 763-0355, Chile}

\begin{abstract}
The evolution of the gas mass of planet-forming disks around young stars is crucial for our understanding of planet formation, yet it has proven hard to constrain observationally, due both to the difficulties of measuring gas masses and the lack of a homogeneous sample. 
Here we present a large grid of thermochemical models which we use to measure protoplanetary gas disk masses of AGE-PRO, the ALMA survey of Gas Evolution in PROtoplanetary disks. AGE-PRO covers a sample of 30 disks around similar spectral type (M3-K6) stars with ages between 0.1 and 10 Myr.
Our approach is to simultaneously fit observations of CO isotopologues and \nhp, a complementary molecule produced when CO freezes out.
We find that the median gas mass of the three regions decreases over time, from $7.0^{+4.4}_{-2.6}\times10^{-3}\ \msun$ in Ophiuchus ($\lesssim$ 1 Myr) to $9.4^{+5.4}_{-3.4}\times10^{-4}\ \msun$ for Lupus ($\sim$1-3 Myr) and $6.8^{+5.1}_{-2.8}\times10^{-4}\ \msun$ for Upper Sco ($\sim$2-6 Myr), with $\sim1$ dex scatter in gas mass in each region. We note that the gas mass distributions for Lupus and Upper Sco look very similar, which could be due to survivorship bias for the latter. 
The median bulk CO abundance in the CO emitting layer is found to be a factor $\sim10$ lower than the ISM value but does not significantly change between Lupus and Upper Sco.
From Lupus to Upper Sco the median gas-to-dust mass ratio increases by a factor $\sim3$ from $\sim40$ to $\sim120$, suggesting efficient inward pebble drift and/or the formation of planetesimals.

\end{abstract}

%
\section{Introduction}
\label{sec: introduction}

The gas mass of planet-forming disks around young stars is a vital yet elusive part of planet formation theories and very much needed for our understanding of observed planetary systems (e.g. \citealt{MorbidelliRaymond2016,Drazkowska2023}). In its simplest form, the disk gas mass represents the total reservoir of material available for the formation of gas giants. Knowing the evolution of protoplanetary gas disk masses is also highly desirable, as it sets how much time giant planets have to accrete their atmosphere. 
In addition, these giant planets can also influence the formation and properties of terrestrial planets (e.g. \citealt{BitschIzidoro2023}).
But gas is also important for all steps of planet formation.
The gas density and local gas-to-dust mass ratio regulate the dynamics of dust grains and larger bodies that could eventually grow to form rocky planets and giant planet cores. Dust growth, settling rates, and the speed at which larger grains radially drift inward all depend on the gas content, as well as the migration of young planets that have already formed (see, e.g., recent reviews by \citealt{Drazkowska2023,Birnstiel2024,Paardekooper2023}).

Measuring gas disk masses has proven difficult (e.g. \citealt{BerginWilliams2018}), primarily because the main constituent of the gas, molecular hydrogen (H$_2$) is both difficult to observe in disks and the few available observations do not trace the bulk gas in protoplanetary disks. H$_2$ is a light symmetric molecule that lacks a permanent dipole, limiting its rotational lines to quadrupole transitions $(\Delta J = 2)$. Even the lowest transition of H$_2$, $S(0)$ $(J=2-0)$ at $28\mu$m, requires exciting the $J=2$ level at $549.2$ K which only happens to any appreciable degree for gas temperatures above $100$ K (e.g. \citealt{Thi2001,Carmona2011}). However, the bulk of the gas in protoplanetary disks resides at temperatures below $\sim30$ K (e.g. \citealt{Pinte2018,Schwarz2018,Law2021bMAPS,Law2022,Paneque-Carreno2023}), meaning that H$_2$ emission does not trace the bulk of the mass (e.g. \citealt{Pascucci2013}).

The gas mass therefore has to be measured using indirect tracers. By far the most numerous have been disk masses derived from observations of the (sub-)millimeter continuum emission of dust grains in protoplanetary disks (e.g. \citealt{Beckwith1990,Dutrey1996,Williams2005,Andrews2013,ansdell2016,Barenfeld2016,Pascucci2016,Cieza2019,Tobin2020,vanTerwisga2022}). Measuring a disk mass from a millimeter continuum flux relies on assumptions of the dust temperature, optical depth, dust opacity, and grain size distribution as well as the inclusion of scattering effects, each with their uncertainties  (e.g. \citealt{WilliamsCieza2011,Zhu2019,miotello2023}). But most importantly, to determine the gas mass, a gas-to-dust mass ratio has to be assumed. It is common to use the interstellar medium (ISM) value of 100 as this is likely the ratio at the formation of the disk, but dust evolution models predict that the gas-to-dust mass ratio would vary across the disk as it evolves due to processes such as photoevaporation, mass loss due to disk winds, grain growth, dust settling and radial drift (e.g. \citealt{Takeuchi2005,Birnstiel2012,Pinilla2012,Testi2014,Rosotti2019}).
Moreover, the presence of substructures in millimeter continuum observations of most disks strongly suggests that the gas-to-dust mass ratio at least varies radially in protoplanetary disks (e.g. \citealt{Andrews2018,Huang2018,Long2019,Cieza2021}; see \citealt{Andrews2020, Bae2023} for recent reviews).

Among the gaseous species, hydrogen deuteride (HD) has been shown as a promising indirect tracer of the gas mass. Due to their chemical similarities HD closely follows H$_2$ (see \citealt{Trapman2017}), but contrary to H$_2$, HD has a small dipole moment and requires much lower temperatures ($\sim30-50$ K; e.g. \citealt{Bergin2013,Trapman2017,calahan2021}) to excite and produce significant emission. HD $J=1-0$, emitted at $112.07 \mu$m, was detected with \emph{Herschel} towards three protoplanetary disks, TW Hya, DM Tau, GM Aur \citep{Bergin2013,McClure2016} and using these detections gas masses were derived for each disk (e.g. \citealt{Bergin2013,McClure2016,Trapman2017,Trapman2022b,calahan2021,Schwarz2021}). 
However, with the end of the \emph{Herschel} mission there is currently no far-infrared observatory available to observe HD in more disks.

Gas masses are therefore most commonly measured from the rotational emission of carbon monoxide (CO), specifically from its more optically thin isotopologues $^{13}$CO and C$^{18}$O (e.g. \citealt{WilliamsBest2014,miotello2017,Long2017}). CO is the second most abundant molecule in the gas of protoplanetary disks and its emission is bright at millimeter wavelengths. Using CO as a gas mass tracer requires knowledge of the CO abundance with respect to H$_2$ (\abu). CO is a relatively chemically stable molecule that is expected to have an abundance of $\abu \approx 10^{-4}$ in the warm molecular layer of protoplanetary disks (e.g. \citealt{Aikawa2002}). Two processes reduce \abu, namely photodissocation in the surface layers of the disk and freeze-out close to the midplane. Both of these processes have been extensively studied in the context of disks (e.g. \citealt{Aikawa2002,Visser2009,Miotello2014,Ruaud2019}) and by including them in thermochemical disk models we can account for these processes and derive a relation between $^{13}$CO and C$^{18}$O fluxes and disk mass (e.g. \citealt{WilliamsBest2014,Miotello2016,Deng2023}). 

Applying this method to derive gas masses for a large number of disks, it has become increasingly clear that are other processes affecting CO in protoplanetary disks that remain unaccounted for (e.g. \citealt{Favre2013,miotello2017,Long2017}). Even after accounting for (isotope-selective) photodissociation and freeze-out, gas masses derived from CO are found to be $\sim5-100\times$ lower than those derived independently from HD (e.g. \citealt{Favre2013, Schwarz2016,Kama2016,McClure2016,Trapman2017,calahan2021}). Similarly, gas-to-dust mass ratios from CO-based gas masses are found to be low, $\sim1-10$, compared to the expected value of $\sim100$ the disk should have inherited from the ISM (e.g., \citealt{ansdell2016,miotello2017,Long2017}). 
Suggestions for these unaccounted processes that reduce \abu\ include the chemical conversion of CO into more complex, less volatile species (e.g., \citealt{Aikawa1997, FuruyaAikawa2014,Yu2016,Yu2017,Bosman2018b,Schwarz2018,Ruaud2022,Furuya2022}) and locking up CO into larger dust bodies that settle towards the midplane (e.g. \citealt{Bergin2010,Bergin2016,Kama2016,Krijt2018,Krijt2020,vanClepper2022,Powell2022}). 

Models using a self-consistent disk structure, and accounting for the photodissociation, freeze out, and CO conversion into CO$_2$ ice can reproduce the observed line fluxes of CO isotopologues as well as the main dissociation product of CO, i.e. atomic carbon, without significantly reducing the gas-to-dust mass ratio or the bulk carbon abundance \citep{Ruaud2022,Deng2023,Pascucci2023}. 

Rather than focusing on one or more of these processes, each with their own uncertainties and assumptions, recent work has proposed a different strategy, namely constraining the bulk \abu\ of disk using observations of diazenylium (\nhp; \citealt{Anderson2019,Anderson2022,Trapman2022b}). \nhp\ is formed through proton transfer of H$_3^+$ to N$_2$. If CO is present it competes with N$_2$ for the available H$_3^+$ to form HCO$^+$. Additionally, the main destruction pathway of \nhp\ is through reacting with CO. As a result of both of these factors \nhp\ is a molecule that is particularly sensitive to the gas-phase CO.
\cite{Trapman2022b} compared the gas masses derived from the CO isotopologue and \nhp\ method with the masses from HD (1-0) line fluxes in three protoplanetary disks. The results demonstrated that, even including the two orders of magnitude uncertainties on the ionization rate,  the two methods give consistent gas mass estimates. Similarly, \cite{Anderson2019, Anderson2022} demonstrated (using an independent thermo-chemical code) that \nhp-to-CO isotopologue line flux ratio is highly sensitive to the gas-phase \abu\ and the inclusion of \nhp\ significantly improves the gas mass accuracy. 

Based on previous successful benchmark tests, in this work we apply this CO and \nhp\ method on new observations of the AGE-PRO ALMA large program \citep{AGEPRO_I_overview} to measure gas masses, gas-to-dust mass ratios and bulk CO abundances for the twenty Lupus and Upper Sco Class II disks in the AGE-PRO sample. The ten Ophiuchus Class I disks in the AGE-PRO sample are young enough that the aforementioned processes affecting CO have not had the time to run their course (e.g. \citealt{Zhang2020b}). When measuring their gas masses, we can therefore limit ourselves to photo-dissociation of CO in the disk surface and freeze-out in the midplane when interpreting the CO isotopologue observations of Ophiuchus, without having to include \nhp\ to constrain their bulk CO abundances.

We run a large grid of thermochemical models, similar in spirit to the grids in \cite{WilliamsBest2014} and \cite{Miotello2016} but now also covering a range of CO abundances. 
The AGE-PRO program observed \co, \xco, \cyo, and C$^{17}$O for its sources in Ophiuchus \citep{AGEPRO_II_Ophiuchus} and \co, \xco, \cyo, and \nhp\ for its sources in Lupus \citep{AGEPRO_III_Lupus} and Upper Sco \citep{AGEPRO_IV_UpperSco}. We extract synthetic observations of these lines from our models and compare them to the AGE-PRO observations to derive gas masses. Combining this with dust masses derived from the millimeter continuum, we can provide gas-to-dust mass ratios for the whole AGE-PRO sample. These newly derived gas masses are an essential input for the disk evolution models of \cite{AGEPRO_VII_diskpop}.

The structure of this paper is a follows: In Section \ref{sec: Model grid} we describe the model grid. In Section \ref{sec: results} we first examine the simulated observables of the model after which we compare them to the observations using an MCMC method to derive gas masses and their associated uncertainties. In Section \ref{sec: discussion} we compare our gas masses to stellar accretion rates and discuss the resulting disk lifetimes. We also look for and discuss correlations between gas mass and other disk and stellar properties. Finally, we discuss caveats that could affect the derived gas masses and we summarize our conclusions in Section \ref{sec: conclusions}.

\section{\texttt{DALI} models}
\label{sec: Model grid}

In this work we used the thermochemical code \texttt{DALI} \citep{Bruderer2012,Bruderer2013}. For a given density structure and stellar spectrum, \texttt{DALI} self-consistently calculates the chemical and thermal structure of the disk. First, the dust temperature and internal radiation field are calculated using a 2D Monte Carlo method to solve the radiative transfer equation. Next, \texttt{DALI} computes the time-dependent chemistry, calculates the atomic and molecular excitation levels, and balances the heating and cooling processes to determine the gas temperature. Due to the inter-dependencies of these processes they are calculated iteratively until a self-consistent solution is found. The resulting temperature structure is then used to run the larger chemical network presented in \cite{Miotello2014}, which includes CO isotope-selective chemistry and photodissociation. We note that in this final step there is no further iteration on the disk temperature. For a more detailed description of \texttt{DALI}, see Appendix A in \cite{Bruderer2012}.

Both chemical networks mentioned previously do not include the chemical conversion of CO into more complex, less volatile species. Instead we opt to mimic these processes in a post-processing step with the peak CO abundance ($x_{\rm CO}$) as a free parameter. Here we calculate the \nhp\ chemistry using the chemical network presented in \cite{vantHoff2017} for a range of CO abundances and cosmic ray ionization rates $(\zeta_{\rm cr})$. The \nhp\ abundance structure is then re-inserted into the DALI model. Similarly, the existing CO abundance structure, and those of its isotopologues are scaled by factor $x_{\rm CO}/10^{-4}$ and re-inserted into the DALI model. Note that at this point the model is not longer fully self-consistent, as CO abundance also affects the chemistry and more importantly the gas temperature. However, in the emitting regions of \xco, \cyo, C$^{17}$O, and \nhp, i.e. the lines of interest for this work, the gas and dust are still well coupled, meaning that the gas temperature is set by dust temperature. 
Finally the excitation is recalculated and the model is ray-traced to produce synthetic line emission. 

For the input surface density we use a tapered powerlaw, the self-similar solution of a viscously evolving disk (e.g. \citealt{LyndenBellPringle1974})
\begin{equation}
\label{eq: surface density}
    \Sigma_{\rm gas} (R) = \frac{\left(2-\gamma\right)\mgas}{2\pi\rc^2} \left(\frac{R}{\rc}\right)^{-\gamma} \exp \left[-\left(\frac{R}{\rc}\right)^{2-\gamma}\right]. 
\end{equation}
Here \mgas\ is the disk mass, \rc\ is the characteristic radius and $\gamma$ is the slope of the surface density, which, for a fully viscous disk coincides with the slope of the kinetic viscosity. Note that for this work we set $\gamma=1$ for all disks. For a typical temperature profile this corresponds to the case where the dimensionless $\alpha$ viscosity parameter stays constant with radius. 

The vertical density structure is given by a Gaussian with a width that increases with radius as a powerlaw to simulate the effect that disks are flared (e.g. \citealt{DullemondDominik2005,Avenhaus2018,Law2021bMAPS,Law2022,Paneque-Carreno2023})
\begin{equation}
\label{eq: disk height}
    H(R) = R\ h_{\rm 100} \left(\frac{R}{\rm 100\ au}\right)^{\psi}.
\end{equation}
Here $h_{\rm 100}$ is the opening angle at 100 au and $\psi$ is the flaring angle.

Dust in the model is split into two populations (following e.g. \citealt{Andrews2012}). Both grain populations follow a MRN size distribution with a slope of $-3.5$ (e.g. \citealt{Mathis1977}). The largest fraction $(f_{\rm large})$ of the total dust mass is in large dust grains [1 $\mu$m - 1 mm]. To simulate the fact that these grains are settled towards the midplane, their scale height is multiplied by a factor $\chi$ with respect to that of the gas. Small grains [0.005 $\mu$m - 1 $\mu$m] make up the remaining fraction $(1-f_{\rm large})$ of the dust mass. As these grains are still well coupled to the gas their scale height is set equal to the gas scale height.

\begin{table}[tbh]
  \centering   
  \caption{\label{tab: model fixed parameters}\texttt{DALI} AGE-PRO model grid parameters.}
  \begin{tabular*}{0.95\columnwidth}{ll}
    \hline\hline
    Parameter & Range\\
    \hline
     \textit{Chemistry}&\\
     Chemical age & 1 Myr\\ 
     {$x_{\rm CO}$} & $[0.003,0.01,0.03,0.1,0.3,1]\times10^{-4}$\\
     \textit{Physical structure} &\\ 
     $\gamma$ &  1.0\\
     \rc & $[1,5,15,30,60,120,180,300]$ AU\\
     $M_{\mathrm{gas}}$ & $[10^{-6},10^{-5},10^{-4},10^{-3},10^{-2},10^{-1}, 0.5]$ M$_{\odot}$ \\
     vertical structure & `flat' ($h_{100}=0.05,\psi=0.1)$\\ 
                        & `flared' ($h_{100}=0.1,\psi=0.25)$\\
     Gas-to-dust ratio & [10, 100, 1000] \\
     \textit{Dust properties} & \\
     dust population & `young' ($\chi=0.6, f_{\rm large}=0.8)$\\ 
                     & `evolved' ($\chi=0.2, f_{\rm large}=0.9)$\\
     composition & standard ISM$^{1}$\\
     \textit{Stellar spectrum} & \\
                     & `pale' ($T_{\rm eff} = 3500\ \mathrm{K}, L_*=0.1\ \mathrm{L}_{\odot},$ \\
                     & \ \ \ \ \ \ \ \ \ \ \ $M_* = 0.1\ \mathrm{M}_{\odot}$,\\
                     & \ \ \ \ \ \ \ \ \ \ \ $\dot{M}_{\rm acc} = 10^{-11} \mathrm{M}_{\odot}\ \mathrm{yr}^{-1}$)\\   
                     & `faint' ($T_{\rm eff} = 4000\ \mathrm{K}, L_*=0.25\ \mathrm{L}_{\odot},$ \\
                     & \ \ \ \ \ \ \ \ \ \ \ $M_* = 0.2\ \mathrm{M}_{\odot}$,\\
                     & \ \ \ \ \ \ \ \ \ \ \ $ \dot{M}_{\rm acc} = 10^{-9} \mathrm{M}_{\odot}\ \mathrm{yr}^{-1}$)\\  
                     & `intermediate' ($T_{\rm eff} = 4000\ \mathrm{K}, L_*=0.5\ \mathrm{L}_{\odot},$ \\
                     & \ \ \ \ \ \ \ \ \ \ \ \ \ \ \ \ \ \ \ \ \ \ \ \ \ $M_* = 0.4\ \mathrm{M}_{\odot}$,\\
                     &  \ \ \ \ \ \ \ \ \ \ \ \ \ \ \ \ \ \ \ \ \ \ \ \ \ $ \dot{M}_{\rm acc} = 10^{-9} \mathrm{M}_{\odot}\ \mathrm{yr}^{-1}$)\\  
                     & `bright' ($T_{\rm eff} = 4000\ \mathrm{K}, L_*=1.0\ \mathrm{L}_{\odot},$ \\
                     & \ \ \ \ \ \ \ \ \ \ \ \ \ \ $M_* = 1.0\ \mathrm{M}_{\odot}$,\\
                     & \ \ \ \ \ \ \ \ \ \ \ \ \ \ $ \dot{M}_{\rm acc} = 10^{-8} \mathrm{M}_{\odot}\ \mathrm{yr}^{-1}$)\\  
     $\zeta_{\rm cr}^{\dagger}$ & $[10^{-19},10^{-18},10^{-17}]\ \mathrm{s}^{-1}$\\
     \textit{ geometry}&\\
     $i$ & [0,\,50,\,80]$^{\circ}$ \\
     PA & 0$^{\circ}$ \\
     $d$ & 150 pc\\
    \hline
  \end{tabular*}
  \begin{minipage}{0.87\columnwidth}
  \vspace{0.1cm}
  {\footnotesize{$^{1}$\citealt{WeingartnerDraine2001}, see also Section 2.5 in \citealt{Facchini2017}. $^{\dagger}$: range of cosmic ray ionization rates only used for computing the \nhp\ chemistry. 
  }}
  \end{minipage}
\end{table}

Finally, the stellar spectrum is assumed to be a blackbody with an effective temperature $T_{\rm eff}$ and a stellar luminosity $L_*$. Added to spectrum is a 10000 K blackbody representing excess ultraviolet (UV) radiation released by stellar accretion. This UV radiation is of particular importance for CO as it sets its photodissociation rate.
The luminosity of this component is determined by stellar mass accretion rate, under the assumption that half of the gravitational potential energy is released as radiation (e.g. \citealt{kama2015}). 

\subsection{The AGE-PRO model grid}
\label{sec: the agepro grid}

While the CO isotopologue lines predominantly trace the gas disk mass, their brightness is also affected by the physical structure of the disk as well as the characteristics of the disk-hosting star. When trying to measure the gas disk mass from \xco\ and \cyo\ line fluxes the disk and stellar properties effectively become nuisance parameters, reducing the accuracy with which disk mass can be obtained. For individual sources it is possible to constrain the disk and stellar properties using additional observations, for example by fitting the observed spectral energy distribution (e.g. \citealt{ChiangGoldreich1997,Hartmann1998,DAlessio1998,Andrews2011,Kama2016,Zhang2019,Zhang2021MAPS,Deng2023}). This is in particular the case for well-known sources, for which a large suite of ancillary observations is available. Alternatively, if these observations are not available or if the number of sources involved makes modeling them individually too time expensive, one can instead create a grid of models that covers a range of values for both the disk mass and the other parameters (e.g., \citealt{WilliamsBest2014,Miotello2016,Miotello2021,Ruaud2022,Pascucci2023}). In this case no model is expected to exactly match the observed source, but the grid as a whole can be used to estimate, within the assumed modeling framework, the uncertainty on the obtained gas mass. Given that the AGE-PRO sample includes both well-known sources, e.g. GW Lup, and sources with limited previous observations, e.g., Sz 95, we opt here for the model grid approach, leaving the modeling of individual sources for future studies (e.g. \citealt{Sierra2024,AGEPRO_XIV_CO_columns}). Below we describe the modeling procedure and discuss the range of each of the parameters varied in our grid.

\subsubsection{Modeling procedure}

The modeling begins by running DALI on a protoplanetary disk structure defined by initial parameters: gas and dust density distributions, stellar spectrum, and elemental abundances for carbon and oxygen at ISM levels ($\sim$10$^{-4}$). From this, DALI calculates fundamental disk properties such as dust temperature, gas temperature, and the intensity of ultraviolet radiation throughout the disk’s radial and vertical extent. It also determines the baseline CO and N\(_2\) abundance profiles, reflecting how molecules distribute and evolve under the influence of stellar irradiation, density gradients, and temperature variations.

\begin{figure}
    \centering
    \includegraphics[width=0.45\textwidth]{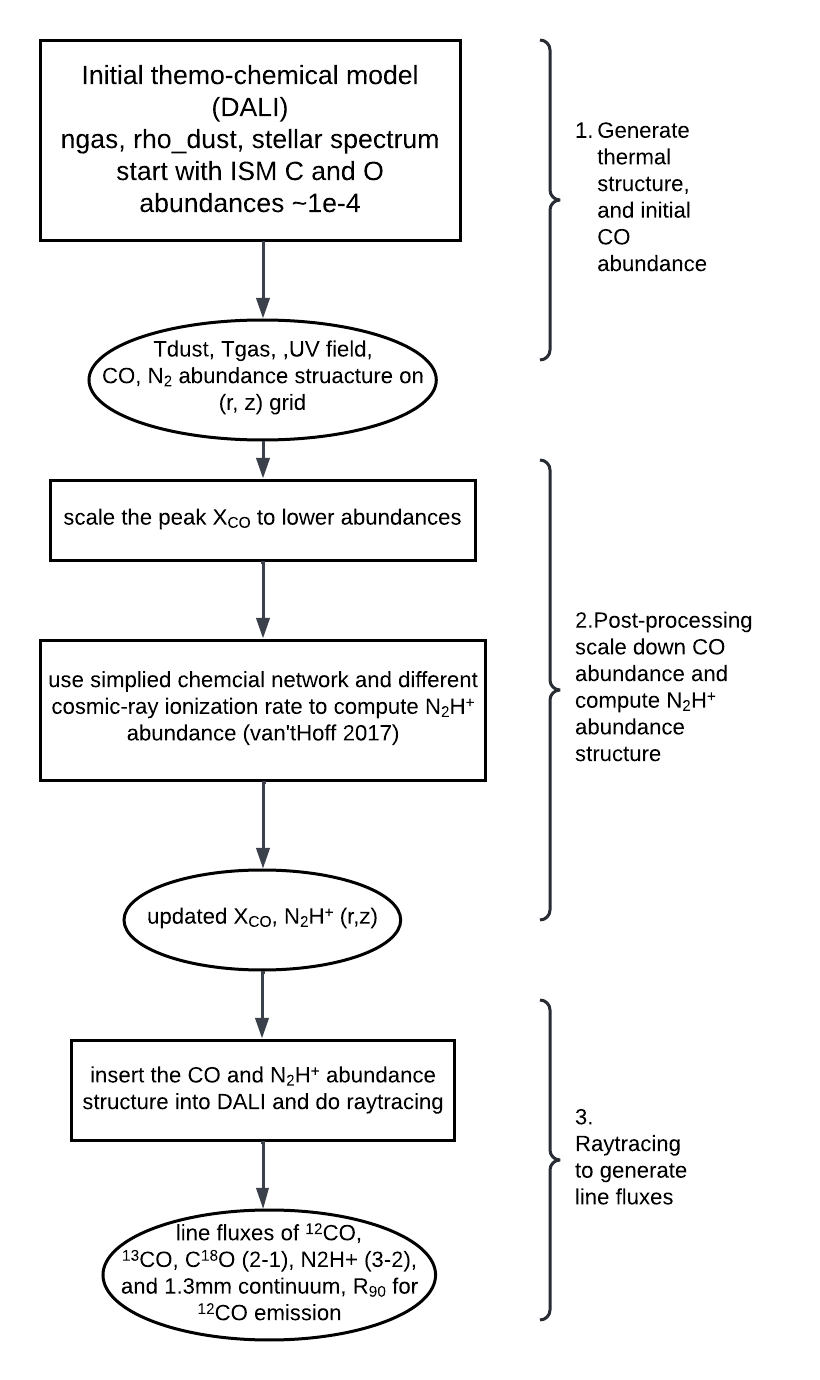}
    \caption{\label{fig:model_processes} Flow chart of processes to generate the AGE-PRO model grid. The square boxes show actions and ellipses show outputs after each action step. }
\end{figure}

In post-processing, the CO abundance is systematically and globally scaled down to mimic various degrees of depletion. We use the simplified chemical network of \citep{vantHoff2017}, paired with different cosmic-ray ionization rates, to calculate the corresponding N\(_2\)H\(^+\) abundance. This updated CO and N\(_2\)H\(^+\) structure is reinserted into the initial DALI model, and then we use ray-tracing function of DALI to produce simulated line fluxes for \(\mathrm{^{12}CO}, \mathrm{^{13}CO}, \mathrm{C^{18}O}\) (2–1), N\(_2\)H\(^+\) (3–2), and the 1.3\,mm dust continuum, as well as the 90\% flux radius of $^{12}$CO\,(2-1) line emission. Figure~\ref{fig:model_processes} shows a flowchart of the modeling processes. 

\subsubsection{Model parameter ranges}

\emph{Gas disk mass:} The gas disk mass (\mdisk) is covered by seven values from $\mdisk = 10^{-6}\ \msun$ up to $0.5\msun$. The range of masses in the grid grew dynamically over the course of this work in order to cover all AGE-PRO sources.

\emph{Disk size:} Similarly, the range for disk size (\rc), seven values between 1 and 300 au, was chosen to match the range of observed sizes within the AGE-PRO sample. We note here that for disks with an exponential taper the observed gas disk size, e.g., the radius that encloses 90\% of the $^{12}$CO, can be up to ten times larger than \rc, depending on the mass of the disk and its radial distribution (e.g. \citealt{Trapman2023}).

\emph{Gas-to-dust mass ratio and dust properties:} For the gas-to-dust mass ratio ($\Delta_{\rm gd}$) we assume three values (10, 100, 1000) representing a one dex spread around $\Delta_{\rm gd}=100$, which is the ratio that disks are expected to inherit from their parent cloud (e.g. \citealt{Goldsmith1997}). We also include two combinations of the mass fraction ($f_{\rm large}$) and degree of settling ($\chi$) of the large grains: a ``young'' population ($f_{\rm large}=0.8, \chi = 0.6$) where large grains have started to form but have not fully settled to the midplane yet (e.g. \citealt{Villenave2023,Lin2023}) and an ``evolved'' population ($f_{\rm large}=0.9, \chi = 0.2$) where the grains have finished settling. In the dust evolution models from \cite{AGEPRO_VI_DustEvolution} these are the approximate global, i.e. dust mass weighted and radially averaged, values of the dust population at 0.1 Myr and 1 Myr, respectively.

\emph{Vertical structure:} The thermal structure of a disk is, to a large degree, set by how much stellar light the disk intercepts, which is determined by its vertical structure. In our grid we cover two cases, a puffed-up disk with a relatively flared surface ($h_{\rm 100}=0.1, \psi=0.25$) and much thinner disk whose surface is only weakly flared ($h_{\rm 100}=0.05, \psi=0.1$). We will refer to these two cases as ``flared'' and ``flat'', respectively. These two cases represent the range of temperature structures we can expect for protoplanetary disks \citep[e.g.,][]{ChiangGoldreich1997,Dullemond2002, Zhang2021MAPS, Law2021bMAPS}.    

\emph{Stellar properties:} The AGE-PRO sample was selected to have a narrow range in stellar spectral types (M3-K6). Within the sample the stellar luminosities range from $L_* = 0.08\ \mathrm{L}_{\odot}$ to $L_* = 1.15\ \mathrm{L}_{\odot}$. 
Because disk properties are known to be stellar-mass dependent (e.g. \citealt{Pascucci2009}) and the disk temperature structure depends on heavily on the stellar luminosity as the main heating source, we cover this stellar luminosity range with four stellar spectra that 
have $L_* = [0.1,\ 0.25,\ 0.5,\ 1.0]\ \mathrm{L}_{\odot}$, which we will refer to as ``pale'', ``faint'', ``intermediate'', and ``bright'', respectively. For each spectrum we use the observations in \cite{alcala2017} to determine a typical stellar mass accretion rate for sources with similar stellar luminosities as the spectrum. Using this accretion rate we add a FUV component to the spectrum, as described at the end of the previous section. 

\emph{CO abundance:} Our model does not include the chemical conversion of CO into CO$_2$ ice or the locking up of CO into larger icy bodies which have been proposed to explain the observed low CO isotopologue fluxes. Rather than including these processes, as was done in e.g. \cite{Krijt2020, Trapman2021, Powell2022, Ruaud2022, vanClepper2022}, here we instead include the peak CO gas abundance as a free parameter. This allows us to cover a much larger parameter space in a reasonable amount of time and it limits number of additional parameters that would need to be included and explored to a single one. In our grid we include six values of the peak CO gas abundance between $\abu = 3\times10^{-7}\ \mathrm{and}\ 10^{-4}$, spanning the range of previous estimates (e.g. \citealt{Bergin2013,Favre2013,Cleeves2015,Trapman2017,calahan2021}). One caveat of the abundance scaling method is that a simple scaling may over-predict the abundances of rare isotopologues like C$^{18}$O due to isotopologue selective photo-dissociation, when the CO gas column becomes very low \citep{Miotello2014}. To evaluate the overprediction levels, in Appendix~\ref{app: flux comparison}, we run additional models with initial low CO abundance and low gas disk masses to predict line fluxes and compare with that of our scaling method. The results show that in the line flux range of the AGE-PRO sample, the line fluxes between the two methods are consistent within 30\%. It should be kept in mind however that this is an approximation of more complex chemical and physical processes occurring in the disk. The CO abundances in this work should therefore be seen as the bulk gas CO abundance in the \xco\ and \cyo\ emitting layers.

\emph{Ionization rate:} 
The formation of \nhp\ requires H$_3^+$ that are products of cosmic ray ionization or X-ray ionization of molecular hydrogen \citep{Oka2006_H3p}. These ionization rates thus directly affect the abundance of \nhp. X-ray is hard to penetrate beyond the upper atmosphere of the disk, In contrast, Cosmic rays penetrate deeper in the disk interior where CO freezes out and \nhp\ is more likely to form. Detailed disk ionization models by \citet{Anderson2022} showed \nhp\ line flux is insensitive to X-ray luminosity below 10$^{30}$ erg s$^{-1}$, and only increases by a factor of two when X-ray luminosity increases from 10$^{30}$ to 10$^{31}$ erg s$^{-1}$.  
The cosmic ray ionization rate is expected to have a larger effect on the abundance of \nhp. The cosmic ray ionization rate in disks is not well constrained. It is often assumed that the rate is similar to that found in molecular clouds ($\zeta_{\rm CR}\approx 10^{-17}\ \mathrm{s}^{-1}$; e.g. \citealt{vdTakvDishoeck2000}), but analysis of individual disks shows that it can be much lower ($\zeta_{\rm CR}\approx 10^{-19}\,-\,10^{-18}\ \mathrm{s}^{-1}$; e.g. \citealt{Cleeves2015,Aikawa2022MAPS}). The sensitivity of \nhp\ to the cosmic ray ionization rate translates into an uncertainty on how well it can be used to constrain the CO abundance and measure the disk mass (e.g. \citealt{vantHoff2017,Trapman2022b,Anderson2022,Sturm2023}). To properly include this uncertainty we include three cosmic ray ionization rates in our grid, $\zeta_{\rm CR} \in [10^{-19}, 10^{-18}, 10^{-17}]\ \mathrm{s}^{-1}$. 

It is worth pointing out that the ionization rate can also play a role in the chemical conversion of CO into CO$_2$ ice and/or hydrocarbons. In the gas phase, CO is destroyed through reaction with He$^+$ created by cosmic rays or X-ray. On the other hand, CO ice on grain surface can be converted into CO$_2$ and/or hydrocarbon via grain surface reactions with OH and H. The amount of OH on the grain-surface depend on photodissociation rate of H$_2$O ice by cosmic-ray as well as UV radiation  
(see Section 2.3 in \citealt{Bosman2018b} for a detailed discussion) See also \citet{Ruaud2019, Ruaud2022}. Previous theoretical studies of CO depletion in disks suggested the chemical conversion timescale of CO is particularly sensitive to the cosmic-ray ionization rate \citep[e.g.][]{Reboussin2015, Bosman2018b,Schwarz2018, Eistrup2018}. As such, the uncertainty on the ionization rate also translates into an uncertainty on the efficiency of the chemical conversion of CO. 

\emph{Model summary:} 
In total, we ran 2232 models\footnote{Note that not all \mdisk-\rc\ combinations are available. For $\mdisk=0.5~\msun$ and $0.1~\msun$ the minimum sizes are $\rc=15~\mathrm{au}$ and $5~\mathrm{au}$, respectively. For $\mdisk=10^{-4}~\msun$, $10^{-5}~\msun$ and $10^{-4}~\msun$ the maximum sizes are $\rc=180~\mathrm{au}$, $120~\mathrm{au}$ and $120~\mathrm{au}$, respectively.}\footnote{Model grid and fitting code are available here: \url{https://bulk.cv.nrao.edu/almadata/lp/AGE-PRO/}.}, each with 18 different \nhp\ chemistry, i.e. different (\abu, \zet), post-processing steps, resulting in 40176 distinct points in our model grid. A summary of the grid parameters is given in Table \ref{tab: model fixed parameters}.

\section{Results}
\label{sec: results}

\subsection{Effect of individual parameters on CO and \nhp\ line fluxes}
\label{sec: effect of parameters}

\begin{figure*}[th]
    \centering
    \begin{minipage}[c]{0.95\textwidth}
    \includegraphics[width=\textwidth]{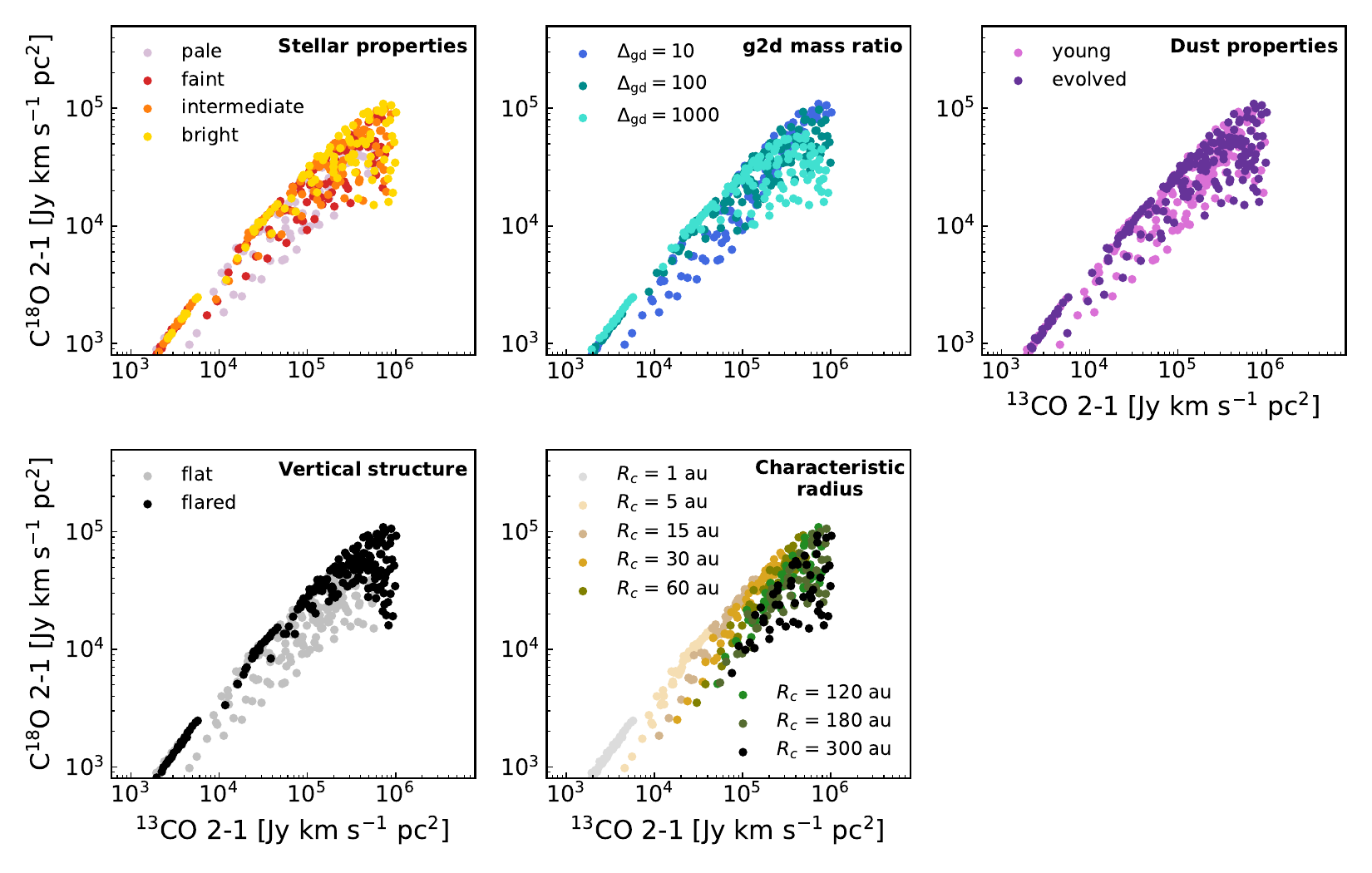}    
    \end{minipage}
    \begin{minipage}[c]{0.95\textwidth}
    \includegraphics[width=\textwidth]{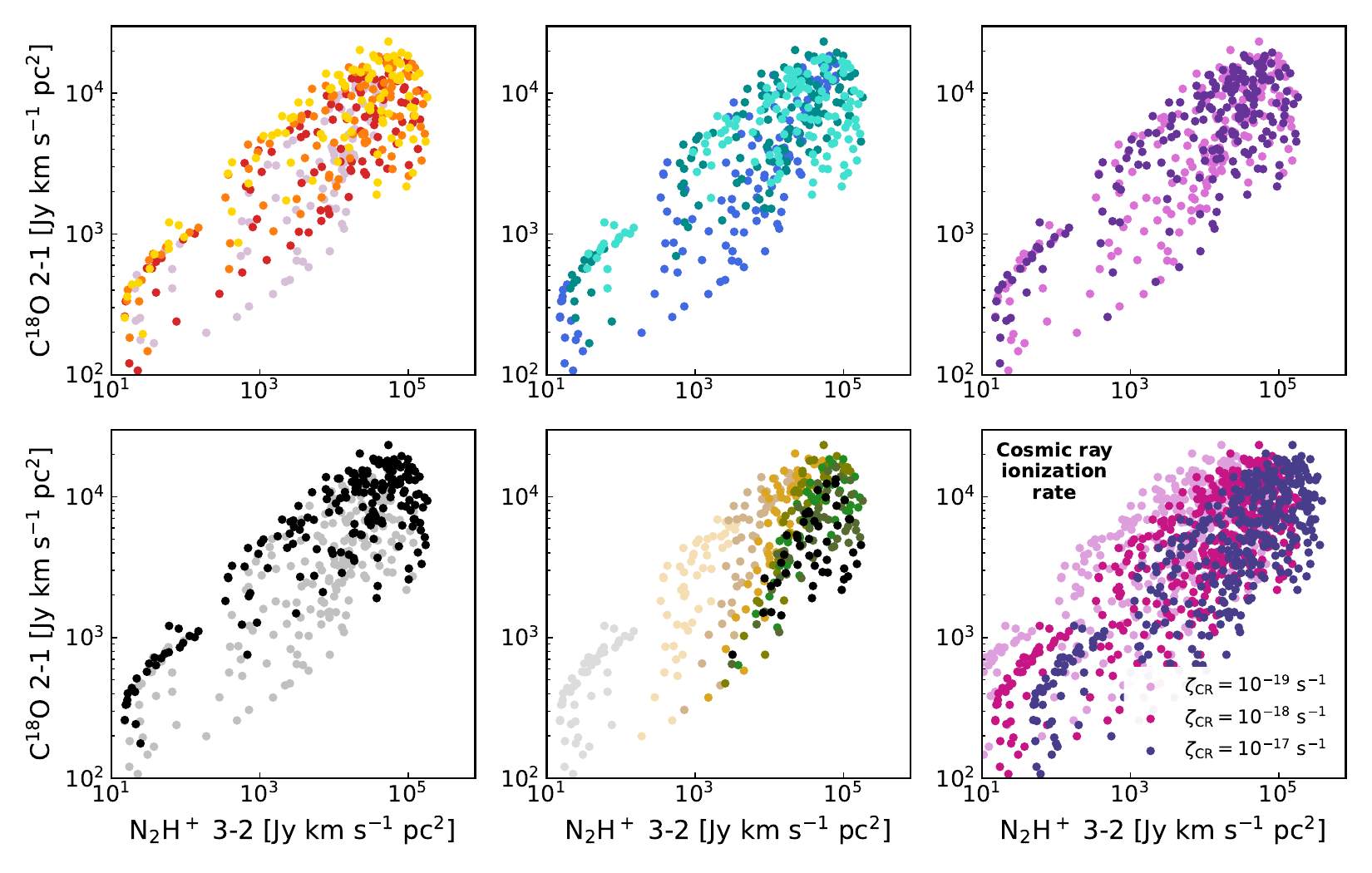}    
    \end{minipage}
    \caption{\label{fig: effect parameters}\textbf{Top panels}: \xco\ and \cyo\ 2-1 line luminosities $(i=0^{\circ})$ for models with a gas mass of $\mgas = 10^{-3}\ \mathrm{M}_{\odot}$. The same models are shown in each panel, with colors highlighting the effect of different disk parameters on the line luminosity (see Table \ref{tab: model fixed parameters}). \textbf{Bottom panels}: as above, but showing \cyo\ 2-1 and \nhp\ 3-2 line luminosities $(i=0^{\circ})$ for models with a gas mass of $\mgas = 10^{-3}\ \mathrm{M}_{\odot}$, $x_{\rm CO} = 10^{-5}$, and $\zeta_{\rm CR} = 10^{-18}\ \mathrm{s}^{-1}$. The bottom right panel instead shows models with $\zeta_{\rm CR} \in [10^{-19}, 10^{-18}, 10^{-17}]\ \mathrm{s}^{-1}$. }
\end{figure*}

\begin{figure*}
    \centering
    \includegraphics[width=\textwidth]{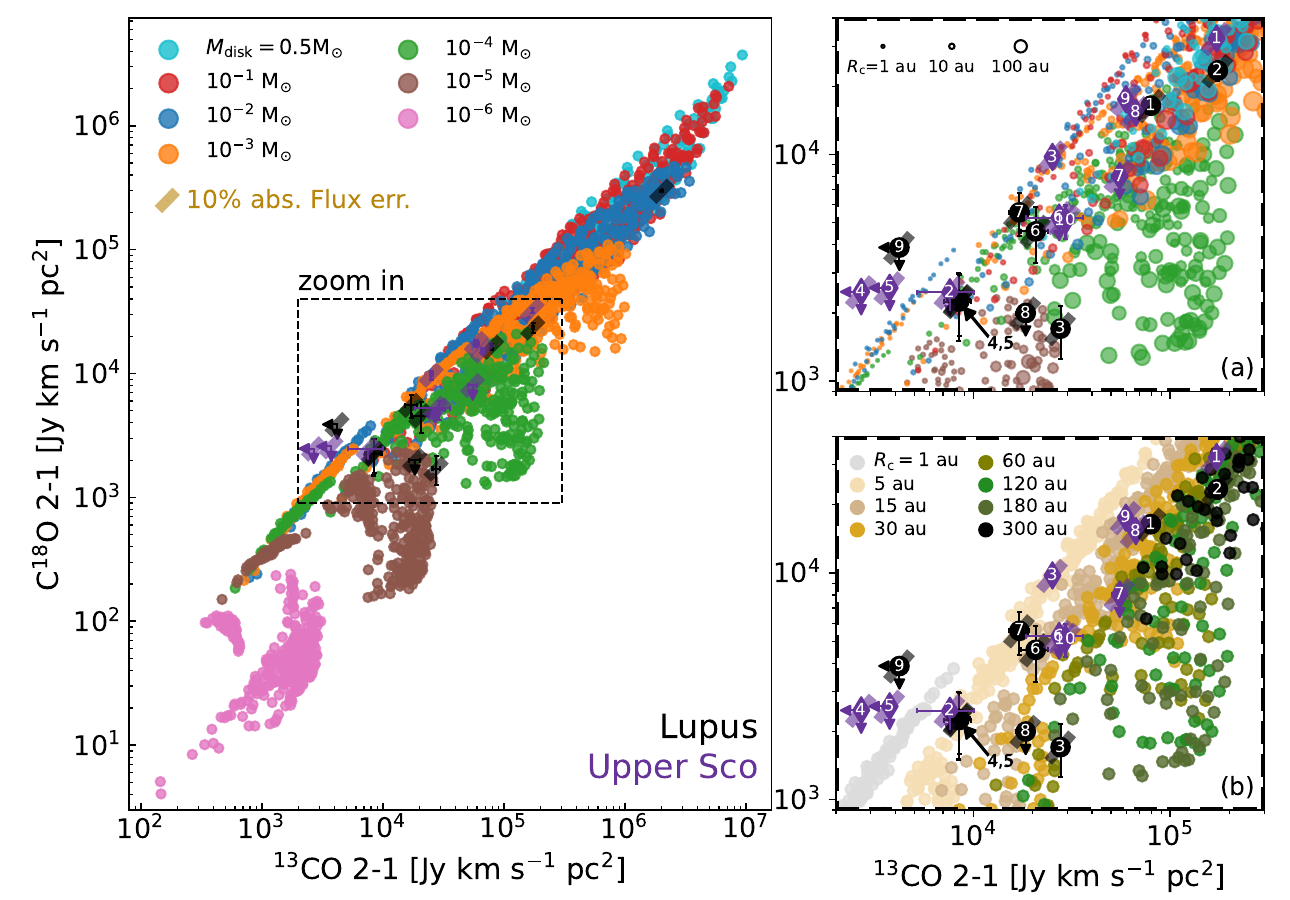}
    \caption{\label{fig: CO comparison}\xco\ and \cyo\ $J=2-1$ line luminosities for the AGE-PRO sources in Lupus and Upper Sco compared to our grid of thermochemical disk models with $\abu=10^{-4}$. The rightmost panels zoom in on the part of the main figure that contains the bulk of the AGE-PRO observations. One source, Lupus 10 (V1094 Sco) lies outside this region. A diagonal shaded region show the absolute flux uncertainty expected for ALMA Band 6. The color of the model points show either the mass (left panel and panel a) or the characteristic radius (panel b). In the top right panel the marker size is scaled based on the $R_{\rm c}$ of the model.}
\end{figure*}

Before comparing our models to the AGE-PRO observations we first examine the models themselves. Figure \ref{fig: effect parameters} shows how disk parameters other than disk mass and CO abundance affect the \xco\ and \cyo\ 2-1 line luminosities. Flared disks have an overall higher disk temperature, which results in having more of the gas above the CO freeze-out temperature and thus more gas-phase CO. This effect increases the \cyo\ luminosity more than \xco, because the more optically thin \cyo\ is a better tracer of the full CO reservoir. For the characteristic size instead the models show that the size has a larger effect on the more optically thick \xco, as its integrated line luminosity more strongly scales with the emitting area $(\propto \rc^2)$. For the stellar parameters, gas-to-dust mass ratios, and dust properties the effects are more mixed. A more luminous star raises the disk temperature, but the FUV luminosity also increases, which photodissociates more CO. Similarly, a disk with a higher gas-to-dust ratio, which for a fixed gas is the same as a lower dust mass, has fewer dust particles that are able to cool the disk but also fewer small grains that absorb FUV photons.
Overall our results are consistent with the similar model grids presented in \cite{Miotello2016} and \cite{Miotello2021}.

Turning now to \nhp, shown in the bottom panels of Figure \ref{fig: effect parameters}, it immediately stands out that the range of \nhp\ line luminosities is somewhat larger than the range in \cyo\ line luminosities. Within the mass bin that is shown in the figure, the \nhp\ line luminosity has a $\sim4$ dex spread, whereas the \cyo\ line luminosities have a spread of $\sim2$ dex. The middle panel at the bottom of Figure \ref{fig: effect parameters} shows that a large part of this spread is driven by the compact disk models ($\rc \leq 15$ au). Closer inspection of these disks reveals that their CO snowline is located beyond \rc, in the exponential taper of the surface density profile. Due to the steep dependence of the surface density on radius, small changes in the exact location of the CO snowline have large effects on the column density of \nhp, which is abundant between the CO and N$_2$ snowlines (e.g. \citealt{Qi2013,vantHoff2017}). For models with $\rc \geq 30$ au the CO snowline lies consistently inside \rc, i.e., in the powerlaw part of the surface density profile. Here the exact location of the CO snowline has much smaller effect on \nhp\ column density, which is reflected in the relatively small spread ($\approx1$ dex) of \nhp\ line luminosities for models with $\rc \geq 30$ au under a single value of cosmic-ray ionization rate.  
Finally, the bottom right panel of Figure \ref{fig: effect parameters} shows the effect of the cosmic ray ionization rate. Increasing or decreasing \zet\ by an order of magnitude increases or decreases the \nhp\ flux by a factor $\sim3$, largely independent of the other model parameters (see also \citealt{Sturm2023}). 

\begin{figure*}[ht]
    \centering
    \includegraphics[width=\textwidth]{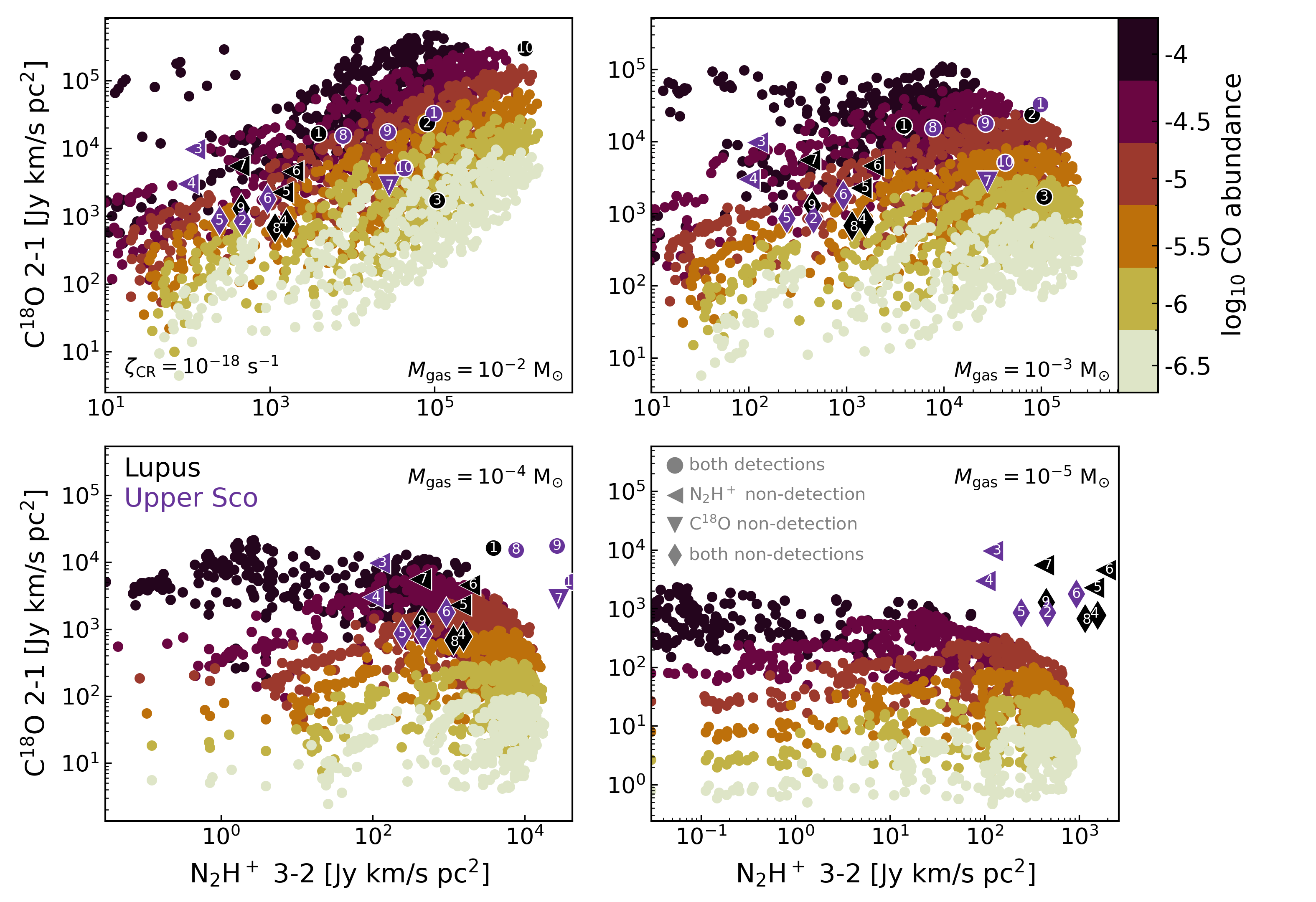}
    \caption{\label{fig: CO abundance} \nhp\ 3-2 and \cyo\ 2-1 line luminosities for our models and the AGE-PRO sources. Observations for Lupus and Upper Sco are shown in black and purple, respectively. Circles denote detections in both lines and diamonds denote $1\sigma$ for non-detections in both lines. In case only one of the lines is detected the observations is shown as a triangle pointing in the direction of the $1\sigma$ upper limit.
    Each panels shows models with a single disk mass, where the color denotes the bulk CO abundance of the model (Note the different ranges on x- and y-axes for each panel).  All models shown here were run using a cosmic ionization rate $\zet=10^{-18}\ \mathrm{s}^{-1}$.}
\end{figure*}

\subsection{AGE-PRO \co, \xco, and \cyo\ observations versus models}
\label{sec: AGEPRO CO comparison}

The AGE-PRO large program observed multiple CO isotopologue lines for thirty disks distributed over three star-forming regions \citep{AGEPRO_I_overview}. In this work we focus on the following subset: the 225 GHz continuum (100\%) and C$^{17}$O~2-1 (80\%) integrated line fluxes for disks in Ophiuchus \citep{AGEPRO_II_Ophiuchus}. Here the percentages refer the fraction of detections ($>3\sigma$) for that observable. For disks in Lupus we use the 234 GHz continuum (100\%), \xco~2-1 (100\%), \cyo~2-1 (70\%) and \nhp~3-2 (40\%) integrated fluxes and \rgas, the radius that encloses 90\% of the $^{12}$CO~2-1 integrated flux (100\%;\citealt{AGEPRO_III_Lupus}). For disks in Upper Sco we use the same set of observables, i.e., the 234 GHz continuum (100\%), \xco~2-1 (80\%), \cyo~2-1 (70\%) and \nhp~3-2 (40\%) integrated fluxes and \rgas~(100\%; \citealt{AGEPRO_IV_UpperSco}).

We can compare the CO isotopologue fluxes of our models to the AGE-PRO observations. Figure \ref{fig: CO comparison} shows the \xco\ and \cyo\ line luminosities of the twenty AGE-PRO sources in Lupus and Upper Sco on top of our models (see \citealt{AGEPRO_III_Lupus} and \citealt{AGEPRO_IV_UpperSco} for details on the observations). Most of the Lupus disks are consistent with $10^{-5}\leq\mgas\leq 10^{-3}\ \msun$ (when assuming an ISM CO abundance), which is in line with the previous gas estimates for these disks from \cite{ansdell2016} and \cite{miotello2017} that were based  solely on CO isotopologue line fluxes. However, the observations are also consistent with much more massive disks with small \rc, as was previously proposed by \citealt{Miotello2021}. 
Fortunately, the AGE-PRO observations also provide some constraints on \rc, which can be used to better estimate disk masses (see Appendix \ref{app: Rc derivation}). We will discuss this further in Section \ref{sec: measuring gas masses with MCMC}.

The disks in Upper Sco cover a similar mass range as the Lupus disks but this includes three sources (Upper Sco 2, 4, and 5) where \xco\ and \cyo\ are either not or only tentatively detected, meaning their masses can be much lower.
Lupus 10 (V1094 Sco) is the clear outlier with respect to the rest of the sample, being an order of magnitude brighter in both \xco\ and \cyo\ compared to the other Lupus and Upper Sco disks (see \citealt{vanTerwisga2018,Pascucci2023,AGEPRO_III_Lupus} for a more detailed discussion on this source). 
It is also worth highlighting that several sources, specifically Upper Sco 4, and 5 and Lupus 9, appear to lie outside of the model grid, but since \xco\ and \cyo\ are not detected, within the observational uncertainties they are therefore consistent with $\mdisk\leq 10^{-5}\ \msun$. Furthermore, if we compare the line luminosity of $^{12}$CO 2-1, which was detected for all three sources, to our models it suggest gas masses between $10^{-6}$ and $10^{-5}$ \msun\ (see Figure \ref{fig: 12CO comparison} in Appendix \ref{app: 12CO comparison}).

\subsection{Measuring CO abundance from \cyo\ and \nhp}
\label{sec: co abundance}

Moving our attention to \nhp\ and the question of the bulk CO abundance of the AGE-PRO disks, Figure \ref{fig: CO abundance} shows the \cyo\ 2-1 and \nhp\ line luminosities for models within four of the gas mass bins, with colors showing the CO abundance (\abu) of the models. On top of these models we show the AGE-PRO observations of the twenty AGE-PRO disks in Lupus and Upper Sco. Focusing on the models first, we see a clear stratification based on CO abundance. Models with a lower CO abundance have fainter \cyo\ line luminosities but a constant or increasing \nhp\ line luminosity, which is consistent with previous work (e.g. \citealt{Anderson2019,Anderson2022,Trapman2022,Sturm2023}). The figure also shows a not insignificant amount of overlap between models with different CO abundances, which reduces how well the bulk CO abundance of a disk can be determined. 

From the twenty AGE-PRO sources in the figure, eight have detections of both \cyo\ and \nhp. With the exception of Lupus 3, these sources overlap with the models that have $\abu =  10^{-5.5} - 10^{-4.5}$  (when assuming $\zet=10^{-18}\ \mathrm{s}^{-1}$), suggesting that the CO abundance in their warm molecular layer has been reduced by approximately one order of magnitude with respect to the ISM value. Two other disks, Lupus 3 and Upper Sco 7, the latter with no \cyo\ detection, are consistent with a lower CO abundance, $\abu\approx10^{-6}$. For the remaining 11 disks \nhp\ is not detected at the 3 mJy km/s level and \cyo\ is only detected for half of them. For these sources it is difficult to determine how much CO, if any, has been removed from their warm molecular layer. It is also worth pointing out that the models shown in Figure \ref{fig: CO abundance} only represent a single cosmic ray ionization rate, which Figure \ref{fig: effect parameters} showed to have a clear effect on the \nhp\ flux. Including this effect, and the effect of other parameters discussed previously, on measuring the gas masses and CO abundances will therefore require a different approach.

\subsection{Measuring gas masses and their uncertainties}
\label{sec: measuring gas masses with MCMC}

\begin{figure}[htb]
    \centering
    \includegraphics[width=\columnwidth]{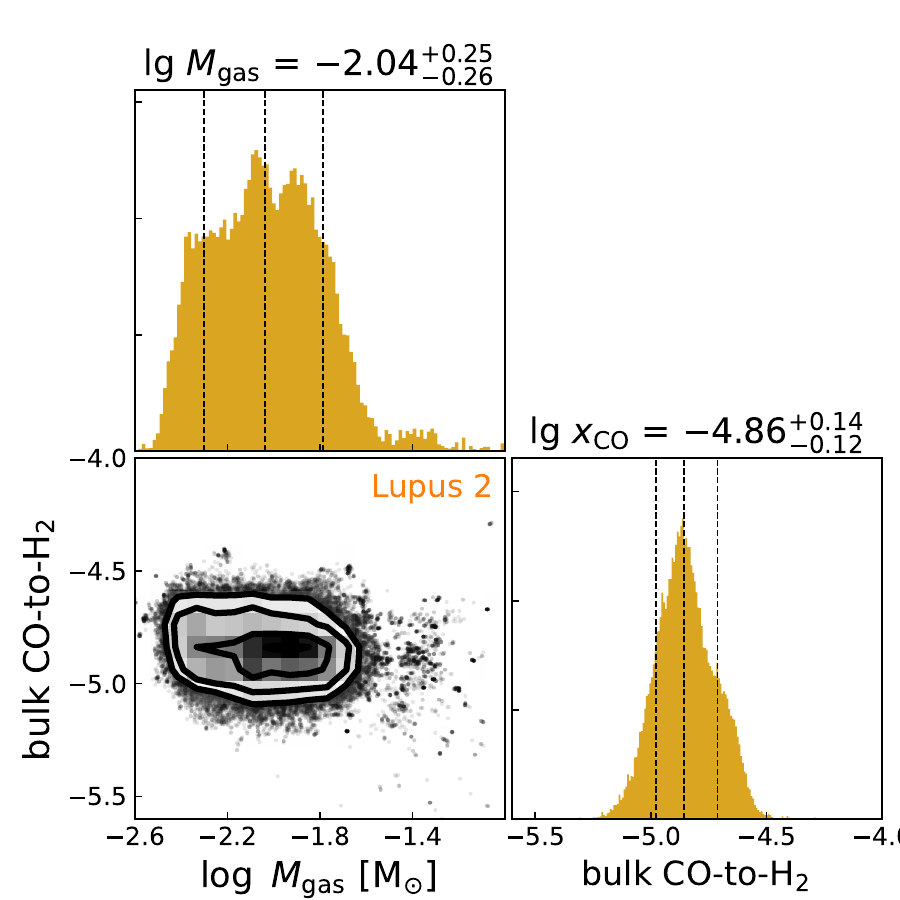}
    \caption{\label{fig: posterior example} Example posterior distributions of \mgas\ and \abu\ for Lupus 2 (Sz 71/GW Lup), using constraints from fluxes of the \xco, \cyo, \nhp~lines, and 1.3\,mm continuum. Shown at the top of each panel and by the vertical dashed line are the 16$^{\rm th}$, 50$^{\rm th}$, and 84$^{\rm th}$ quantile of each distribution.}
\end{figure}

In the previous two sections we have compared the observations of the AGE-PRO disks to our models, which gives us a rough estimate of their mass, bulk CO abundance and characteristic size. However, these comparisons also revealed an inherent problem with deriving gas masses. Depending on their disk parameters, which are often unknown or poorly constrained, models with substantially different disk masses and CO abundances can produce similar \xco, \cyo, and \nhp\ fluxes (see, e.g. panel (a) in Figure \ref{fig: CO comparison}). Here we can turn to Bayesian statistics, which provides a framework for answering the question: what is the probability of \mgas, given the observations and our prior knowledge of the disk parameters, i.e. $P(\mgas | F_{\xco},..,\rc,..)$. This can be obtained constructing a joint posterior probability distribution $P(\mgas,\rc,.. | F_{\xco},..)$ and marginalizing over all disk parameters but the gas mass. 

To sample the probability distribution $P(\mgas,\rc,.. | F_{\xco},..)$ using our models we have to make one assumption, namely that our grid is fine enough (i.e., not too coarse) that a piecewise linear interpolation accurately captures the dependence of flux on each of our disk parameters. This is a reasonable assumption for some parameters, such as \mdisk\ and \rc, where the full domain is covered in multiple steps. However, for parameters with only two values, like flaring or dust properties, it is unclear how well a linear interpolation represents the effect of these parameters on the line fluxes over the full domain included in the fit. We will discuss this further in Appendix \ref{app: linear interpolation assumption}.

For each disk our process to measure the gas mass is the following: We first obtain integrated \xco\ 2-1, \cyo\ 2-1, \nhp 3-2, and 1.3 mm continuum fluxes and the radius that encloses 90\% of the \co\ 2-1 line flux (\rgas) from our models for the inclination that most closely match the inclination of the source ($i_{\rm model} = 0^{\circ}$ if $i_{\rm obs} < 30^{\circ}$; $i_{\rm model} = 50^{\circ}$ if $ 30^{\circ} \leq i_{\rm obs} < 65^{\circ}$; $i_{\rm model} = 80^{\circ}$ if $i_{\rm obs} \geq 65^{\circ}$). 

Our analysis constrains the disk gas mass by leveraging multiple lines. The \xco\ line flux provides an initial gas mass range based on comparison with our model grid. If \xco\ is optically thick, its saturation leads to a very high upper limit on the possible mass. The \cyo\ flux narrows the possible mass range. In the case of a \cyo\ non-detection, the non-detection sets an upper limit of the gas mass, as an optically thick \cyo\ line would be detectable. Furthermore, fitting the emitting radius of the $^{12}$CO constrains the compactness of the disk and thus the emitting areas of \xco\ and \cyo, thereby refining the gas mass estimate. An important additional constraint comes from the flux of \nhp, which is sensitive to the CO abundance. In Appendix~\ref{app:mass_uncertainty_comparison}, we show tests on the effect of including \nhp\ on the gas mass constraints. We compare the resulting gas masses based on CO isotopologues alone with masses from the combined results of CO and \nhp\ line fluxes. The results show that including \nhp\ significantly improves both the accuracy and precision of the gas mass determination.

In the case of \nhp\ non-detection, the minimum gas mass is set by the lowest \xco+\cyo\ gas mass corresponding to the ISM CO abundance, while the maximum is determined by the highest gas mass allowed under the maximum CO depletion consistent with the \nhp\ non-detection.

To calculate the probability of observed line fluxes, we model each measurement as a Gaussian distribution with the observed flux as the mean and the corresponding uncertainty as the standard deviation, regardless of whether the flux exceeds a 3\(\sigma\) threshold. However, for cases where the observed flux is negative, we set the value to zero. This approach is motivated by two considerations. First, for disks where we report \cyo\ upper limits, most measured fluxes fall between the 2\(\sigma\) and 3\(\sigma\) levels, making a Gaussian treatment more representative than a hard cutoff. Second, the \nhp\ detections exhibit a bimodal behavior: the line is either robustly detected (\(>\)4.5\(\sigma\)) or detected at a low level (around 1\(\sigma\)) or even appears negative. We compared two methods for constraining the gas mass: (1) adopting a hard 3\(\sigma\) upper limit, and (2) using the full Gaussian probability distribution defined by the flux and its uncertainty. Although both methods yield similar gas mass estimates, the hard upper limit method tends to predict slightly higher masses because it assumes that all models below the upper limit are equally probable, thereby allowing for higher average \nhp\ fluxes. Consequently, we adopt the Gaussian distribution approach for all cases, as it better captures the statistical nature of our flux measurements.

For \rgas\ we also have to consider the spatial resolution of the \co\ observations, especially since some sources are only marginally resolved. Across the AGE-PRO sample the \co\ resolution is $\sim0\farcs34$ with some outliers at $0\farcs16$ and $0\farcs4$. For each model in our grid we created synthetic \co\ 2-1 channel maps and convolved them with circular Gaussian beams of varying size $({\rm BMAJ} \in [0.1,0.2,0.3,0.4,0.5])$. We integrate these convolved channel maps over the velocity axis to create integrated intensity (moment zero) maps, from which we measure \rgas\ in the same manner as was done for the observations (see \citealt{AGEPRO_III_Lupus, AGEPRO_IV_UpperSco} for details).  
For each source we then select the convolved model radii that were created with circular Gaussian beam whose size $({\rm BMAJ}_{\rm model})$ is closest but smaller than the \co\ resolution of that source $({\rm BMAJ}_{\rm obs})$. To correct for the small difference between ${\rm BMAJ}_{\rm model}$ and ${\rm BMAJ}_{\rm obs}$ we increase the model \rgas\ by $\tfrac{1}{2 \sqrt{2\ln 2}}\left({\rm BMAJ}_{\rm obs} - {\rm BMAJ}_{\rm model}\right)^2$. Here we have made use of the fact that a convolution with a Gaussian with variance $\sigma^2_{\rm wide}$ be written as consecutive convolutions with two Gaussians with smaller variances $\sigma^2_1,\sigma^2_2$ where $\sigma^2_{\rm wide} = \sigma^2_1 + \sigma^2_2$.

Using MCMC as implemented in the python package \texttt{emcee} \citep{Foreman-Mackey2013} we sample the posterior probability distribution $P(\mgas,\rc,.. | F_{\xco},..)$ by minimizing 

\begin{equation}
\label{eq: loglikelihood}
    \ln P(Y_{\rm model} | X_{\rm obs}) = -\frac{1}{2} \sum_{x\in X_{\rm obs}} \left[ \frac{\left(x_{model}(Y_{\rm model}) - x_{obs}\right)^2}{\sigma_{x,obs}^2} \right]
\end{equation}
Here $X_{\rm obs} = \{F_{\xco},F_{\cyo},F_{\nhp}, F_{\rm 1.3mm},\rgas\}$ is the set of observables for each disk and $Y_{\rm model} = \{\mgas,\rc,\Delta_{\rm gd},\psi,f_{\rm large},L_*,x_{\rm CO},\zet\}$ is the set of model parameters that are varied in the grid. 

Note that we sample \mgas, $\Delta{\rm gd}$, \abu, and \zet\ in logspace while running the MCMC.
Also note that $\psi$, $f_{\rm large}$, and $L_*$ are used as numerical proxies for the vertical structure, the dust properties, and the stellar properties, respectively. For example, the flux of an ``intermediately flared'' ($\psi=0.175$) model is calculated by linearly interpolating between fluxes of a ``flat''$(\psi=0.1, h_{100}=0.05)$ and ``flared''$(\psi=0.25, h_{100}=0.1)$ disk model (see Table \ref{tab: model fixed parameters}).

\begin{figure*}
    \centering
    \includegraphics[width=\textwidth]{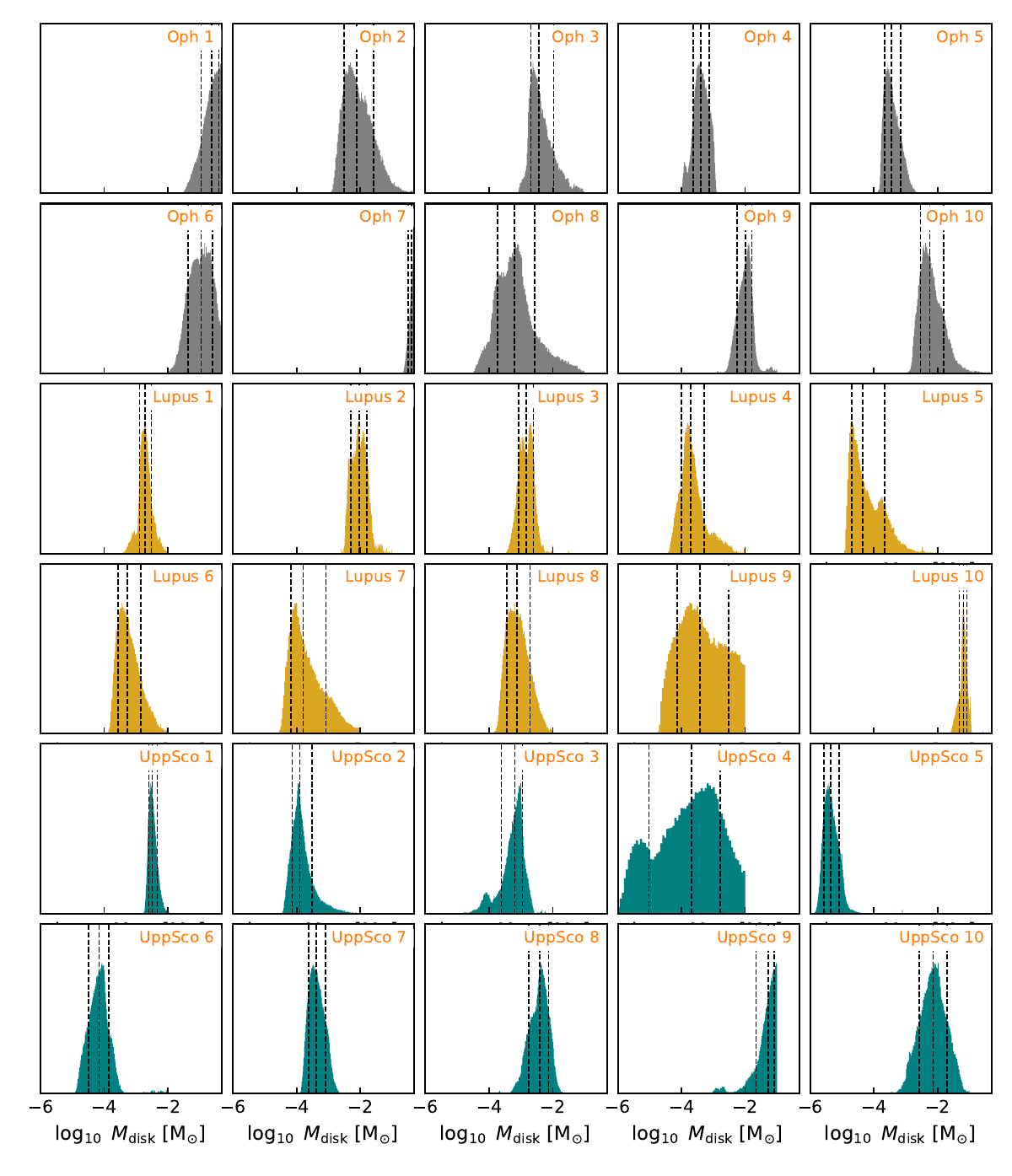}
    \caption{\label{fig: mass histograms}Posterior probability distribution of \mgas\ obtained from fitting either the observed integrated \xco\ 2-1, \cyo\ 2-1, \nhp 3-2, and 1300$\mu$m continuum fluxes (Lupus and Upper Sco) or the C$^{17}$O 2-1 and 1.300 $\mu$m continuum fluxes (Ophiuchus) using the MCMC routine outline in Section \ref{sec: measuring gas masses with MCMC}. In case of non-detections, upper limits are used for the MCMC constraints. Orange dashed lines show the 16$^{\rm th}$, 50$^{\rm th}$, and 84$^{\rm th}$ quantile of the distribution.}
\end{figure*}

We assume uniform priors for $\rc, \psi, f_{\rm large}$ and loguniform priors for $\mgas,\Delta_{\rm gd},\abu,\zet$ (see Table \ref{tab: MCMC priors} for the ranges used for each source). For the stellar luminosity where instead we assume a Gaussian prior based on the observed stellar luminosity and assuming a 30\% uncertainty (see, e.g. \citealt{alcala2014,alcala2017,Alcala2019,Manara2020,UpperSco_followup}).
The MCMC is computed using 1152 walkers and 4000 steps, which after discarding the first 250 steps and thinning by samples by 40, gives a set of 107136 samples from which we compute the posterior probability distribution. Example posterior probability distributions of \mgas\ and \abu\ for Lupus 2 (Sz 71/GW Lup) are shown in Figure \ref{fig: posterior example}, and Figure \ref{fig: mass histograms} shows the \mgas\ posterior distributions for all 20 AGE-PRO disks in Lupus and Upper Sco (see Figure \ref{fig: 2D histograms} in Appendix \ref{app: joint posteriors} for joint posterior probability distributions of \mgas\ and \abu\ for these sources. Full corner plots can be found in Appendix \ref{app: MCMC corner plots }).

As a test of the reliability of the MCMC method we used it to measure the gas masses for the three disks where HD was detected and compared our gas masses estimates with the independent HD-based gas masses. We find that the MCMC-estimated gas masses agree with HD-based gas masses within their respective uncertainties, similar to what was found by \cite{Trapman2022b}. A full description of this test can be found in Appendix \ref{app: HD comparison}.

\subsection{Gas masses for the Ophiuchus disks}
\label{sec: ophiuchus gas masses}

The young Ophiuchus disks in the AGE-PRO sample differ from the older disks in Lupus and Upper Sco in several ways. Firstly, these younger disks are still partially embedded in the remnants of their envelope. From the observational side, the presence of this envelope makes it harder to separate the emission originating from the disk from the contribution of the envelope (see, e.g. \citealt{vtHoff2018} or \citealt{AGEPRO_II_Ophiuchus,AGEPRO_XIII_size_embedded} for details specific to the AGE-PRO Ophiuchus sources). Furthermore, the Ophiuchus star-forming region has a larger extinction than both Lupus and Upper Sco. As a result of this, cloud emission and absorption have a significant effect on not only \co, but also on the \xco\ and \cyo\ emission (see \citealt{AGEPRO_II_Ophiuchus} for details).

From the disk structure side, accretion from the envelope onto the disk, higher stellar mass accretion, and backwarming from the envelope through micron sized grains scattering photons down to the midplane, mean that young embedded disks have a higher overall disk temperature structure than older class II protoplanetary disks (see, e.g. \citealt{DAlessio1998,Whitney2013,vtHoff2018,vtHoff2020,Kuznetsova2022}). Due to their young age the Ophiuchus disks are also expected to have undergone less grain growth, dust settling, and chemical conversion of CO than those in Lupus and Upper Sco, meaning that their bulk CO abundance is expected to still be close to the ISM value (e.g. \citealt{Zhang2020b}). 

With these points in mind we make a few changes to the gas mass measurement method described in Section \ref{sec: measuring gas masses with MCMC}. First, we only fit the C$^{17}$O $2-1$ integrated line and 1.3 millimeter flux as C$^{17}$O is the most optically thin CO line in the AGE-PRO data and the one least affected by cloud emission/absorption (cf. \citealt{AGEPRO_II_Ophiuchus}). To further focus on the disk emission, we use the C$^{17}$O $2-1$ measured from a tight Keplerian mask around the disk to mask out any significant emission not originating from the disk, such as the infalling material seen in Oph 7 (e.g. \citealt{Flores2023}). In some cases absorption can also be seen in channels close to the cloud velocity. If present these channels are masked to prevent them from lowering the C$^{17}$O $2-1$ flux. We do note that even when masked out in this manner the presence of absorption means that the integrated C$^{17}$O and the gas mass derived from it will be underestimated.
For Oph 2, 6, and 9 the \cyo\ emission also appears reasonably clear of cloud contamination. For these sources we repeat the fit adding the \cyo\ emission and find that the resulting gas masses are in good agreement with those obtained from only fitting C$^{17}$O. 

From the side of the models, we fix the dust population to ``young'' (see Table \ref{tab: model fixed parameters}) to represent the fact that the large dust grains have not fully settled yet. Similarly, we limit $\rc \leq 60$ au as the Ophiuchus sources appear to be relatively compact and previous studies have found young Keplerian disks to be compact (see e.g. \citealt{NajitaBergin2018,Maret2020}). Finally, we limit the CO abundance to a narrow range of $3\times10^{-5}\leq\abu\leq10^{-4}$. This represents the fact that, as mentioned earlier, the bulk CO abundances in the Ophiuchus is expected to be approximately the ISM value but allows for the possibility that some of the processes that lower this value have already started.

\begin{figure}
    \centering
    \includegraphics[width=\columnwidth]{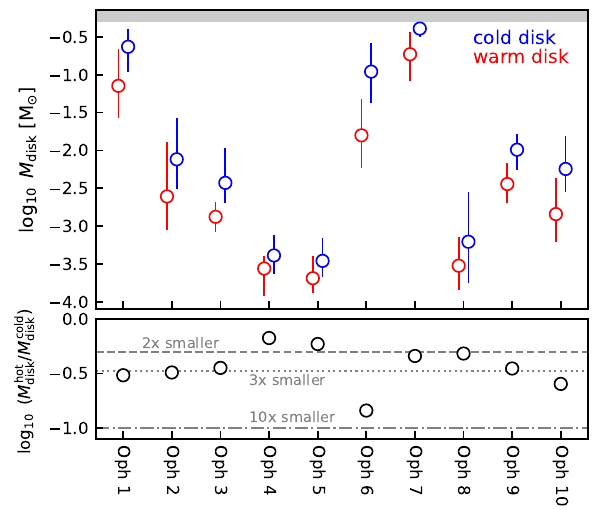}
    \caption{\label{fig: ophiuchus effect of temperature}Comparison of the Ophiuchus gas disk masses derived using cold ($L_* = L_{\rm *,\ obs}$) and hot ($L_* = 10\times L_{\rm *,\ obs}$) models. A comparison of the posterior distributions is presented in Figure \ref{fig: ophiuchus cold warm posteriors}.}
\end{figure}

To account of the overall higher disk temperature of young disks we run two sets of gas mass fits for each source. The first, which we will refer to as the ``cold model'' uses only the models with a stellar luminosity closest to the observed stellar luminosity (see Table 1 in \citealt{AGEPRO_II_Ophiuchus}). In this model the only increase in disk temperature comes from the overall higher stellar luminosity of the Ophiuchus sources compared to the other two regions. However, as mentioned previously, the envelope will also increase the disk temperature (e.g. \citealt{DAlessio1998,Whitney2013,Kuznetsova2022,Han2023}. The Ophiuchus sources are all Class I or Flat-Spectrum type objects and are therefore likely in the process of losing their envelope but can still have enough of an envelope present to affect the disk temperature. We therefore run a second fit, which we will refer to as ``warm model'', using models with a stellar luminosity that is $\sim10\times$ the observed stellar luminosity. Given that the disk temperature is expected to scale as $T\propto L_*^{0.25}$, the ``warm models'' have average disk temperature that is $\sim75\%$ higher compared to the ``cold models''. From the modeling side, the ``cold models'' can be considered an upper limit on the gas mass and the ``warm models'' a lower limit. We should note here however that increasing disk temperature through increasing the stellar luminosity is not strictly equivalent to including heating by an envelope, as the latter is expected to mostly affect larger radii from the star. 

Figure \ref{fig: ophiuchus effect of temperature} shows the effect of this disk temperature increase on the derived gas disk masses for the Ophiuchus sources. As expected, gas masses are lower if the warmer disk models are used. Based on the bottom panel of the figure, an increase of $\sim75\%$ in disk temperature leads to a $\sim2-3\times$ decrease in gas disk mass. It should be noted here that for Ophiuchus 1 and 7 the ``cold model'' gas mass posterior peaks at the upper edge of our model grid, meaning that these gas masses should be viewed as upper limits (see, e.g. \ref{fig: ophiuchus cold warm posteriors}). For their ``warm model'' this is no longer the case, which means that the gas mass ratio seen in Figure \ref{fig: ophiuchus effect of temperature} underestimates the true gas mass ratio for these sources.

While the comparison of the ``cold'' and ``warm'' models gives a quantitative handhold on the effect of disk temperature on the derived gas mass, it is unclear if the Ophiuchus disks in the AGE-PRO sample warmer than expected from just their stellar luminosity. For the rest of this work we therefore adopt the ``cold model'' gas masses with the explicit caveat that the gas masses could be lower if the disks turn out be warmer.

\subsection{AGE-PRO gas masses}
\label{sec: agepro gas masses}

\begin{table*}[htb]
\centering
\caption{\label{tab: agepro gas masses} Derived gas masses and bulk CO abundances}
\def\arraystretch{1.1}
\begin{tabular*}{0.95\textwidth}{ll|c|c|c|cccccc}
\hline\hline
Name  &  AGEPRO ID &  $M_{\rm dust}$ & lg$_{10}$ $M_{\rm gas}$ & lg$_{10}$ $x_{\rm CO}$ & \multicolumn{6}{c}{observational constraints}\\
& & & & & \multicolumn{6}{c}{(d=detection, u=upper limit ($<3\sigma$))}\\
      &            &  [M$_{\oplus}$]  &  lg$_{10}$ [M$_{\odot}$]  &  & 1.3~mm & \xco~2-1 & \cyo~2-1 & C$^{17}$O~2-1 & \nhp~3-2 & \rgas\ \\
\hline
J1626-2424 & Oph 1 & 63.6 & $-0.63^{+0.23}_{-0.33}$ & & d  & - & - & d & - & -\\
J1627-2420 & Oph 2 & 16.9 & $-2.12^{+0.53}_{-0.39}$ & &  d & - & - & d & - & -\\
J1627-2428 & Oph 3 & 12.5 & $-2.43^{+0.45}_{-0.26}$ & &  d & - & - & d & - & -\\
J1627-2427 & Oph 4 & 1.1 & $-3.39^{+0.26}_{-0.24}$  & &  d & - & - & u & - & -\\
J1627-2442 & Oph 5 & 2.6 & $-3.46^{+0.29}_{-0.20}$  & &  d & - & - & u & - & -\\
J1627-2440 & Oph 6 & 21.2 & $-0.96^{+0.37}_{-0.40}$ & &  d & - & - & d & - & -\\
J1631-2401 & Oph 7 & 188.2 & $-0.39^{+0.06}_{-0.10}$& &  d & - & - & d & - & -\\
J1623-2302 & Oph 8 & 2.5 & $-3.21^{+0.64}_{-0.53}$  & &  d & - & - & d & - & -\\
J1626-2441 & Oph 9 & 4.7 & $-1.99^{+0.20}_{-0.26}$  & &  d & - & - & d & - & -\\
J1627-2439 & Oph 10 & 16.3 & $-2.24^{+0.43}_{-0.30}$& &  d & - & - & d & - & -\\
 IK Lup     & Lupus 1 &21.2 &$-2.72^{+0.20}_{-0.17}$ &$-4.57^{+0.16}_{-0.16}$ & d & d & d & - & d & d\\
 GW Lup     & Lupus 2 &50.1 &$-2.04^{+0.25}_{-0.26}$ &$-4.86^{+0.14}_{-0.12}$ & d & d & d & - & d & d\\
 J1612-3815 & Lupus 3 &9.0 &$-2.83^{+0.22}_{-0.24}$ &$-5.79^{+0.14}_{-0.24}$  & d & d & u & - & d & d\\
 HM Lup     & Lupus 4 &4.1 &$-3.72^{+0.42}_{-0.28}$ &$-5.21^{+0.26}_{-0.25}$  & d & d & d & - & u & d\\
 Sz 77      & Lupus 5 &1.4 &$-4.35^{+0.68}_{-0.35}$ &$-4.80^{+0.35}_{-0.32}$  & d & d & u & - & u & d\\
 J1608-3914 & Lupus 6 &6.4 &$-3.27^{+0.43}_{-0.29}$ &$-4.76^{+0.29}_{-0.27}$  & d & d & d & - & u & d\\
 Sz 131     & Lupus 7 &2.6 &$-3.79^{+0.71}_{-0.38}$ &$-4.22^{+0.15}_{-0.21}$  & d & d & d & - & u & d\\
 Sz 66      & Lupus 8 &4.7 &$-3.12^{+0.40}_{-0.33}$ &$-4.93^{+0.23}_{-0.22}$  & d & d & d & - & u & d\\
 Sz 95      & Lupus 9 &1.4 &$-3.42^{+0.90}_{-0.71}$ &$-5.62^{+0.51}_{-0.55}$  & d & d & u & - & u & d\\
 V1094 Sco  & Lupus 10 &135.1 &$-1.19^{+0.11}_{-0.13}$ &$-4.76^{+0.14}_{-0.12}$ & d & d & d & - & d & d\\
 J1612-3010 & UppSco 1 &8.5 &$-2.49^{+0.15}_{-0.11}$ &$-4.82^{+0.11}_{-0.18}$ & d & d & d & - & d & d\\
 J1605-2023 & UppSco 2 &2.5 &$-3.90^{+0.38}_{-0.24}$ &$-4.61^{+0.27}_{-0.27}$ & d & d & d & - & u & d\\
 J1602-2257 & UppSco 3 &1.0 &$-3.19^{+0.24}_{-0.43}$ &$-4.04^{+0.03}_{-0.07}$ & d & d & d & - & u & d\\
 J1611-1918 & UppSco 4 &0.08 &$-3.69^{+0.92}_{-1.33}$ &$-4.28^{+0.20}_{-0.26}$ & d & u & u & - & u & d\\
 J1614-2332 & UppSco 5 &0.4 &$-5.36^{+0.27}_{-0.20}$ &$-4.78^{+0.28}_{-0.36}$ & d & u & u & - & u & d\\
 J1616-2521 & UppSco 6 &0.8 &$-4.16^{+0.30}_{-0.33}$ &$-4.45^{+0.25}_{-0.23}$ & d & d & u & - & u & d\\
 J1620-2442 & UppSco 7 &1.1 &$-3.38^{+0.29}_{-0.23}$ &$-5.02^{+0.18}_{-0.20}$ & d & d & d & - & d & d\\
 J1622-2511 & UppSco 8 &7.4 &$-2.41^{+0.27}_{-0.35}$ &$-4.22^{+0.15}_{-0.19}$ & d & d & d & - & u & d\\
 J1608-1930 & UppSco 9 &9.3 &$-1.28^{+0.20}_{-0.38}$ &$-4.88^{+0.21}_{-0.24}$ & d & d & d & - & d & d\\
 J1609-1908 & UppSco 10 &12.9 &$-2.14^{+0.43}_{-0.45}$ &$-5.52^{+0.22}_{-0.32}$ & d & d & d & - & d & d\\
\hline\hline
\end{tabular*}
\begin{minipage}{0.92\textwidth}
\vspace{0.1cm}
{\footnotesize{\textbf{Notes:} From left to right, the first five columns show the dust mass, obtained from converting the 1.3 millimeter flux under the assumptions of optically thin emission and a dust temperature of 20 K, the median gas mass, and the median bulk CO abundance, the latter two obtained from the MCMC fit
(see details in Section \ref{sec: gas mass uncertainties}). For Ophiuchus disks, the gas masses are derived by assuming ISM level CO abundance. Uncertainties on \mgas\ (and \abu\ for disks in Lupus and Upper Sco) correspond to the 16$^{\rm th}$ and 84$^{\rm th}$ quartile of their respective posterior distributions. 
The remaining 5 columns list the observational constraints used in the MCMC fit and whether it is detected at $\geq3\sigma$ for a given source.
}}
\end{minipage}
\end{table*}

\begin{figure*}[thb]
    \centering
    \begin{minipage}{\textwidth}
    \includegraphics[width=\textwidth,clip,trim={0 1.4cm 0 0}]{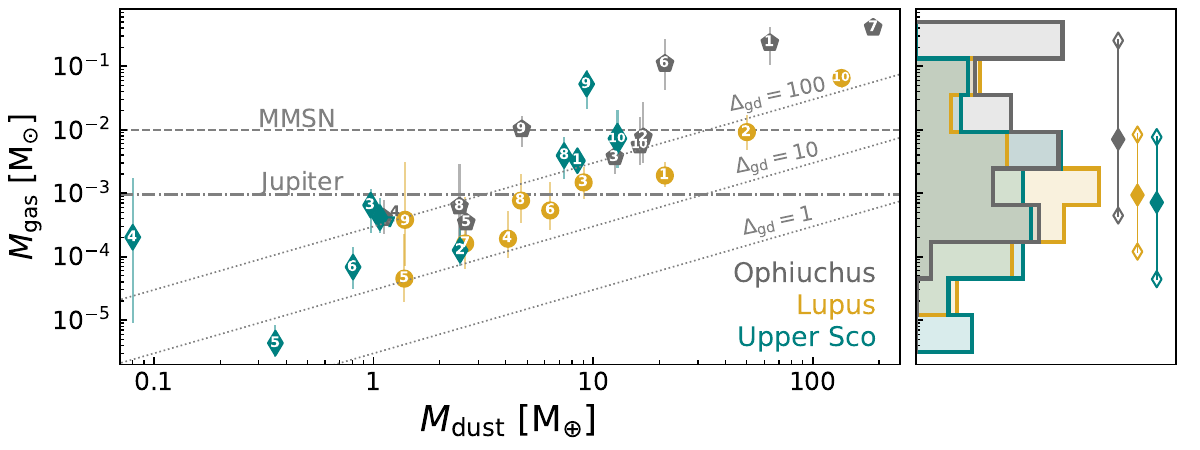}    
    \end{minipage}
    \begin{minipage}{\textwidth}
    \includegraphics[width=\textwidth,clip,trim={0 1.4cm 0 0}]{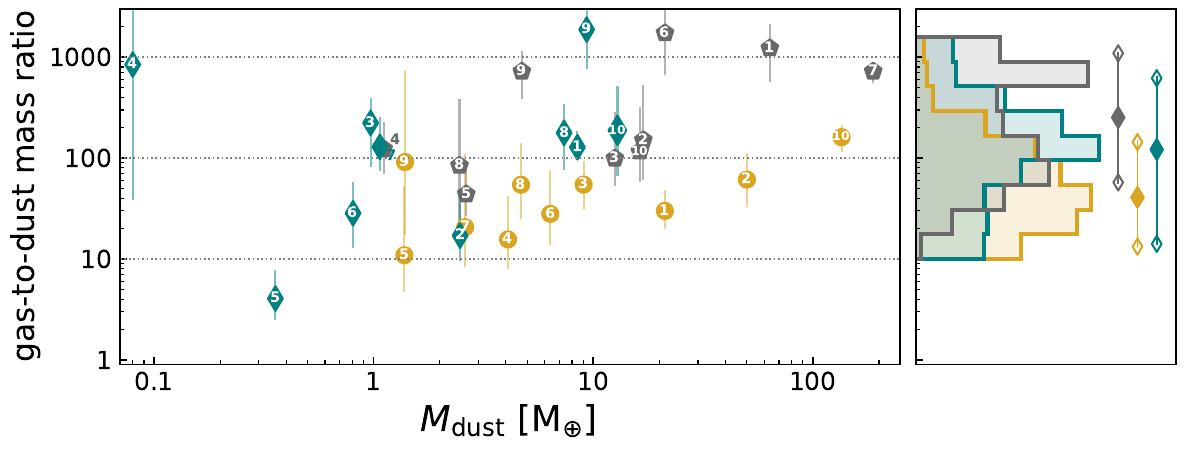}    
    \end{minipage}
    \begin{minipage}{\textwidth}
    \includegraphics[width=\textwidth]{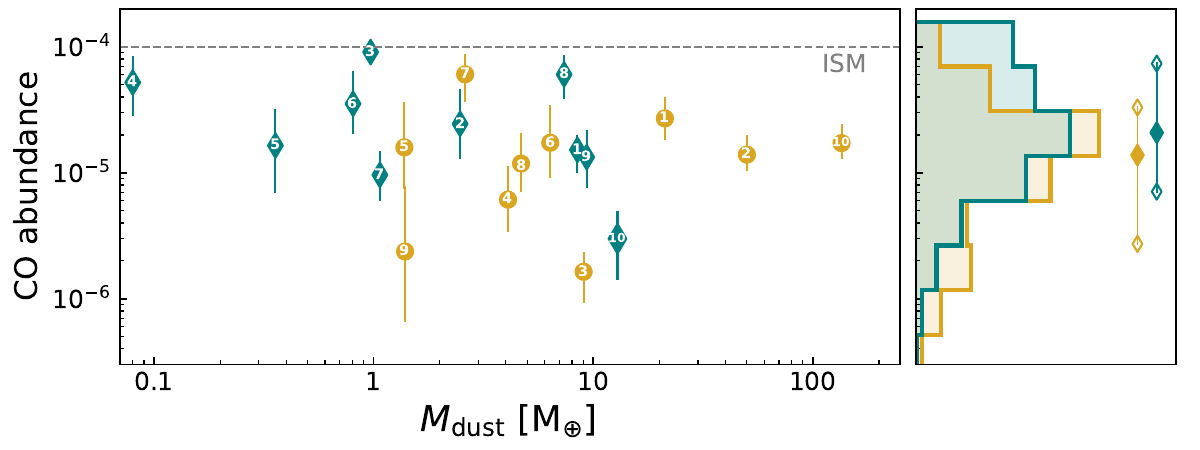}    
    \end{minipage}
    \caption{\label{fig: final masses} From top to bottom, left panels show the gas disk masses, gas-to-dust mass ratios and bulk CO gas abundance for sources in Ophiuchus (gray pentagons), Lupus (brown circles), and Upper Sco (green diamonds), all set against their dust disk mass. The markers show the median of the posterior distribution and the vertical lines denote its 16$^{\rm th}$ and 84$^{\rm th}$ quantile (see Section \ref{sec: measuring gas masses with MCMC}). 
    The horizontal gray dashed lines in the top panel denotes the minimum mass solar nebula (MMSN; \citealt{Hayashi1981}) and the mass of Jupiter. The diagonal dotted lines show a constant gas-to-dust mass ratio. 
    The right panels show histograms of the gas disk masses, gas-to-dust mass ratios and bulk CO gas abundance, obtained by summing the posterior distributions of individual sources and normalizing the histogram so that the sum over it equals ten.}
\end{figure*} 

The top panels of Figure \ref{fig: final masses} show the gas masses for the AGE-PRO disks in Ophiuchus, Lupus and Upper Sco (see also Table \ref{tab: agepro gas masses}). 
The gas mass scales with the dust mass, taken from \cite{ManaraPPVII}. Throughout this work we will use these dust masses, as they facilitate easy comparisons with previous work.
The small exception is Upper Sco, where dust masses are available only for half of the source and these are based on $\sim335$ GHz observations rather than the $\sim225$ GHz observations used for Ophiuchus and Lupus. To facilitate a consistent comparison, we compute the dust masses for our ten Upper Sco sources using the 225\,GHz AGE-PRO continuum fluxes from \cite{AGEPRO_IV_UpperSco} and the same equation and assumptions as were used in \cite{ManaraPPVII} (see also \citealt{ansdell2016}). The dust masses are tabulated in Table \ref{tab: agepro gas masses}.
It should be kept in mind, however, that these dust masses are simple conversions of the millimeter flux assuming a single temperature and dust opacity. Uncertainties in, e.g., the dust composition and grain size distribution, mean that the dust masses can only be assumed to be correct to within an order of magnitude (see, e.g. \citealt{Ricci2010,BalleringEisner2019,Tychoniec2020,Ribas2020,Macias2021}). 

\begin{figure}
    \centering
    \includegraphics[width=\columnwidth]{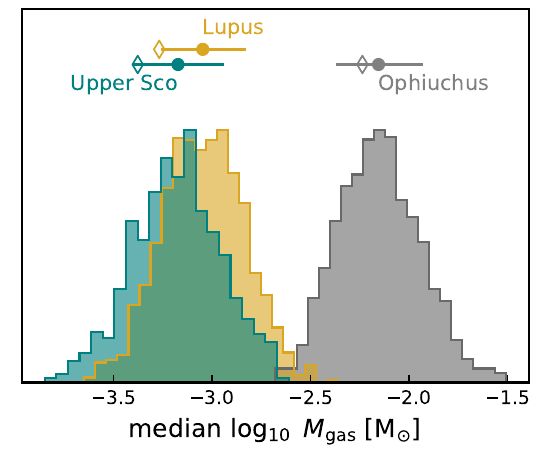}
    \caption{\label{fig: median gas masses} Distribution of median gas masses for the three regions, obtained using the Monte Carlo process described in Section \ref{sec: agepro gas masses}. Circles above each distribution shows its median and the vertical line stretches from its 16$^{\rm th}$ to its 84$^{\rm th}$ quantile. Open diamonds show the median values from \cite{AGEPRO_I_overview}. Note that the uncertainties shown here only include the observational uncertainties on the individual gas masses and not any statistical uncertainties due to the limited sample size of ten sources.}
\end{figure}

To measure the median gas mass in region while including the uncertainties on each individual gas mass we use a Monte Carlo approach. Note that \cite{AGEPRO_I_overview} calculate the median gas mass using a Kaplar-Meier estimator instead but find similar values. 
For each disk in the region we randomly draw a gas mass from its posterior distribution and compute the median for these gas masses. We repeat this process 1000 times for each star-forming region to produce a distribution of median gas masses, which are shown in Figure \ref{fig: median gas masses}. The median gas disk mass is $7.0^{+4.4}_{-2.6}\times10^{-3}\ \msun$ in Ophiuchus ($\sim0.5$ Myr; e.g. \citealt{Evans2009,AGEPRO_II_Ophiuchus}), $9.4^{+5.4}_{-3.4}\times10^{-4}\ \msun$ for Lupus ($\sim 1-3$ Myr; \citealt{comeron2008,Galli2020,AGEPRO_III_Lupus}) and $6.8^{+5.1}_{-2.8}\times10^{-4}\ \msun$ for Upper Sco ($\sim2-6$ Myr; e.g. \citealt{Pecaut2012,Briceno-MoralesChaname2023,Ratzenbock2023,AGEPRO_IV_UpperSco,AGEPRO_VIII_ext_phot_evap}), with approximately a one dex scatter in gas disk mass in all three regions. Note that the uncertainties on the medians only include the observational uncertainties on the individual gas masses and not any statistical uncertainties due to the limited sample size of ten sources and systematic uncertainties due to the models (cf Section \ref{sec: gas mass uncertainties}).

Comparing the three regions suggest that median gas mass decreases over time, dropping by a factor of $\sim10$ over 10 Myr. Most of this decrease happens between the ages of Ophiuchus and Lupus, as the median gas masses of Lupus and Upper Sco agree with each other within their uncertainties. In addition to that Figure \ref{fig: final masses} shows that the gas mass distribution of disks in Lupus looks very similar to the distribution for disks in Upper Sco. At face value this suggest that very little evolution of the gas reservoir. However, such a direct comparison between the two regions does not include survivorship bias. During the evolution of a population the low mass disks could disperse first, meaning that our sample of Upper Sco disks likely represents the large massive disks that have survived up to this point (see \citealt{AGEPRO_VII_diskpop} for a detailed discussion).  

\begin{figure}[htb]
    \centering
    \includegraphics[width=\columnwidth]{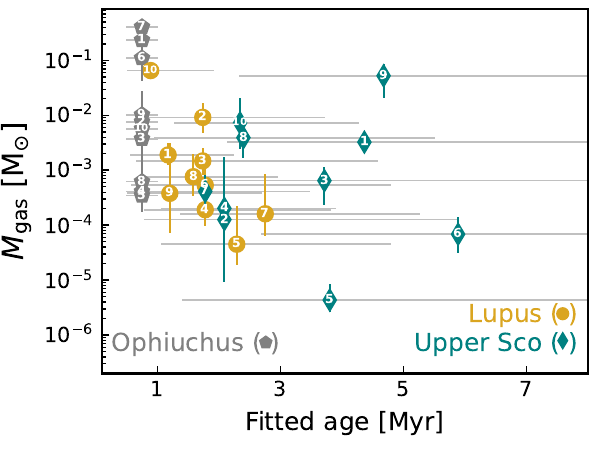}
    \caption{\label{fig: mass vs stellar age}Gas disk mass versus stellar age for individual disks}
\end{figure}

Continuing on the topic of age, Figure \ref{fig: mass vs stellar age} shows the derived gas mass set against the derived stellar ages of the individual sources (see \citealt{AGEPRO_I_overview,AGEPRO_II_Ophiuchus,AGEPRO_III_Lupus,AGEPRO_IV_UpperSco} for details on the age derivation).
Here we see a clear evolution of the median gas mass with time from Ophiuchus ($<$1\,Myr) to the older samples from the Lupus and Upper Sco regions. However, between Lupus and Upper Sco sources, there is only a slight decrease in the median gas mass (from 6.5 to 4.2$\times10^{-4}$M$_\odot$) that is within the uncertainties of the median gas masses.
Contributing to the overlap in the median gas masses is the large spread of individual disks, including a few relatively old ($>$4 Myr) Upper Sco disks with high gas masses. There is also a significant overlap in stellar age between the sources in Lupus and some of the sources in Upper. These younger Upper Sco sources tend to be more massive than the older ones. A partial age overlap can therefore help explain why the median disk masses of Lupus and Upper Sco are so similar. Considering the individual stellar ages have large uncertainties, especially some of the Upper Sco sources, constraining the gas mass evolution in disks older than 1\,Myr will need a larger sample and particularly more disks older than 5\,Myr.

Comparing the obtained gas masses to the dust masses, the middle panels of Figure \ref{fig: final masses} show the gas-to-dust mass ratios of the thirty disks. 
In Ophiuchus, we find gas-to-dust ratios from $\sim40$ to $\sim 1000$, with a median value of $\Delta_{\rm gd} \approx 250$. Notably, four Ophiuchus sources show very high gas-to-dust ratio, $\sim$1000. However, we caution that dust masses for these sources may be underestimated, as all dust masses are derived under the assumption of optically thin emission and using the same dust opacity across the three regions for a homogeneous analysis. In the most massive disks, the dust emission may be optically thick, and dust opacity in the younger disks may be different from older Class ones, as shown in \citet{AGEPRO_VI_DustEvolution} which compares dust evolution simulations with AGE-PRO continuum observations. Therefore, these high gas-to-dust ratios in the most massive disks need to be looked at with caution.  

Gas-to-dust mass ratios are lower in Lupus, ranging from $\sim13$ to $\sim 140$ with a median value of $\Delta_{\rm gd} = 40$. In Upper Sco we find higher ratios again, between $\sim13$ and $500$, with a median gas-to-dust ratio $\Delta_{\rm gd} = 116$. The increase in gas-to-dust mass ratio with age between Lupus and Upper Sco is in line with predictions from dust evolution models with no dust traps, where over time radial drift removes dust from the disk (e.g. \citealt{Weidenschilling1977,Birnstiel2012,Drazkowska2023,Birnstiel2024}). It could also be indicative of the formation of planets, where dust traced by millimeter emission has been incorporated into planetesimals and planets, thereby removing it from the disk.
For a more detailed analysis of dust evolution in the AGE-PRO sample we refer the reader to \cite{AGEPRO_VI_DustEvolution}. 

As a whole, the oscillation  of the median gas-to-dust mass ratio, i.e. a decrease from Ophiuchus to Lupus followed by a increase when moving to Upper Sco, is a somewhat puzzling surprising. It suggests that there are different timescales for the evolution of the gas and the dust. While the dust disk mass continuously decreases across different evolutionary stages, most of the gas disk mass evolution happens between Ophiuchus and Lupus, and at a faster rate than the dust mass evolution. At later stages, there is no significant change between Lupus and Upper Sco. As a result, the gas-to-dust mass ratio changes from ISM values in Ophiuchus, to lower values in Lupus, to ISM values in Upper Sco.

The bottom panels of Figure \ref{fig: final masses} show bulk CO abundances in the \xco\ and \cyo\ emitting layers. Most disks in Lupus and Upper Sco have CO abundances between $0.1\times$ and $1\times$ the ISM value ($\sim10^{-4}$; e.g. \citealt{Lacy1994}). Both regions have very similar median CO abundances, $1.4\times10^{-5}$ and $2.0\times10^{-5}$ for Lupus and Upper Sco, respectively. Given the scattering of the CO abundances in these two regions, the median values are statistically indistinguishable.  As with the gas masses, the similarities between the CO abundances in both regions are surprising. It suggests that the removal and production of 
CO from the gas reaches a steady-state after $\lesssim$1\,Myr. For example, if the removal of CO is driven by cosmic rays (e.g. \citealt{Bosman2018b, Schwarz2018}), the rate impinging on the disk would need to be high in the first Myr of the disk evolution, followed by an extended period of low cosmic ray rates. 
Significant cosmic ray production by protostellar jets (e.g. \citealt{{Padovani2016}}) during the class I stage would fit this description. However, CO removal could also driven by another process, such as UV processing of ices (e.g. \citealt{Ruaud2019, Furuya2022}) or sequestering of CO in large dust grains (e.g. \citealt{Krijt2020}). However, a full quantitative comparison between these processes and how well they describe the observations lies beyond the scope of this work.

\section{Discussion}
\label{sec: discussion}

\subsection{Correlations between disk and stellar parameters}
\label{sec: correlations between parameters}

\begin{table}[htb]
\centering
\caption{\label{tab: correlations} Correlations between disk and stellar properties }
\def\arraystretch{1.2}
\begin{tabular*}{0.95\columnwidth}{ll|c|c|c}
\hline\hline
x  &  y &  Spearman $\rho$ & $p-$value & correlation? \\
\hline
 $M_{\rm gas}$ & $T_{\rm eff}$ & $0.35^{+0.19}_{-0.09}$ &$0.07^{+0.11}_{-0.04}$ & positive (tentative)\\
 $M_{\rm gas}$ & $L_{*}$ & $0.37^{+0.18}_{-0.09}$ &$0.05^{+0.09}_{-0.03}$ & positive (tentative)\\
 $M_{\rm gas}$ & $M_{*}$ & $0.38^{+0.16}_{-0.08}$ &$0.05^{+0.08}_{-0.03}$ & positive (tentative)\\
 $M_{\rm gas}$ & $\dot{M}_{\rm acc}$ & $0.16^{+0.22}_{-0.11}$ &$0.28^{+0.26}_{-0.12}$ &-\\
 $M_{\rm gas}$ & $R_{\rm CO\ 90\%}$ & $0.68^{+0.17}_{-0.10}$ &$0.0005^{+0.0032}_{-0.0004}$ & \textbf{positive}\\
 $M_{\rm gas}$ & $R_{\rm dust\ 90\%}$ & $0.49^{+0.15}_{-0.07}$ &$0.01^{+0.03}_{-0.01}$ & \textbf{positive}\\
 $M_{\rm gas}$ & $R_{\rm c}$ & $0.46^{+0.20}_{-0.11}$ &$0.02^{+0.06}_{-0.02}$ & positive (tentative)\\
\hline
 $\Delta_{\rm gd}$ & $T_{\rm eff}$ & $0.32^{+0.24}_{-0.13}$ &$0.08^{+0.18}_{-0.06}$ & positive (tentative)\\
 $\Delta_{\rm gd}$ & $L_{*}$ & $0.10^{+0.26}_{-0.13}$ &$0.67^{+0.38}_{-0.21}$ &-\\
 $\Delta_{\rm gd}$ & $M_{*}$ & $0.38^{+0.23}_{-0.12}$ &$0.05^{+0.11}_{-0.03}$ & positive (tentative)\\
 $\Delta_{\rm gd}$ & $\dot{M}_{\rm acc}$ & $-0.07^{+0.26}_{-0.13}$ &$0.41^{+0.35}_{-0.17}$ &-\\
 $\Delta_{\rm gd}$ & $R_{\rm CO\ 90\%}$ & $0.35^{+0.23}_{-0.12}$ &$0.06^{+0.14}_{-0.04}$ & positive (tentative)\\
 $\Delta_{\rm gd}$ & $R_{\rm dust\ 90\%}$ & $0.48^{+0.21}_{-0.11}$ &$0.98^{+0.05}_{-0.04}$ &-\\
 $\Delta_{\rm gd}$ & $R_{\rm c}$ & $0.22^{+0.28}_{-0.14}$ &$0.19^{+0.31}_{-0.12}$ &-\\
\hline
 $x_{\rm CO}$ & $T_{\rm eff}$ & $-0.09^{+0.27}_{-0.13}$ &$0.36^{+0.40}_{-0.18}$ &-\\
 $x_{\rm CO}$ & $L_{*}$ & $-0.37^{+0.25}_{-0.12}$ &$0.05^{+0.13}_{-0.04}$ & negative (tentative)\\
 $x_{\rm CO}$ & $M_{*}$ & $-0.09^{+0.26}_{-0.13}$ &$0.35^{+0.39}_{-0.18}$ &-\\
 $x_{\rm CO}$ & $\dot{M}_{\rm acc}$ & $-0.36^{+0.28}_{-0.13}$ &$0.09^{+0.18}_{-0.06}$ & negative (tentative)\\
 $x_{\rm CO}$ & $R_{\rm CO\ 90\%}$ & $0.03^{+0.25}_{-0.12}$ &$0.54^{+0.40}_{-0.20}$ &-\\
 $x_{\rm CO}$ & $R_{\rm dust\ 90\%}$ & $-0.04^{+0.25}_{-0.12}$ &$0.43^{+0.39}_{-0.18}$ &-\\
 $x_{\rm CO}$ & $R_{\rm c}$ & $-0.05^{+0.30}_{-0.15}$ &$0.66^{+0.55}_{-0.31}$ &-\\
\hline\hline
\end{tabular*}
\begin{minipage}{0.85\columnwidth}
\vspace{0.1cm}
{\footnotesize{\textbf{Notes:} Correlations with $p \leq0.01$ are assumed to be statistically significant. Those with $p - \sigma_p \leq 0.05 \leq p+\sigma_p$ are denoted as tentative}}
\end{minipage}
\end{table}

\begin{figure*}[ht]
    \centering
    \begin{minipage}{0.49\textwidth}
    \includegraphics[width=\textwidth]{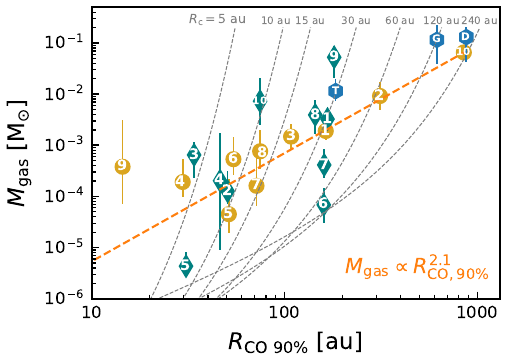}    
    \end{minipage}
    \begin{minipage}{0.49\textwidth}
    \includegraphics[width=\textwidth]{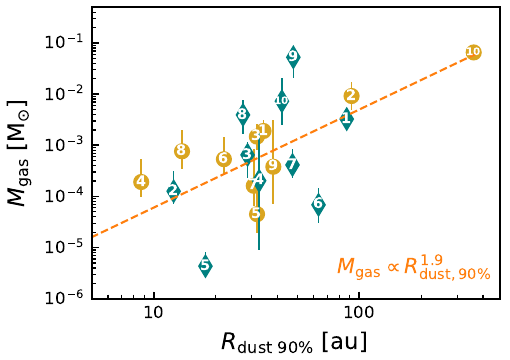}    
    \end{minipage}
    \begin{minipage}{0.49\textwidth}
    \includegraphics[width=\textwidth]{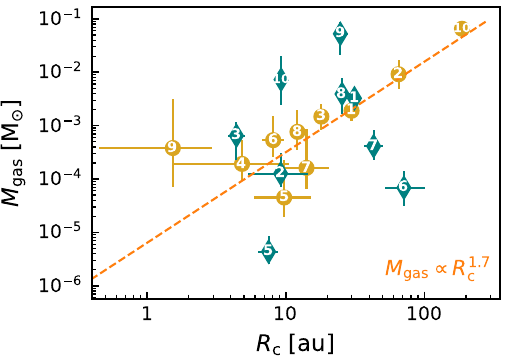}    
    \end{minipage}
    \caption{\label{fig: Mgas-Rco-Rc correlation}\textbf{Top left:} The correlation between \mgas\ and \rgas, the latter being obtained from the best fit Nuker profile of the \co\ 2-1 emission (see \citealt{AGEPRO_XI_gas_disk_sizes} for details). Black circles and purple diamonds represent sources in Lupus and Upper Sco, respectively. Blue hexagons shows the three disks with HD-based gas masses, TW Hya, DM Tau, and GM Aur, denoted by ``T'', ``D'', and ``G''. The orange dashed line shows a powerlaw fit to the observations. Shown in gray are analytically calculated \rgas\ for different combinations of \mgas\ and \rc\ using the analytical equation for \rgas\ from \cite{Trapman2023}. Here we used $L_*=0.3\ \mathrm{L}_{\odot}$, $\gamma=1$, and $\xi =0$ (see \citealt{Trapman2023} for details).
    \textbf{Top right:} \rdust\ versus \mgas\ for the same sources. Note that the \rdust\ shown here were measured from the intensity profiles fitted to the visibilities (see \citealt{AGEPRO_X_substructures} for details).
    \textbf{Bottom:} The correlation between \mgas\ and \rc\ analytically derived from \mgas\ and \rgas\ following the method from \cite{Trapman2023} (see Appendix \ref{app: Rc derivation} for details.) }
\end{figure*}

Having derived bulk disk properties $(\mgas,\Delta_{\rm gd}, \abu)$, an interesting question to ask is how these correlate with other properties of the disk and of the host star. 
Here we compare the gas masses derived for Lupus and Upper Sco to $T_{\rm eff}, L_*, M_*, \dot{M}_{\rm acc}$, $R_{\rm CO,\ 90\%}$, and $R_{\rm dust,\ 90\%}$ (\citealt{alcala2014,alcala2017,Alcala2019,Manara2020,UpperSco_followup}; see \citealt{AGEPRO_I_overview} for details).
Note that here we take the \rdust\ from \cite{AGEPRO_X_substructures} and the \rgas\ from \cite{AGEPRO_XI_gas_disk_sizes}, who correct for the convolution with the observational beam by modeling the emission in the visibility and image plane, respectively.

Visual inspection reveals positive correlations between \mgas\ and the two radii (\rgas,\rdust) (see Figure \ref{fig: Mgas-Rco-Rc correlation}), but for the other parameters it is not clear if they correlate with \mgas\ (cf. Figure \ref{fig: correlations} in Appendix \ref{app: correlations}). To test this statistically, we carried out Spearman rank test for correlations between $x\in\{\mgas,\Delta_{\rm gd}, \abu\}$ and $y\in\{T_{\rm eff}, M_*, L_*, R_{\rm CO,\ 90\%},R_{\rm dust,\ 90\%}\}$. To include the, in some cases substantial, uncertainties on \mgas, $\Delta_{\rm gd}$ and \abu, we used a Monte Carlo approach. Briefly, for each source we draw a value for $x$ from their posterior distribution of the relevant parameter (see Section \ref{sec: measuring gas masses with MCMC}) and use these plus the $y$ for each source to compute the Spearman $\rho$ and the associated p-value. By repeating this exercise 10000 times we build up a distribution of Spearman $\rho$s from which we can obtain its median value and uncertainties, taken to be the 16$^{\rm th}$ and 84$^{\rm th}$ quantile (see Figures \ref{fig: Mgas MC correlation}, \ref{fig: gdratio MC correlation}, and \ref{fig: xco MC correlation} in Appendix \ref{app: correlations}).

Table \ref{tab: correlations} summarizes the results of the Spearman rank tests. In total we find two strong ($\langle p \rangle \lesssim 0.01$) positive correlations, between \mgas and \rgas, and between \mgas\ and \rdust. If we fit the relation between \mgas and $R_{\rm CO,\ 90\%}$ with a powerlaw we obtain $\mgas \propto R_{\rm CO,\ 90\%}^{2.1}$. That more massive disks have a larger observed gas disk size is to some extent expected, as the gas mass is one of the parameters that determines the observed gas disk size (see, e.g. \citealt{Trapman2020,Toci2023,Trapman2023,Zagaria2023}). Figure \ref{fig: Mgas-Rco-Rc correlation} shows the \mgas-\rgas\ dependence, calculated for different values of \rc\ using the analytical formula for \rgas\ presented in \cite{Trapman2023}. Based on the figure the disks in the sample have \rc\ between $\sim5$ and $\sim60$ au, with the exception of Lupus 10, which has a much larger \rc. In appendix \ref{app: Rc derivation} we use the same analytical formula to derive \rc\ for each of the disks in our sample. Figure \ref{fig: Mgas-Rco-Rc correlation} shows that \mgas\ is also positively correlated with \rc, suggesting that more massive disks are also physically more extended, i.e., their mass reservoir is spread out over a larger area. 

In the right panel of Figure \ref{fig: Mgas-Rco-Rc correlation} we also see a positive correlation between \mgas\ and \rdust. 
\cite{AGEPRO_VI_DustEvolution} demonstrate that to explain the continuum observations of the AGE-PRO sample substructures in the disks are needed. If substructures are present in the AGE-PRO sources (see also \citealt{AGEPRO_X_substructures}), \rdust\ traces the location of the farthest away dust trap. The positive correlation between \mgas\ and \rdust\ is therefore indirect evidence that more massive disks can develop substructures farther away from the star.

Several other parameter combinations show tentative positive or negative correlations (see Table \ref{tab: correlations}). For example, there is some evidence for a positive correlation between \mgas\ and stellar mass, but its median $p-$value is 0.06, meaning there is a reasonable chance that the correlation is not statistically significant. Is is not unsurprising, as by design the AGE-PRO sample has a narrow stellar mass and spectral type range and lacks the baseline with which previous correlations were found (e.g. \citealt{Pascucci2016,Testi2022}).
The forthcoming ALMA cycle 9 large program DECO (2022.1.00875.L, PI: Ilse Cleeves), which covers a similar spectral set-up as AGE-PRO, will significantly expand the stellar mass and spectral type baselines, meaning a combined analysis of these two programs will significantly increase our understanding of the dependence of disk properties such as gas disk mass on the properties of their host star.

\subsection{Estimating disk lifetimes}
\label{sec: disk lifetimes}

\begin{figure}[ht]
    \centering
    \includegraphics[width=\columnwidth]{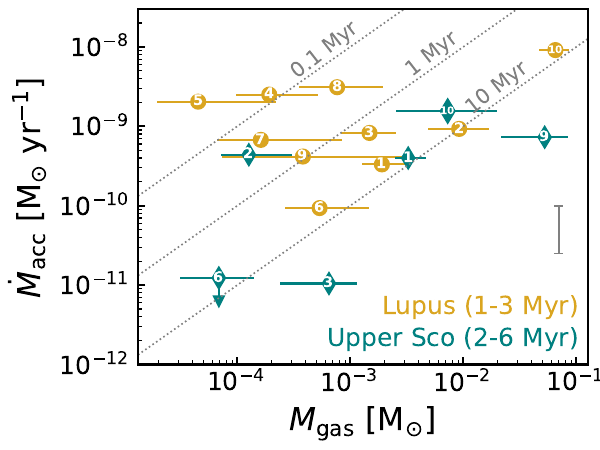}
    \caption{\label{fig: mass vs macc}Disk mass versus stellar mass accretion rate for disks in Lupus (black) and Upper Sco (purple). The gray errorbar shows typical uncertainty of the accretion rate (see \citealt{alcala2017,Manara2020}). Dotted gray lines show different disk lifetimes, defined as $\mdisk/\dot{M}_{\rm acc}$. }
\end{figure}

One of key parameters for planet formation is the lifetime of the protoplanetary disk. A rough estimate of the disk lifetime can be obtained from the ratio of the disk mass and the stellar accretion rate, $\mgas/\macc$ (see, e.g., \citealt{Jones2012,Rosotti2017,Lodato2017}). Of the 30 disks in the AGE-PRO sample, all ten disks in Lupus have measured accretion rates from near-UV/optical spectroscopy \citep{alcala2014,alcala2017,Alcala2019} and six of the ten disks in Upper Sco have reported values as well \citep{Manara2020,Fang2023}. 

Figure \ref{fig: mass vs macc} sets \mgas\ from this work against the observed stellar mass accretion rate. Six of the ten disks in Lupus occupy the part of the \macc-\mdisk\ plane with disk lifetimes of 1-10 Myr, which is consistent with the age of the star-forming region. Two sources, Lupus 7 and 8 lie between 0.1 and 1.0 Myr and the remaining two sources, Lupus 4 and Lupus 5, have estimated disk lifetimes $\lesssim 0.1$ Myr, much shorter than the age of the region. Especially for the latter two sources, independent on the mechanism driving their evolution, these disks should have already been dispersed based on their current gas mass and stellar accretion rate, assuming that the accretion rate has been constant during the disk's evolution. For Upper Sco five of the six disks have disk lifetimes consistent with the age of Upper Sco, but one disk, Upper Sco 2, has a disk lifetime of $\lesssim1$ Myr and should have already been dispersed. 

This raises the question, how are the three disks with lifetimes shorter than $\sim10\times$ their star-forming region age still around? We will discuss three possible sets of explanations: the gas masses are underestimated, the stellar accretion rates are overestimated, and the lifetime computed from the disk mass and stellar accretion disk mass does not accurately represent the true lifetime of the disk. \cite{Grant2023} recently showed that a similar disk lifetime problem exists for disks Herbig Ae/Be stars, which they suggest could partially be explained by a bias in the classification of Herbig stars. A similar bias is less likely for T-Tauri stars, but it could be a contributing factor.  

The first possibility is that the gas masses for these sources are underestimated. We have dedicated Section \ref{sec: gas mass uncertainties} to discussing different scenarios by which the gas masses presented in this work could be underestimated. Looking at Figure \ref{fig: mass vs macc}, requiring that the disk lifetime be equal or longer than the star-forming region age would require increasing the gas mass by a factor $\sim 10-50$. It is interesting to note here that all of these disks are also compact ($\rgas \lesssim 100$ au; \citealt{AGEPRO_III_Lupus,AGEPRO_IV_UpperSco}). Whether this indicates a different evolutionary path or simply highlights a difficulty with measuring gas masses and/or stellar mass accretion rates for compact disks remains unclear.  

Alternatively, the stellar mass accretion rates could be overestimated. The accretion rate is based on the accretion luminosity ($L_{\rm acc}$), which was obtained by fitting the observed UV spectrum of each source with a template spectrum of a non-accreting class III star plus a slab of hydrogen (see \citealt{Manara2016a,Hartmann2016,alcala2017} for details on how accretion columns are modeled). The mass accretion is then computed using $\macc\approx 1.25 L_{\rm acc} R_* / G \mstar$, where $R_*$ and \mstar\ are the stellar radius and mass, respectively (see, e.g., \citealt{alcala2017,Manara2020}).

The main sources of uncertainty for \macc\ are the extinction, the choice of stellar template, and the uncertainties on the stellar parameters, the latter of which depends on the assumed stellar evolutionary track. In aggregate, \citealt{alcala2014} estimate that this leads to a $\sim0.35$ dex, or a factor of $\sim2.3$, total uncertainty on \macc\ (see also \citealt{alcala2017,ManaraPPVII}). Also, we note that \cite{Pittman2022} have used near-UV data to show that accretion rates derived with the before discussed methodology can be underestimated by a factor of 3-4, which would lower the disk lifetime even more. This suggests that the uncertainty on the stellar mass accretion rate alone cannot explain the short disk lifetimes for five of our sources.

Finally, the fault could be in our interpretation of \mdisk/\macc\ as the disk lifetime. Stellar accretion is known to be a stochastic rather than a smooth process (see \citealt{Fischer2022} for a recent review), meaning the currently observed stellar accretion rate need not be the same as the disk-lifetime integrated accretion rate. In addition, accretion bursts of a factor $\sim10$, the approximate factor the lifetime-averaged \macc\ would need to be lower than the current value to provide a disk lifetime of $\sim 1$ Myr, are within the expected range of variability (e.g. \citealt{Fischer2022}). However, we do not expect accretion bursts to be the general explanation, as they are uncommon and our samples was not selected to favor strong accretors.

\subsection{Could the gas masses be underestimated?}
\label{sec: gas mass uncertainties}

\begin{figure}[ht]
    \centering
    \includegraphics[width=\columnwidth]{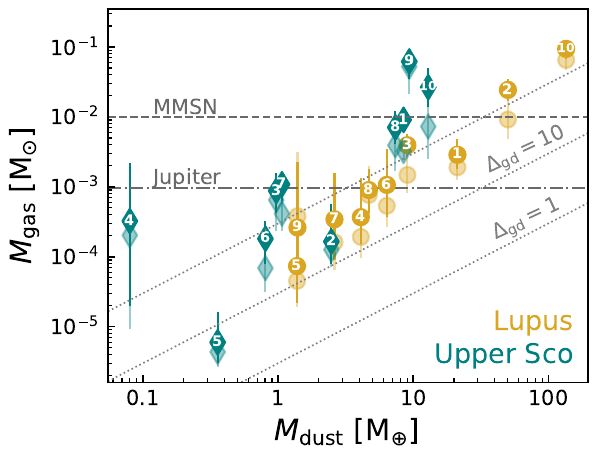}
    \caption{\label{fig: effect of low CR} Lupus and Upper Sco gas masses derived using either $\zet \in [10^{-19},10^{-17}] \ \mathrm{s}^{-1}$ (clear markers) or $\zet \in [10^{-20},10^{-19}] \ \mathrm{s}^{-1}$ (solid markers) as a prior for the cosmic ray ionization rate (see Section \ref{sec: measuring gas masses with MCMC} for details). Note that in most cases decreasing \zet\ increases \mgas.}
\end{figure}

As discussed previously in Section \ref{sec: disk lifetimes} disk lifetimes (based on estimated disk masses and accretion rates) for some disks are lower than expected and apparently inconsistent with the age of the star-forming region in which these disks reside. It is therefore worthwhile to discuss whether the disks masses measured in this work are underestimated. In this section we will examine a few possible reasons.

\textbf{Low cosmic ionization rate (\zet):} as an ion, the abundance of \nhp\ is directly affected by the amount of ionization in the disk. For \nhp, which is abundant close to the disk midplane, the ionization is dominated by cosmic rays. This creates a link between \zet\ and \mgas: a disk with a lower \zet\ requires a larger \mgas\ to reproduce the observed \nhp\ emission. Or in the case of an upper limit, a lower \zet\ would mean that a higher gas mass is still consistent with the observed upper limit. In our analysis we have assumed a wide range of cosmic ray ionization rates, $\zet \in [10^{-19}$, $10^{-17}]\ \mathrm{s}^{-1}$. However, work by \cite{Cleeves2015} on the disk around TW Hya suggest that the cosmic ray ionization rate in disks could be even lower, $\zet\sim10^{-20}\ \mathrm{s}^{-1}$. If this is also the case in some or all of the AGE-PRO disks our gas masses could be underestimated. 

In order to quantify this effect we compute \nhp\ fluxes for all models assuming $\zet=10^{-20}\ \mathrm{s}^{-1}$ and derive the gas masses using a low \zet\ prior, $\zet \in [10^{-20}$, $10^{-19}]\ \mathrm{s}^{-1}$, consistent with the range found by \cite{Cleeves2015}. These new gas masses are compared to our canonical gas masses in Figure \ref{fig: effect of low CR}. In all cases the low-\zet-gas mass is indeed larger the canonical gas mass, usually by a factor $\sim 2-4$. This is in particular the case for disks where \nhp\ was detected, as a detection provides a harder constraint while fitting the observations. 
As discussed in Section \ref{sec: measuring gas masses with MCMC}, when \nhp\ is not detected the gas mass upper limit is set by the maximum CO depletion that is consistent with the \nhp\ non-detection. With a lower cosmic ray ionization rate the overall \nhp\ abundance is lower, meaning that more massive disks with a lower CO abundance still produce a \nhp\ flux that is consistent with the observed upper limit.
To summarize, assuming a lower cosmic ray ionization rate does increase \mgas, but it cannot fully explain the short disk lifetimes that were found in Section \ref{sec: disk lifetimes}.

\textbf{Low N$_2$ abundance:} One of the underlying assumptions in our \nhp\ chemical network is that N$_2$ is the main carrier of Nitrogen in protoplanetary disks. If this is not the case or if N$_2$ is underabundant in a manner similar to CO then our models would be overestimating the emission of \nhp, meaning that the gas masses derived here are underestimated. So far studies seem to suggest that disks are not underabundant in nitrogen (e.g., \citealt{Cleeves2018}) and that other species like ammonia are not a dominant nitrogen carrier (e.g., \citealt{Salinas2016}). These results are consistent with recent modeling results, which show that degree of nitrogen depletion from the warm molecular layer is much lower than that of carbon and oxygen \citep{Furuya2022}.
However, the sample of sources with a well-studied nitrogen reservoir is still small and there is some evidence from solar system bodies that some fraction of the available nitrogen could be locked up in ammonia salts (e.g. \citealt{Boogert2015,altwegg2020}).

Lowering the N$_2$ abundance is similar to lowering \zet, in that both reduce the amount of \nhp, which in turn means that a larger gas mass is needed to match the observations. Therefore, if the N$_2$ abundance is indeed lower than what is assumed in our models we would expect that the effect on the gas mass is similar to the low \zet\ experiment discussed above. For disks with \nhp\ detections the gas mass will increase by a factor inversely proportional to the amount of N$_2$ removed, but for disks with \nhp\ upper limits the effect are likely minimal. Since the disks with the lowest disk lifetimes also do not have detected \nhp\ emission, neither lowering the cosmic ray ionization rate nor reducing the bulk N$_2$ abundance is expected to solve the discrepancy between the derived disk lifetimes and the age of the star-formation region.

\textbf{Model assumptions:} The disk models used in this work to derive gas masses from the observations depend on a large number of parameters and assumptions. A number of these, for example the disk size, were varied within the model grid and their effect was therefore included in the gas mass measurement, but many others were not. For example, all models have an exponentially tapered gas surface density , whereas we know at least Upper Sco 1 to have a gas gap \citep{Sierra2024}. This gas surface density profile is expected for turbulently evolving disks (e.g. \citealt{LyndenBellPringle1974,Hartmann2000,Tabone2022a}), but observational constraints on the gas surface density in the outer disk are limited (e.g., \citealt{Dullemond2020,Zhang2021MAPS}). Moreover, this theoretically predicted gas surface density profile neglects the effects of external photo-evaporation, an assumption which likely cannot be made for the disks in Upper Sco (see \citealt{AGEPRO_VIII_ext_phot_evap} for a detailed discussion on the effect on external photo-evaporation on the AGE-PRO Upper Sco disks.)

As another example, the models use a Gaussian vertical density distribution expected for vertically isothermal disks, but it is well known that disks are not vertically isothermal. \cite{Ruaud2022} recently showed that this assumption can have a significant effect on the CO isotopologue line fluxes, although, as \cite{Bosman2022} argue, their analysis is also not free of its own underlying assumptions. With these points in mind, it is therefore interesting to note that for the Lupus disks in AGE-PRO \cite{AGEPRO_III_Lupus} show that gas masses obtained using the \cite{Ruaud2022} models are within a factor $\sim3-5$ from the values reported in this work (see \citealt{AGEPRO_III_Lupus} for details). 

Added up, these and the other unmentioned model assumptions will have an effect on the measured gas masses. While it is hard to determine how much they would affect the gas mass estimates or in what direction they would change them, assuming an additional uncertainty of a factor $\sim 2-3$ is likely advisable based on previous benchmarking of similar codes (e.g. \citealt{Rollig2007}).

\textbf{Compact disks:} A unifying characteristic of the disks with short disk lifetimes is that all of them are compact ($\rgas\lesssim 100$ au, a measurement which still includes convolution with a beam that is $\sim56$ au in diameter). \cite{Miotello2021} proposed that compact, massive disks provided an alternative explanation for the faint \co\ and \xco\ line emission observed in Lupus (see \citealt{ansdell2016,ansdell2018}). 
Fitting the observations with our model grid, which includes disks with $\rc = 1$ au and \mgas\ up to 0.1 \msun, shows that this cannot be the sole explanation. Examining the \rc-posteriors of the individual disks reveals that the fit does favor compact disks ($\rc \approx 1-25$ au; see Appendix \ref{app: MCMC corner plots }), but compact massive disks can be ruled out either by the need to reproduce \rgas\ (which sets a minimum on \rc) and/or the requirement for \nhp\ to remain undetected (which set a maximum on the gas mass). The latter can also be seen in the fact that the \zet-posterior highly favors a low cosmic ray ionization rate for these disks. 

That being said, however, these compact disks have also been considerably less well studied than their larger, brighter counterparts (e.g. \citealt{Pegues2021,Kurtovic2021}). Most of our current understanding about protoplanetary disks is therefore based on the properties of large disks and the models that try to reproduce those. For example, the \nhp\ + \cyo\ gas mass measurement was tested against the three disks with HD detections, all of which are large disks \citep{Trapman2022b}. If compact disks are not simply scaled-down large disks and instead have different physical structure and/or chemical composition, their gas masses estimates would be inaccurate and could be underestimated. 
This suggests the need for future continuum and gas observations focused on more compact disks $(\rc \lesssim 15\ \mathrm{au})$. Such compact disks are much more common than previously thought and crucial to our understanding of planet formation in nearby star-forming regions.

\section{Conclusions}
\label{sec: conclusions}

In this work we examine line observations of $^{12}$CO, \xco, \cyo, C$^{17}$O 2-1, and \nhp\ 3-2 of the AGE-PRO large program and compared them to a large grid of protoplanetary disk models. We combine a piece-wise linear interpolation of the model \xco, \cyo, and \nhp\ line fluxes and \co\ gas disk size with MCMC routine to derive gas masses and their associated uncertainties for the twenty AGE-PRO disks in the Lupus and Upper Sco star-forming regions. For the ten AGE-PRO disks in the Ophiuchus star-forming region we compare the observed C$^{17}$O 2-1 line fluxes to a wide range of models to estimate their gas masses. Our conclusions are as follows:

\begin{itemize}
    \item Using our MCMC routine we measure the gas masses of class II disks with a typical precision of $\sim0.35 {\rm~dex}$. Uncertainties for the younger class I sources are much larger due to uncertainties in their temperature structure and difficulties due to cloud contamination and envelope contribution to the line fluxes. 
    The accuracy of the gas masses is harder to quantify, but analysis in \cite{AGEPRO_III_Lupus} suggests an overall good agreement, i.e. within a factor of $\sim3-5$, between gas masses derived using different models and methods.
    
    \item The median gas mass of the three regions decreases over time, from $7.0^{+4.4}_{-2.6}\times10^{-3}\ \msun$ in Ophiuchus ($\lesssim$ 1 Myr) to $9.4^{+5.4}_{-3.4}\times10^{-4}\ \msun$ for Lupus ($\sim$1-3 Myr) and $6.8^{+5.1}_{-2.8}\times10^{-4}\ \msun$ for Upper Sco ($\sim$2-6 Myr), with $\sim1$ dex scatter in gas mass in each region. Almost all of the gas mass evolution occurs between the ages of Ophiuchus and Lupus, given that the median gas masses of Lupus and Upper Sco agree within their uncertainties and the gas mass distributions of Lupus and Upper Sco appear very similar. This suggest limited evolution of the gas mass distribution at later times, but it might be due to survivorship bias, i.e., disk dispersal selects for the most massive disks.

    \item The median gas-to-dust mass ratio of disks in Upper Sco $(\langle\Delta_{\rm gd}\rangle \approx 120)$ is on a factor $\sim 3$ higher than that of disks in Lupus $(\langle\Delta_{\rm gd}\rangle \approx 40)$. Given that the gas masses of Lupus and Upper Sco are similar, this evolution in the gas-to-dust mass ratio is driven by the evolution of the dust, likely efficient inward drift of pebbles and/or the formation of planetesimals and planets. 

    \item Median bulk CO abundances in the warm molecular layer are about an order of magnitude below the ISM value for both disks in Lupus and Upper Sco. This suggests that CO is efficiently removed from the warm molecular layer in the first $\sim1$ Myr, followed by a steady state that lasts until the end of the gas disk lifetime.

    \item We find strong correlations between \mgas\ and the gas disk size \rgas\ and between \mgas\ and the dust disk size \rdust. The former is in some ways expected as \mgas\ plays an important role in setting \rgas, but further analysis shows that there is also a correlation between \mgas\ and the characteristic radius \rc, suggesting that more massive disk are physically more extended than lower mass disks.
    The correlation between \mgas\ and \rdust\ could be indirect evidence that more massive disk can develop substructures farther away from the star.

    \item Most AGE-PRO disk have disk lifetimes, computed as \mgas/\macc, that are similar or longer than the age of the star-forming region in which they reside. However, a number of disks with measured \macc\ have a disk life time that is $\gtrsim10\times$ shorter than their star-forming region age. This suggests that they have not maintained their current accretion rate throughout their lifetime, that their gas mass is underestimated, or that they will disperse in the near future.
\end{itemize}

\begin{acknowledgements}
We thank the anonymous reviewer for their helpful comments. L.T. and K. Z. acknowledge the support of the NSF AAG grant \#2205617. 
G.R. acknowledges funding from the Fondazione Cariplo, grant no. 2022-1217, and the European Research Council (ERC) under the European Union’s Horizon Europe Research \& Innovation Programme under grant agreement no. 101039651 (DiscEvol). Views and opinions expressed are however those of the author(s) only, and do not necessarily reflect those of the European Union or the European Research Council Executive Agency. Neither the European Union nor the granting authority can be held responsible for them.
P.P. and A.S. acknowledge the support from the UK Research and Innovation (UKRI) under the UK government’s Horizon Europe funding guarantee from ERC (under grant agreement No 101076489).
A.S. also acknowledges support from FONDECYT de Postdoctorado 2022 $\#$3220495.
B.T. acknowledges support from the Programme National “Physique et Chimie du Milieu Interstellaire” (PCMI) of CNRS/INSU with INC/INP and co-funded by CNES.
I.P. and D.D. acknowledge support from Collaborative NSF Astronomy \& Astrophysics Research grant (ID: 2205870).
C.A.G. and L.P. acknowledge support from FONDECYT de Postdoctorado 2021 \#3210520.
L.P. also acknowledges support from ANID BASAL project FB210003.
L.A.C and C.G.R. acknowledge support from the Millennium Nucleus on Young Exoplanets and their Moons (YEMS), ANID - Center Code NCN2021\_080 and 
L.A.C. also acknowledges support from the FONDECYT grant \#1241056.
N.T.K. acknowledges support provided by the Alexander von Humboldt Foundation in the framework of the Sofja Kovalevskaja Award endowed by the Federal Ministry of Education and Research.
K.S. acknowledges support from the European Research Council under the Horizon 2020 Framework Program via the ERC Advanced Grant Origins 83 24 28.
All figures were generated with the \texttt{PYTHON}-based package \texttt{MATPLOTLIB} \citep{Hunter2007}. This research made use of Astropy,\footnote{http://www.astropy.org} a community-developed core Python package for Astronomy \citep{astropy:2013, astropy:2018}, and Scipy \citep{2020SciPy-NMeth}.
\end{acknowledgements}

\bibliographystyle{aasjournal}
\bibliography{references}

\begin{appendix}

\section{Line flux comparisons}
\label{app: flux comparison}

In the \S 2.1, we first compute gas temperature and CO isotopologue abundances from an ISM level of C/H and O/H elemental ratio. We then mimic the CO depletion by scaling down the peak CO abundance and calculating line fluxes. This cost-effective approach allows for efficient exploration of a broad range of parameter space. Although isotopologue-selective photodissociation is included in the initial models, when the CO gas column becomes very low \citep{Miotello2014}, simply scaling the abundances may over-predict the abundances of rare isotopologues like C$^{18}$O, because these rare isotopologues experience less self-shielding than the more abundant ones. To assess this overestimation, we run additional models with initial low CO abundance and low gas disk masses and compare the line fluxes with those from the scaling method. Specifically, we adopt C/H ratios of 1.4$\times10^{-5}$ and 1.4$\times10^{-6}$, 
representing CO depletion by a factor of 10 and 100 times relative to the ISM levels. The O/H ratio in these models are also scaled similarly. We test disk gas masses of 10$^{-4}$- 10$^{-3}$\,M$_\odot$,  characteristic radii R$_c$ of 5-60\,au, a L$_\star$=0.5\,L$_\odot$, a gas-to-dust mass ratio of 100, a flat disk structure, and an evolved dust population, which spans the bulk parameter range of the AGE-PRO Lupus and Upper Sco sample. These models use cosmic-ray ionization rate of 10$^{-17}$ s$^{-1}$ and run for 1\,Myr. 

\begin{figure*}
\centering
\includegraphics[width=0.7\textwidth]{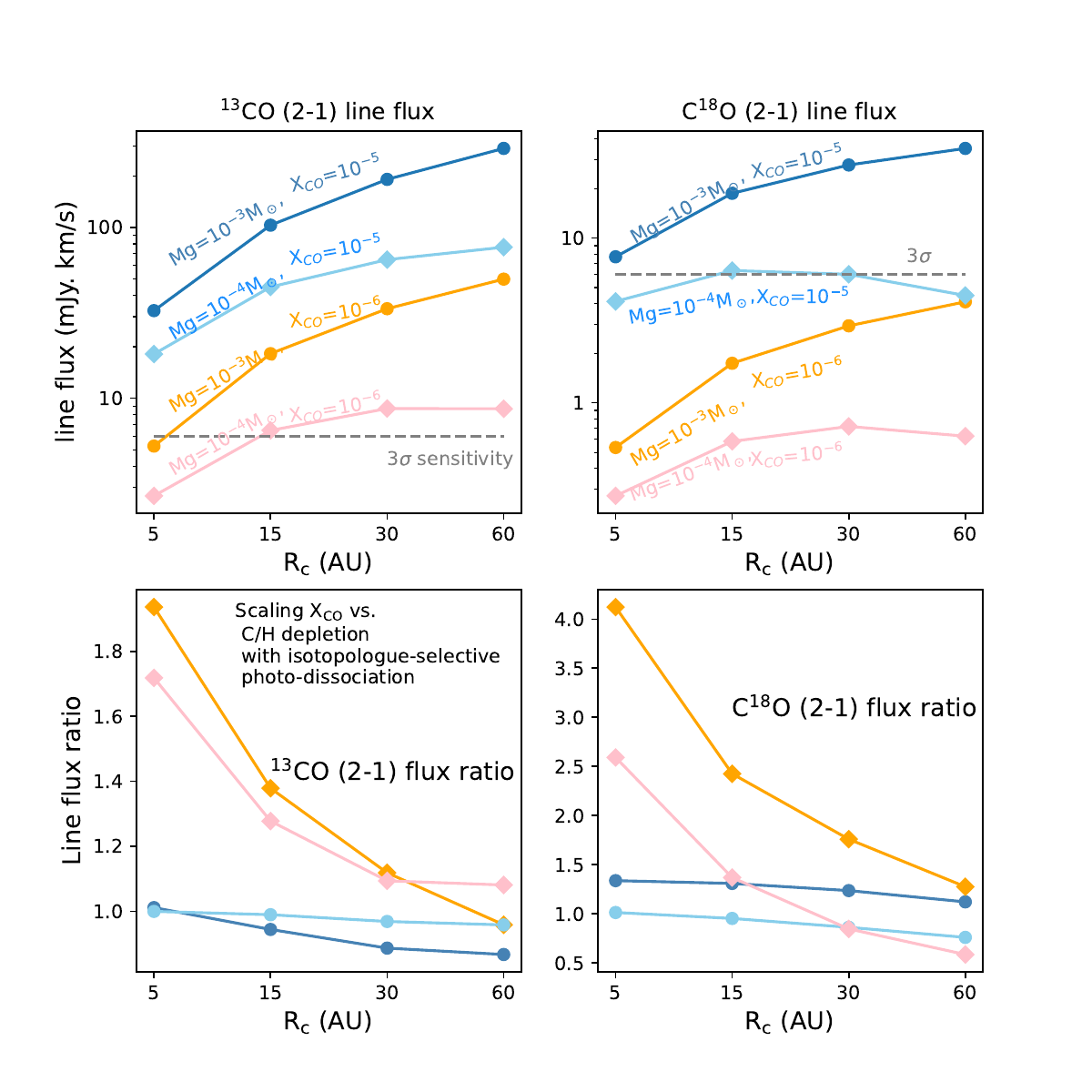}    
\vspace{-2cm}
\caption{Top row: line fluxes of models with low C/H and O/H elemental ratios. The grey dash line indicates the 3$\sigma$ detection level of AGE-PRO observations. Bottom row: line flux ratios between the models with initial low elemental ratios with models with scaled CO abundance structures. \label{fig:flux_comparison}}
\end{figure*}

Figure~\ref{fig:flux_comparison} shows the line fluxes from the additional models, and the line flux ratios of \xco\ and \cyo\ (2-1), calculated by dividing fluxes from the initial low abundance models by those from the scaling models. For $x_{\rm CO}$ = 10$^{-5}$ case, both \xco\ and \cyo\ flux ratios are consistent with 30\% for gas disk masses of 10$^{-4}$ and 10$^{-3}$\,M$_\odot$. For the extremely depleted $x_{\rm CO}$ = 10$^{-6}$, the \xco\ flux ratios stay within 30\%, except for the most compact disks (R$_c$=5\,au). For \cyo\ in $x_{\rm CO}$ = 10$^{-6}$ models, the line ratios vary significantly, ranging between 1-4. However, these cases have \cyo\ line fluxes much lower than the 3$\sigma$ flux detection limits of our AGE-PRO measurements. In these cases, their gas disk masses are mainly constrained by their \xco\ flux. Therefore, for the line flux ranges probed by the AGE-PRO sample, we consider the line flux predictions from abundance scaling method valid.

\section{Disk mass retrieval comparison with HD-based gas mass measurements}
\label{app: HD comparison}

\begin{figure*}
\centering
\includegraphics[width=\textwidth]{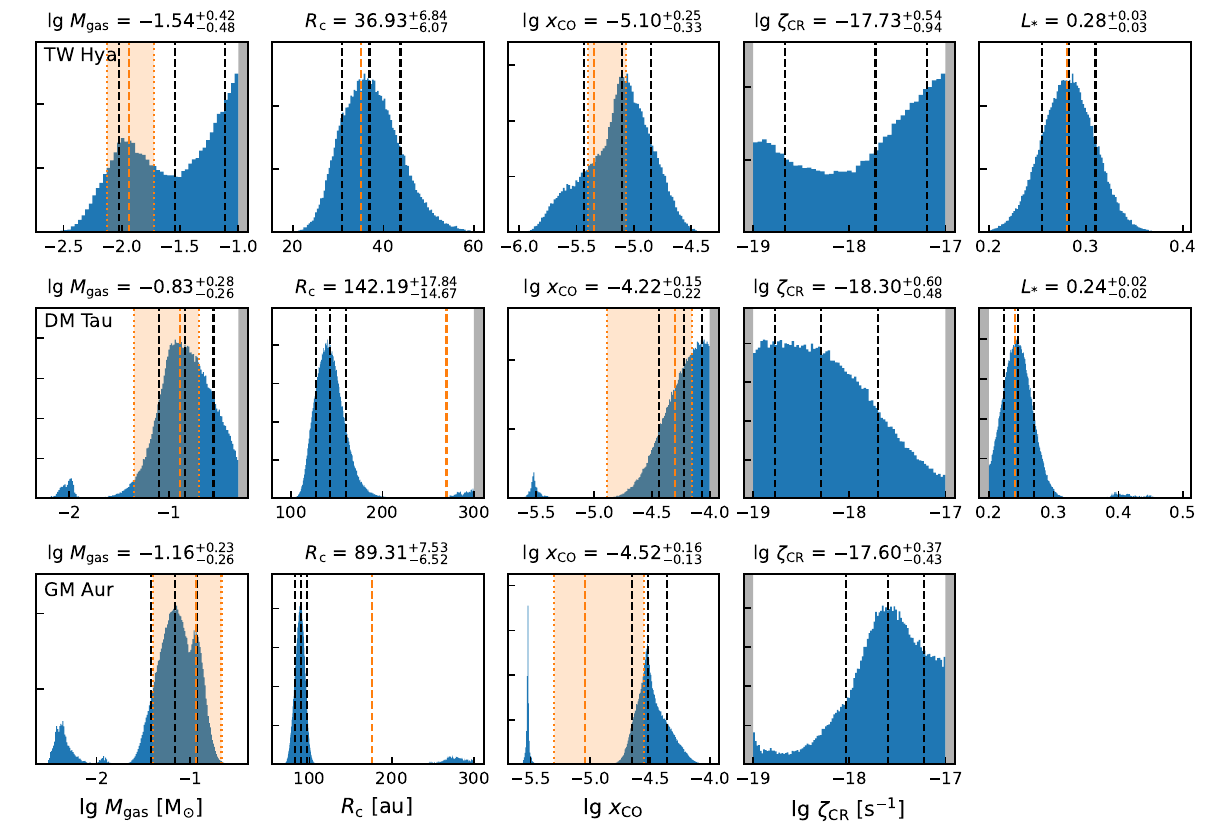}    
\caption{\label{fig: HD comparison} Posterior distribution of \mgas, \rc, \abu, $\zeta_{\rm cr}$, and $L_*$ for TW Hya (top), DM Tau (middle), and GM Aur (bottom). Black vertical dashed lines show the 16$^{\rm th}$, 50$^{\rm th}$, and 84$^{\rm th}$ quantile of the distribution. Shown in orange are the HD-based gas mass and the corresponding \abu\ from \cite{Trapman2022b}, with the dashed and dotted lines showing the measured value and its uncertainties, respectively. The orange dashed lines in the \rc\ and $L_*$ panels show the values used in the source specific models used to derive the HD-based gas mass (cf \citealt{Kama2016,Zhang2019,Zhang2021MAPS,Schwarz2021}). 
}
\end{figure*}

\begin{table}[htb]
\centering
\caption{Source information}
\begin{tabular*}{0.87\columnwidth}{lcc|cccc|c}
\hline
\hline
Source & $L_{\star}$&  dist & 1.3 mm & \xco\ $2-1$  & \cyo\ $2-1$ & \nhp\ $3-2$ & HD-based \mgas \\
&  [L$_{\odot}$] & [pc] &  [mJy] & [Jy km/s] & [Jy km/s] & [Jy km/s]  & [\msun]\\
\hline
TW Hya &  0.28&   59 & - & $1.89\pm0.18$ & $0.57\pm0.06$ & $2.0\pm0.21$ & $1.15\times10^{-2}\ \ (7.5\times10^{-3}, 1.9\times10^{-2})$ \\
DM Tau &  0.24&  145 & - & $4.2\pm0.076$ & $1.11\pm0.11$ & $1.286\pm0.17$ & $1.3\times10^{-1}\ \ (4.5\times10^{-2}, 2\times10^{-1})$ \\
GM Aur &  1.2 &  159 & - & $5.028\pm0.048$ & $1.092\pm0.039$ &$1.487\pm0.18$ & $1.15\times10^{-1}\ \ (4\times10^{-2}, 2.15\times10^{-1})$ \\
\hline
Refs &  \multicolumn{2}{c}{(1,2,3)} & & (4,5) & (6,7,5) & (8,9) & 10 \\
\end{tabular*}
\label{tab: observations HD test}
\vspace{0.1cm}
\begin{minipage}{0.8\textwidth}
{\em Notes}:  ALMA line fluxes include a 10\% systematic flux uncertainty.
\em{References: (1) \cite{Andrews2012}, 
                (2) \cite{KenyonHartmann1987},
                (3) \cite{GaiaDR2_2018},
                (4) \cite{Favre2013},
                (5) \cite{Oberg2021MAPS},
                (6) \cite{calahan2021},
                (7) \cite{Bergner2019},
                (8) Qi et al., in prep.,
                (9) \cite{Qi2019},
                (10) \cite{Trapman2022b}
                } 
\end{minipage}
\end{table}

In this work we demonstrated an approach for deriving gas disk masses using an MCMC fit of the \xco\ 2-1, \cyo\ 2-1, \nhp\ 3-2 line fluxes and the 1.3 millimeter continuum flux. Since this is the first application of this approach, it is highly worth testing it against an independent method of measuring gas masses, such as using hydrogen deuteride. The HD 1-0 line at $112 \mu$m has been detected in three protoplanetary disks with \emph{Herschel} and has proven to be a robust measurement of the gas mass (e.g. \citealt{Bergin2013,McClure2016,Schwarz2016,Trapman2017,calahan2021,Schwarz2021}). A comparison between HD-based gas masses and gas masses derived from the combination of \cyo\ and \nhp\ was recently presented by \cite{Trapman2022b}. The test presented here is similar, but instead of using the source specific models used in the aforementioned work we instead derive gas masses using the MCMC method outlined in this work and compare them to the HD-based gas masses from \cite{Trapman2022b}. 

We ran the MCMC method as outlined in Section \ref{sec: measuring gas masses with MCMC} for TW Hya, DM Tau, and GM Aur using the integrated fluxes collated in Table \ref{tab: observations HD test}. For GM Aur a small modification was made. Its stellar luminosity lies outside of our grid of models (the ``bright'' star had $L_* = 1.0\ \mathrm{L}_{\odot}$, see Table \ref{tab: model fixed parameters}) so we limit the models to only ``bright'' stars. 
Figure \ref{fig: HD comparison} shows the resulting posterior distributions for \mgas, \rc, \abu, $\zeta_{\rm cr}$, and $L_*$. For DM Tau and GM Aur we find good agreement between the two gas mass estimates, with the median gas mass obtained from the MCMC lying with the uncertainties on the HD-based gas mass.
For TW Hya the gas mass posterior distribution is bimodal, with lower mass peak (around $\sim10^{-2}\ \msun$) coinciding with the HD-based gas mass and a high mass peak that lies outside of our prior ($\mgas > 0.1\ \msun$). Closer inspection reveals that this bimodality is correlated with the cosmic ray ionization rate. The lower gas mass corresponds to $\zeta\approx10^{-17}\ \mathrm{s}^{-1}$, whereas the higher gas mass corresponds to a lower rate of $\zet \lesssim 10^{-18}\ \mathrm{s}^{-1}$. 

These comparisons show that the grid model approach recovers mass well but do not sufficiently constrain the \rc. The previous source-specific models had various assumptions and the \rc\ estimations do not have any uncertainties, which suggest that the \rc\ solution may not be unique in these models. Therefore, surface mass distribution needs further investigation.  TorresVillanueva et al. 2025 in prep does source specific models to match the observed radial profiles of \xco\ and \cyo\ line emission, which will provide more insights on the mass distribution.

\FloatBarrier

\section{Testing gas mass constraints with and without \nhp\ }\label{app:mass_uncertainty_comparison}

\begin{figure}
    \centering
    \includegraphics[width=0.7\textwidth]{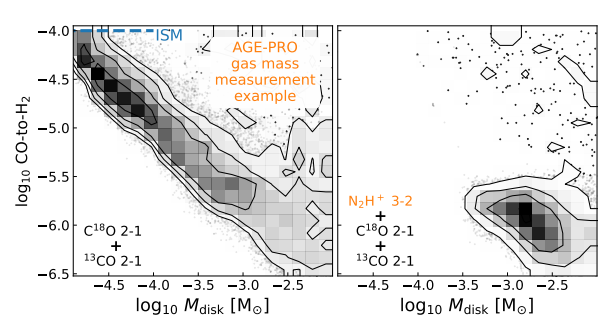}
    \caption{\label{fig:mass_wo_n2hp} Comparison of the accuracy of gas mass constraints without (left) and with (right) \nhp\ for the Lupus 3 source.}
\end{figure}

To test the effect of including \nhp\ on gas mass constraints, 
we compare MCMC results of gas masses derived from CO isotopologues alone with those derived from both CO isotopologues and \nhp\ line fluxes. Figure~\ref{fig:mass_wo_n2hp} shows the resulting gas mass constraints without (left) and with (right) \nhp\ for the Lupus 3 source as an example. After marginalizing over all other disk properties -- including the uncertainty of cosmic ray ionization rate -- we find that including \nhp\ significantly increases both the accuracy and precision of the gas mass estimate.


\section{Testing the (piece-wise) linear interpolation assumption}
\label{app: linear interpolation assumption}

\begin{figure*}
    \centering
    \begin{minipage}{0.49\textwidth}
    \includegraphics[width=\columnwidth]{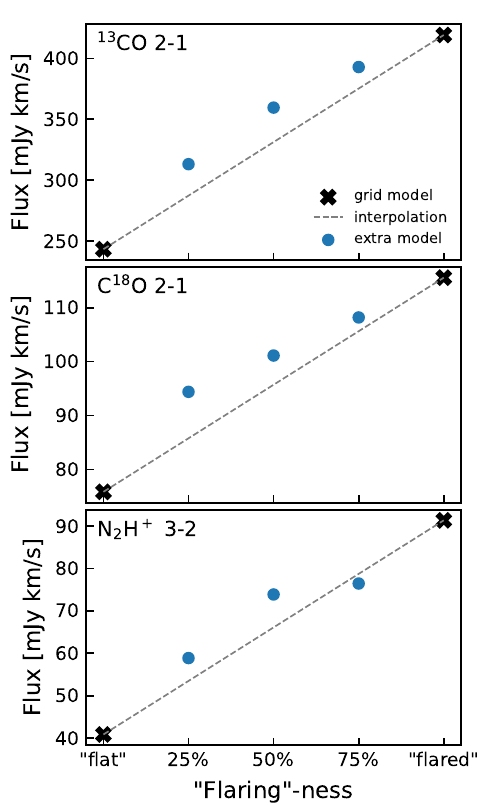}    
    \end{minipage}
    \begin{minipage}{0.48\textwidth}
    \includegraphics[width=\columnwidth]{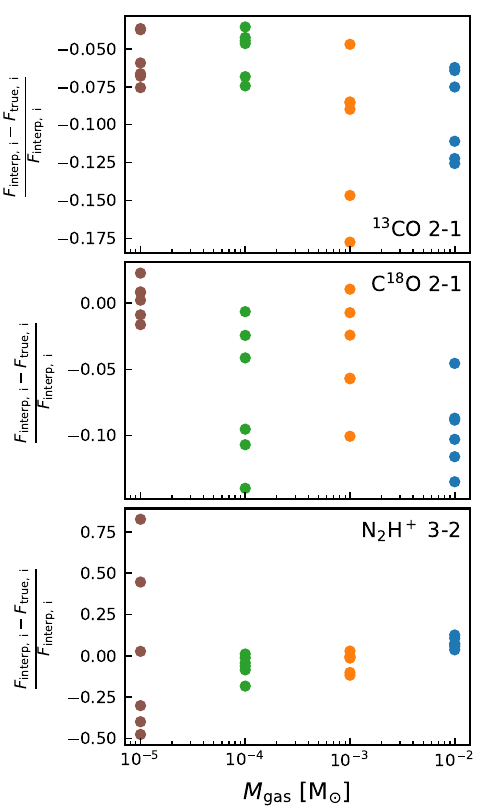}    
    \end{minipage}
    
    \caption{\label{fig: interpolation test} \textbf{left panels:} \xco\ 2-1, \cyo\ 2-1, and \nhp\ 3-2 line fluxes as a function of ``flaring''-ness (i.e. $(\psi_{x\%},h_{100,x\%})$) for an example disk with $\mgas=10^{-3}\ \msun$ and $\rc=60$ au. Black crosses denote ``flat'' and ``flared'' models that are in the model grid, with the dashed gray line showing the linear interpolation used in the MCMC. The blue circles show models with intermediate vertical structures. 
    \textbf{right panels:} The difference between the line flux obtained from linearly interpolating between ``flat'' and ``flared'' models and the true line flux for that vertical structure, expressed as the fraction of the interpolated flux. The gas mass of each model is shown on the x-axis, which each gas mass bin containing six models: two disk sizes and three intermediate vertical structures.  }
\end{figure*}

One of the main assumptions we have to make in order to fit the AGE-PRO observations to our models using MCMC is that we can linearly interpolate fluxes between models with different parameters. For most of the parameters the model grid covers the parameter space in several steps (for example, there are three gas-to-dust mass ratios), but for two parameters, the vertical structure and the dust properties, are only covered by two steps, one at each extreme of the parameters range. Without any constrain in between there is a possibility that our linear interpolation is misrepresenting the actual dependence of, for example, the \xco\ 2-1 flux on the vertical structure of the disk.  

To investigate this possibility, we ran a series of models with vertical structures in between ``flat'' $(h_{100}=0.05, \psi=0.1)$ and ``flared'' $(h_{100}=0.1, \psi=0.25)$ and with dust populations between ``young'' $(\chi=0.6,f_{\rm large}=0.8)$ and ``evolved'' $(\chi=0.2,f_{\rm large}=0.9)$ as defined in Section \ref{sec: Model grid}. Using the vertical structure as an example, we scale the relevant parameters as $h_{100,\ x\%} = x\times(h_{\rm  100, flared} - h_{\rm  100, flat}) + h_{\rm  100, flat}$ and $\psi_{x\%} = x\times(\psi_{\rm flared} - \psi_{\rm flat}) + \psi_{\rm flat}$. These models were run for $\mgas \in \{10^{-5},10^{-4},10^{-3},10^{-2}\}\ \msun$, $\rc \in \{15, 60\}$ au, an ``intermediate'' star and $\Delta_{\rm gd} = 100$. 

The left side of Figure \ref{fig: interpolation test} shows the \xco\ 2-1, \cyo\ 2-1, and \nhp\ 3-2 line fluxes as a function of vertical structure for a disk with $\mgas = 10^{-3}\ \msun$ and $\rc = 60$ au. The line fluxes follow an approximately linear trend with increasing $\psi$ and $h_{100}$, which supports our use of a linear interpolation between ``flat'' and ``flared'' disks, but it is also clear that the linear interpolation consistently underestimates the \xco\ and \cyo\ 2-1 fluxes. 
To quantify this we calculated the difference between the flux obtained from the linear interpolation, $F_{\rm interp}$, and true flux $F_{\rm true}$ of a model with that specific vertical structure. The right panels of Figure \ref{fig: interpolation test} show this difference, expressed as a fraction of $F_{\rm interp}$, as a function of gas disk mass. We find that the linear interpolation consistently underestimates the line fluxes by $\sim5-20\%$. This suggests that we are underestimating line fluxes in our MCMC fits, which could potentially result in a small overestimation of the disk mass, as a slightly lower disk mass with the same vertical structure can now also reproduce the observations. However, this offset is much smaller than the effect changes of \mgas\ and \abu\ have on the line fluxes, e.g. an increase in gas mass by a factor of three will increase the \cyo\ 2-1 line by a similar factor. 

\section{MCMC gas mass fitting priors}
\label{app: priors}

\begin{table}[htb]
\centering
\caption{\label{tab: MCMC priors} MCMC priors}
\def\arraystretch{1.1}
\begin{tabular*}{0.85\columnwidth}{l|cccccccc}
\hline\hline
AGEPRO ID &  lg$_{10}$ $M_{\rm gas}$ & $R_{\rm c}$ & lg$_{10}$ $\Delta_{\rm gd}$ & $\psi$ & $f_{\rm large}$ & $L_*^{\dagger}$ & lg$_{10}$ $x_{\rm CO}$ & lg$_{\rm 10}$ $\zeta_{\rm CR}$ \\
\hline
 Lupus 1 &-5 - -1 &15 - 60 &1 - 3 &0.1 - 0.25 &0.8 - 0.9 &(0.87,0.26) &-6.5 - -4.0 &-19 - -17 \\
 Lupus 2 &-5 - -1 &15 - 120 &1 - 3 &0.1 - 0.25 &0.8 - 0.9 &(0.33,0.10) &-6.5 - -4.0 &-19 - -17 \\
 Lupus 3 &-4 - -1 &15 - 180 &1 - 3 &0.1 - 0.25 &0.8 - 0.9 &(0.39,0.12) &-6.5 - -4.0 &-19 - -17 \\
 Lupus 4 &-6 - -2 &1 - 60 &1 - 3 &0.1 - 0.25 &0.8 - 0.9 &(0.27,0.08) &-6.5 - -4.0 &-19 - -17 \\
 Lupus 5 &-6 - -2 &1 - 60 &1 - 3 &0.1 - 0.25 &0.8 - 0.9 &(0.59,0.18) &-6.5 - -4.0 &-19 - -17 \\
 Lupus 6 &-6 - -2 &1 - 60 &1 - 3 &0.1 - 0.25 &0.8 - 0.9 &(0.20,0.06) &-6.5 - -4.0 &-19 - -17 \\
 Lupus 7 &-6 - -2 &1 - 60 &1 - 3 &0.1 - 0.25 &0.8 - 0.9 &(0.15,0.04) &-6.5 - -4.0 &-19 - -17 \\
 Lupus 8 &-6 - -2 &1 - 60 &1 - 3 &0.1 - 0.25 &0.8 - 0.9 &(0.22,0.07) &-6.5 - -4.0 &-19 - -17 \\
 Lupus 9 &-5 - -2 &1 - 120 &1 - 3 &0.1 - 0.25 &0.8 - 0.9 &(0.27,0.08) &-6.5 - -4.0 &-19 - -17 \\
 Lupus 10 &-3 - -0.3 &60 - 300 &1 - 3 &0.1 - 0.25 &0.8 - 0.9 &1.15$^{\ddagger}$ &-6.5 - -4.0 &-19 - -17 \\
 UppSco 1 &-5 - -1 &15 - 120 &1 - 3 &0.1 - 0.25 &0.8 - 0.9 &(0.25,0.07) &-6.5 - -4.0 &-19 - -17 \\
 UppSco 2 &-6 - -2 &1 - 60 &1 - 3 &0.1 - 0.25 &0.8 - 0.9 &(0.09,0.03) &-6.5 - -4.0 &-19 - -17 \\
 UppSco 3 &-5 - -2 &1 - 60 &1 - 3 &0.1 - 0.25 &0.8 - 0.9 &(0.08,0.02) &-6.5 - -4.0 &-19 - -17 \\
 UppSco 4 &-6 - -2 &1 - 60 &1 - 3 &0.1 - 0.25 &0.8 - 0.9 &(0.32,0.10) &-6.5 - -4.0 &-19 - -17 \\
 UppSco 5 &-6 - -2 &1 - 60 &1 - 3 &0.1 - 0.25 &0.8 - 0.9 &(0.09,0.03) &-6.5 - -4.0 &-19 - -17 \\
 UppSco 6 &-6 - -2 &1 - 120 &1 - 3 &0.1 - 0.25 &0.8 - 0.9 &(0.17,0.05) &-6.5 - -4.0 &-19 - -17 \\
 UppSco 7 &-4 - -1 &15 - 180 &1 - 3 &0.1 - 0.25 &0.8 - 0.9 &(0.18,0.05) &-6.5 - -4.0 &-19 - -17 \\
 UppSco 8 &-5 - -1 &15 - 120 &1 - 3 &0.1 - 0.25 &0.8 - 0.9 &(0.10,0.03) &-6.5 - -4.0 &-19 - -17 \\
 UppSco 9 &-5 - -1 &5 - 120 &1 - 3 &0.1 - 0.25 &0.8 - 0.9 &(0.32,0.10) &-6.5 - -4.0 &-19 - -17 \\
 UppSco 10 &-5 - -1 &5 - 120 &1 - 3 &0.1 - 0.25 &0.8 - 0.9 &(0.32,0.10) &-6.5 - -4.0 &-19 - -17 \\
\hline\hline
\end{tabular*}
\begin{minipage}{0.85\columnwidth}
\vspace{0.1cm}
{\footnotesize{$^{\dagger}$ values for the Gaussian prior for stellar luminosity are the mean and standard deviation, respectively. 
$\ddagger:$For Lupus 10 no Gaussian prior is used but instead only models with $L_* = 1.0\ \mathrm{L}_{\odot}$ were used in the fit. }}
\end{minipage}
\end{table}

\section{Comparison of the AGE-PRO \co\ observations and the models}
\label{app: 12CO comparison}

\begin{figure*}[hbt]
    \centering
    \includegraphics[width=\textwidth]{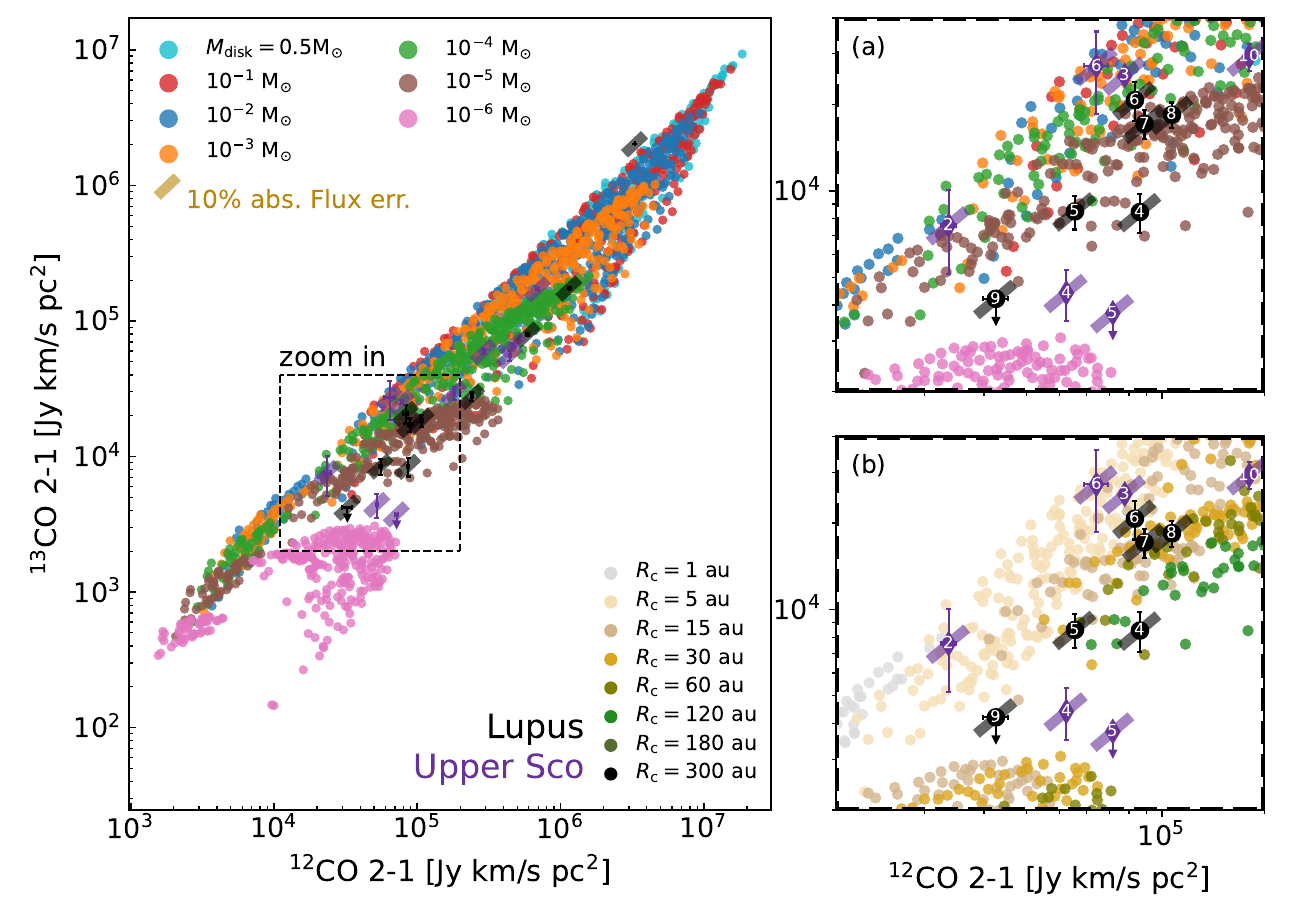}
    \caption{\label{fig: 12CO comparison}As Figure \ref{fig: CO comparison}, but now showing the \co\ and \xco\ $J=2-1$ line luminosities. Note that due to cloud absorption seen towards most sources in Lupus their observed \co\ line luminosities should be considered as lower limits (see \citealt{AGEPRO_III_Lupus} for details). Accounting for this effect would move the points to the right of the figure. The zoom-in shown in panels (a) and (b) now focuses on the sources in Lupus and Upper Sco where \xco\ and \cyo\ are not detected. }
\end{figure*}

\FloatBarrier
\section{Joint posterior \mgas\ and \abu\ probability distributions}
\label{app: joint posteriors}

\begin{figure*}[hbt]
    \centering
    \includegraphics[width=0.9\textwidth,clip,trim={0 0.5cm 0 0}]{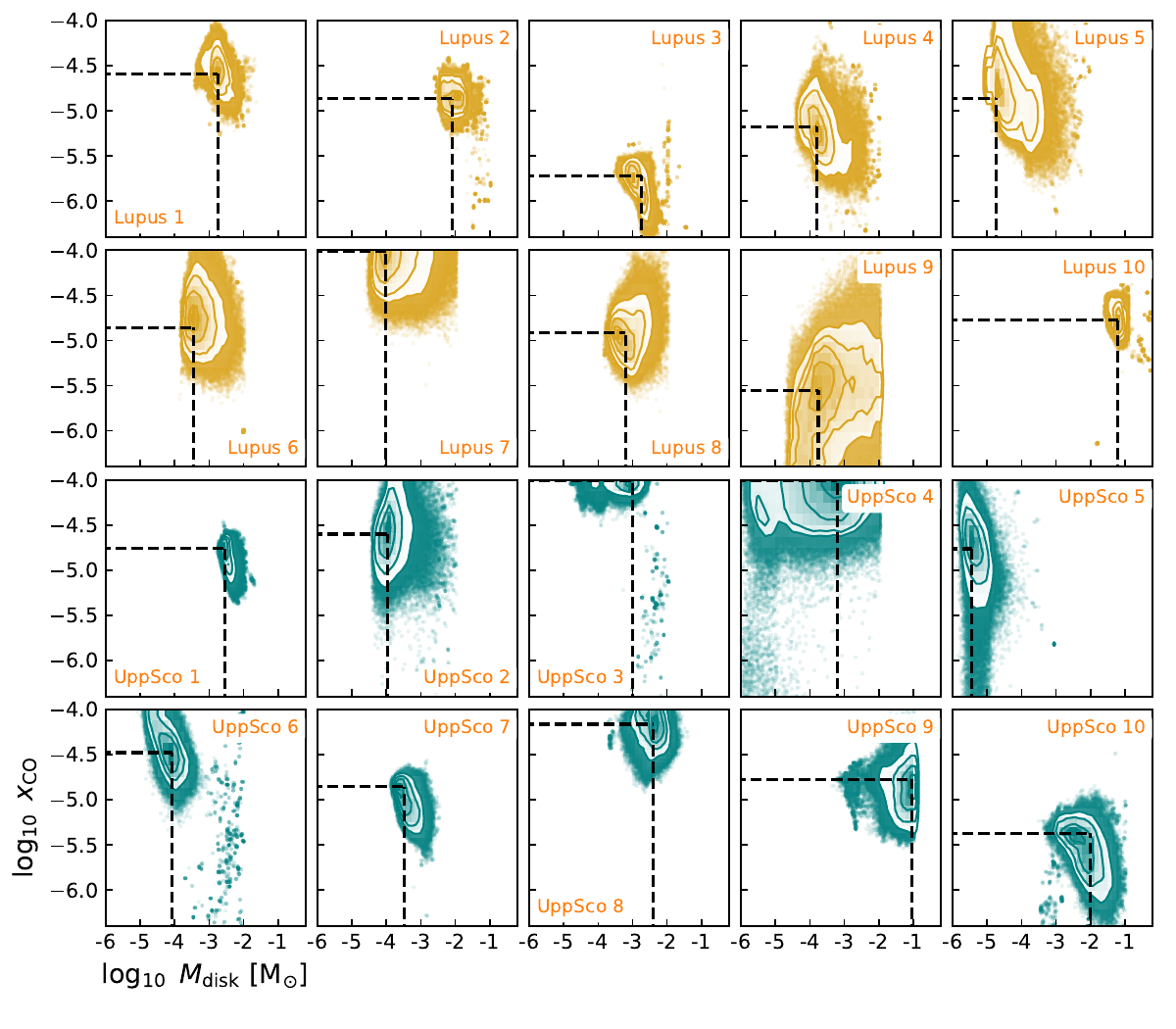}
    \caption{\label{fig: 2D histograms}Joint posterior probability distribution of \mgas\ and \abu\ obtained from fitting the observed integrated \xco\ 2-1, \cyo\ 2-1, \nhp 3-2, and 1300$\mu$m continuum fluxes using MCMC. Orange dashed lines show the best fit \mgas\ and \abu\ for each disk.}
\end{figure*}

\section{Ophiuchus gas mass posteriors from cold and warm disk models}
\label{app: ophiuchus cold warm posteriors}

\begin{figure*}
    \centering
    \includegraphics[width=0.9\textwidth]{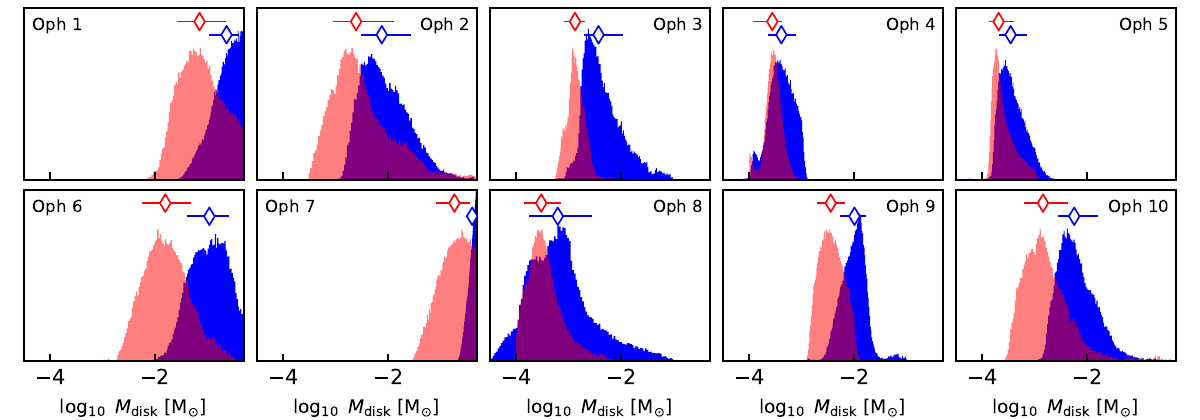}
    \caption{\label{fig: ophiuchus cold warm posteriors} Comparison of the gas mass posterior distribution obtained with fitting the Ophiuchus observations using cold $(L_* = L_{\rm *,\ obs}$; in blue) and warm $(L_* = 10\times L_{\rm *,\ obs}$; in red) disk models. See Section \ref{sec: ophiuchus gas masses} for details.}
\end{figure*}

\FloatBarrier
\newpage
\section{Derivation of the characteristic radius}
\label{app: Rc derivation}

\begin{figure*}[ht]
    \centering
    \includegraphics[width=\textwidth]{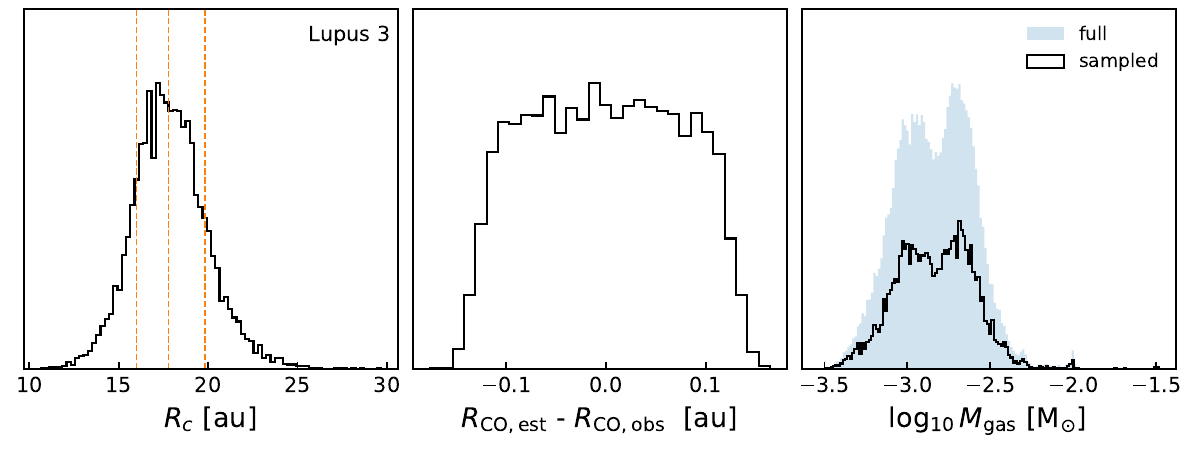}
    \caption{\label{fig: Rc derivation Lupus 2} \textbf{Left:} distribution of derived \rc\ for Lupus 3 based its \rgas, $\mgas$, and $L_*$. \textbf{Middle:} Difference between the best fitting disk size ($R_{\rm CO,\ 90\%,est}$) and the observed value ($R_{\rm CO,\ 90\%,obs}$). \textbf{Right:} Comparison of the full posterior distribution of \mgas\ and the portion sampled during the Monte Carlo process used to build up the distribution of \rc\ in the left panel. }
\end{figure*}

Recently \cite{Toci2023} and \cite{Trapman2023} demonstrated that the observed gas disk size \rgas\ is directly related to a critical surface density and can be written down analytically as a function of \mgas, \rc, ($L_*$), and the slope of the surface density $\gamma$. \citealt{Trapman2023} used this expression and an estimate of the gas mass based on the dust mass to derive \rc\ from the observed \rgas\ of disks in  Lupus, Taurus, Upper Sco, and the DSHARP sample. Here we repeat that exercise for the AGE-PRO disks in Lupus and Upper Sco using the gas masses derived in this work. Briefly, for each source we compute $N$ estimates for $R_{\rm CO,\ 90\%,est}$ using Eq. 9 from \cite{Trapman2023} using the gas mass of the source and a range of $N$ \rc\ values. Here we do also include the stellar luminosity dependence of critical surface density (Eq. 7 in \citealt{Trapman2023}), taking the stellar luminosities of each source from the literature (\citealt{Alcala2019,Manara2020,UpperSco_followup}; see \citealt{AGEPRO_I_overview} for details.). The best fitting \rc\ is then selected by minimizing $|R_{\rm CO,\ 90\%,est}$ - $R^{\rm fit}_{\rm CO,\ 90\%,obs}|$, where $R^{\rm fit}_{\rm CO,\ 90\%,obs}$ is the gas disk size from \cite{AGEPRO_XI_gas_disk_sizes}, who fitted the \co\ moment zero map with a beam-convolved Nuker profile to obtain a $R_{\rm CO,\ 90\%}$ corrected for the resolution of the observations (see \citealt{AGEPRO_XI_gas_disk_sizes} for details).  

To include the uncertainties on \mgas\ and $R^{\rm fit}_{\rm CO,\ 90\%, obs}$ this procedure is repeated $M$ times in a Monte Carlo fashion, each time drawing a random $M_{\rm gas, i}$ from its posterior probability distribution (see Section \ref{sec: measuring gas masses with MCMC}) and a random $R^{\rm fit,\ i}_{\rm CO,\ 90\%, obs}$ from within its uncertainties (see Table 2 in \citealt{AGEPRO_XI_gas_disk_sizes}). 
Figure \ref{fig: Rc derivation Lupus 2} shows an example of the resulting distribution of \rc\ for Lupus 3, and estimates of \rc\ for all twenty sources can be found in Table \ref{tab: rc}. 

Another estimate for \rc\ comes from the MCMC fit to the CO isotopologue line fluxes and \rgas\ discussed in Section \ref{sec: measuring gas masses with MCMC}. The characteristic radius is one of the model parameters we marginalize over to obtain the gas mass posterior distribution, but we can also marginalize over all parameter but \rc\ to obtain its posterior distribution (see Table \ref{tab: rc}). 

Before comparing the two characteristic radii it is useful to first briefly discuss the main difference between the analytical and the MCMC derived \rc. Both are, in their own way, linked to the observed gas disk size \rgas. In the case of the analytical \rc\ \rgas\ is obtained from a best fit model intensity profile fitted to the \co\ 2-1 moment zero map. For fit the model is convolved with the clean beam of the observations. \rgas\ measured from the unconvolved model to correct for the effect of the resolution of the observations on the measurement of \rgas\ (see \citealt{AGEPRO_XI_gas_disk_sizes} for details). For the MCMC \rc\ a the approach is more forward-modeling, where the synthetic observations of the \texttt{DALI} models where convolved with the clean beam of the observations before being compared to the observed \rgas\ (see Section \ref{sec: measuring gas masses with MCMC}). While these two approaches are different, to first order we expect them to provide a similar correction of \rgas\ for the resolution of the observations. 

The largest difference lies with the fact that the MCMC-\rc\ is compared the \xco\ and \cyo\ line fluxes. As can be seen in Panel b in Figure \ref{fig: CO comparison}, in particular the \xco/\cyo\ 2-1 line ratio is sensitive to \rc\ for small \rc, with the line ratio decreasing towards one as \rc\ decreases. The main cause for this is optical depth. For these disks the \xco\ and \cyo\ integrated line fluxes are dominated by optically thick emission, meaning that the line ratio scales as the ratio of their emitting areas. Due to its lower abundance of \cyo, its emission will become optically thin at a higher column density, and therefore a smaller radius, than the more abundant \xco, meaning that the optically thick emitting area of \cyo\ is always smaller than that of \xco. However, if the slope of the surface density profile is steep, such as in the exponential taper of the surface density profile, the radii at which \cyo\ and \xco\ become optically thick will lie close together and the line ratio will be close to one. Decreasing \rc\ moves these two radii further out into the exponential taper and thereby closer together, resulting in the line ratio decreasing towards one.  

\begin{figure}
    \centering
    \includegraphics[width=0.6\columnwidth]{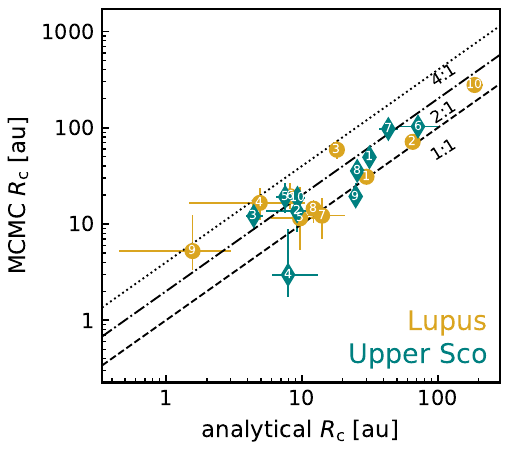}
    \caption{\label{fig: comparison Rc MCMC vs analytical} Comparison of the characteristic radii (\rc) obtained from the MCMC fit described in Section \ref{sec: measuring gas masses with MCMC} to the analytically computed \rc\ discussed in Appendix \ref{app: Rc derivation}.}
\end{figure}

Figure \ref{fig: comparison Rc MCMC vs analytical} compares the MCMC-derived \rc\ to the analytically computed \rc\ discussed previously. In almost all cases the two radii agree to within a factor of two, although for the more compact disks $(\rc \lesssim 10\ \mathrm{au})$ the two estimates start to diverge. Among the larger disks Lupus 3 is a clear outlier with MCMC-\rc\ that is $\sim3\times$ larger than its analytical \rc. This source has a \xco\/\cyo 2-1 line flux ratio of $\sim16$ \citep{AGEPRO_III_Lupus}, which is too large to be consistent with the smaller \rc\ obtained from the analytical calculation. The discrepancy between these two characteristic radii could be an indication that the surface density of this disk does not follow an exponential taper but has a more shallow decline with radius.

\begin{table}[htb]
\centering
\caption{\label{tab: rc} Characteristic radii for Lupus and Upper Sco disks  }
\def\arraystretch{1.2}
\begin{tabular*}{0.52\columnwidth}{lccc|cc}
\hline\hline
name  &  \mgas   & $L_*$ & \rgas & \rc \\
comments & & & & (analytical) & (MCMC) \\
      &  lg [\msun] & [L$_{\odot}$] & [au] & [au] & [au] \\
\hline
 Lupus 1 &$-2.72^{+0.20}_{-0.17}$ & 0.87 &$169.8^{+1.4}_{-2.1}$ &$32^{+2}_{-2}$ &$31^{+4}_{-4}$ \\
 Lupus 2 &$-2.04^{+0.25}_{-0.26}$ & 0.33 &$312.7^{+0.8}_{-0.6}$ &$65^{+8}_{-6}$ &$71^{+10}_{-9}$ \\
 Lupus 3 &$-2.83^{+0.22}_{-0.24}$ & 0.39 &$106.6^{+3.7}_{-9.1}$ &$18^{+2}_{-2}$ &$59^{+14}_{-11}$ \\
 Lupus 4 &$-3.72^{+0.42}_{-0.28}$ & 0.27 &$43.1^{+29.8}_{-39.2}$ &$8^{+8}_{-5}$ &$17^{+7}_{-5}$ \\
 Lupus 5 &$-4.35^{+0.68}_{-0.35}$ & 0.59 &$53.7^{+17.5}_{-9.9}$ &$10^{+5}_{-4}$ &$11^{+12}_{-6}$ \\
 Lupus 6 &$-3.27^{+0.43}_{-0.29}$ & 0.20 &$56.6^{+1.8}_{-1.6}$ &$9^{+1}_{-1}$ &$19^{+7}_{-5}$ \\
 Lupus 7 &$-3.79^{+0.71}_{-0.38}$ & 0.15 &$88.0^{+2.1}_{-1.8}$ &$20^{+4}_{-4}$ &$12^{+6}_{-5}$ \\
 Lupus 8 &$-3.12^{+0.40}_{-0.33}$ & 0.22 &$58.6^{+3.0}_{-1.6}$ &$9^{+1}_{-1}$ &$14^{+3}_{-4}$ \\
 Lupus 9 &$-3.42^{+0.90}_{-0.71}$ & 0.27 &$7.2^{+7.2}_{-7.2}$  &$2^{+1}_{-1}$  &$5^{+7}_{-2}$ \\
 Lupus 10 &$-1.19^{+0.11}_{-0.13}$ & 1.15 &$838.5^{+55.3}_{-58.8}$ &$184^{+24}_{-21}$ &$278^{+15}_{-21}$ \\
 UppSco 1 &$-2.49^{+0.15}_{-0.11}$ & 0.25 &$167.5^{+7.8}_{-2.2}$ &$31^{+2}_{-2}$ &$49^{+7}_{-5}$ \\
 UppSco 2 &$-3.90^{+0.38}_{-0.24}$ & 0.09 &$25.7^{+30.8}_{-19.4}$ &$4^{+5}_{-3}$ &$14^{+6}_{-5}$ \\
 UppSco 3 &$-3.19^{+0.24}_{-0.43}$ & 0.08 &$34.9^{+1.8}_{-4.5}$ &$5^{+1}_{-1}$ &$12^{+3}_{-3}$ \\
 UppSco 4 &$-3.69^{+0.92}_{-1.33}$ & 0.32 &$46.3^{+1.1}_{-1.5}$ &$8^{+5}_{-2}$ &$3^{+6}_{-1}$ \\
 UppSco 5 &$-5.36^{+0.27}_{-0.20}$ & 0.09 &$30.4^{+1.4}_{-1.0}$ &$7^{+1}_{-1}$ &$19^{+7}_{-6}$ \\
 UppSco 6 &$-4.16^{+0.30}_{-0.33}$ & 0.17 &$155.9^{+16.7}_{-13.9}$ &$69^{+28}_{-19}$ &$103^{+12}_{-18}$ \\
 UppSco 7 &$-3.38^{+0.29}_{-0.23}$ & 0.18 &$160.5^{+1.4}_{-1.5}$ &$44^{+7}_{-6}$ &$97^{+27}_{-23}$ \\
 UppSco 8 &$-2.41^{+0.27}_{-0.35}$ & 0.10 &$144.1^{+1.7}_{-2.5}$ &$26^{+3}_{-2}$ &$36^{+5}_{-5}$ \\
 UppSco 9 &$-1.28^{+0.20}_{-0.38}$ & 0.32 &$181.8^{+5.2}_{-8.5}$ &$25^{+3}_{-2}$ &$19^{+7}_{-4}$ \\
 UppSco 10 &$-2.14^{+0.43}_{-0.45}$ & 0.32 &$75.0^{+1.8}_{-4.0}$ &$9^{+1}_{-1}$ &$19^{+6}_{-4}$ \\
\hline\hline
\end{tabular*}
\begin{minipage}{0.85\columnwidth}
\vspace{0.1cm}
\end{minipage}
\end{table}

\FloatBarrier

\section{Extra figures of the correlations between disk and stellar properties}
\label{app: correlations}

\begin{figure*}[ht]
    \centering
    \includegraphics[width=\textwidth]{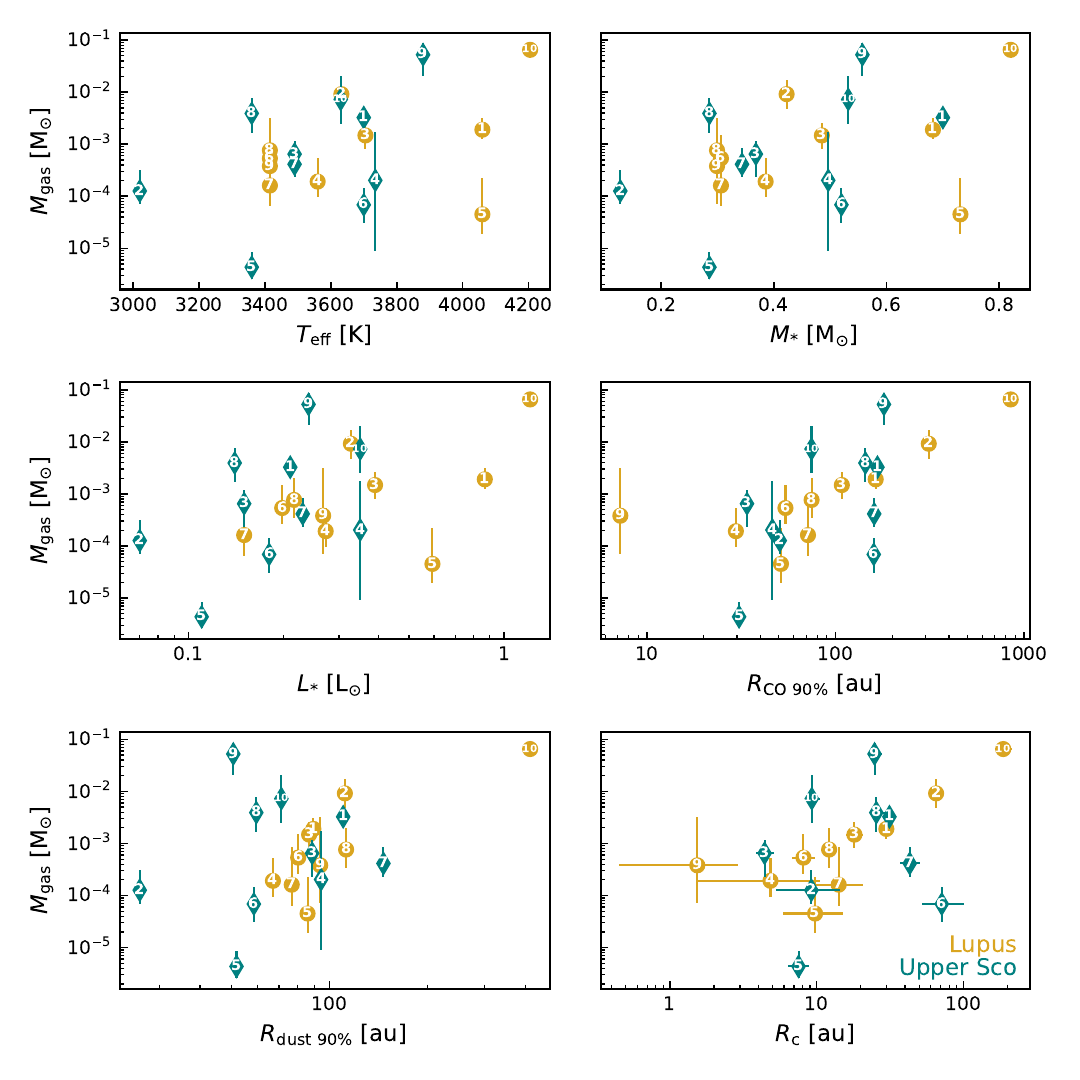}    
    \caption{\label{fig: correlations} Gas mass versus effective temperature $(T_{\rm eff})$, stellar mass $(M_*)$, stellar luminosity $(L_*)$, gas disk outer radius $(R_{\rm CO,\ 90\%})$, dust disk outer radius $(R_{\rm dust,\ 90\%})$, and the characteristic radius \rc\ for the twenty disks in Lupus (black) and Upper Sco (purple). See Appendix \ref{app: Rc derivation} for details on how \rc\ was derived for each source.}
\end{figure*}

\begin{figure*}[ht]
    \centering
    \includegraphics[width=\textwidth]{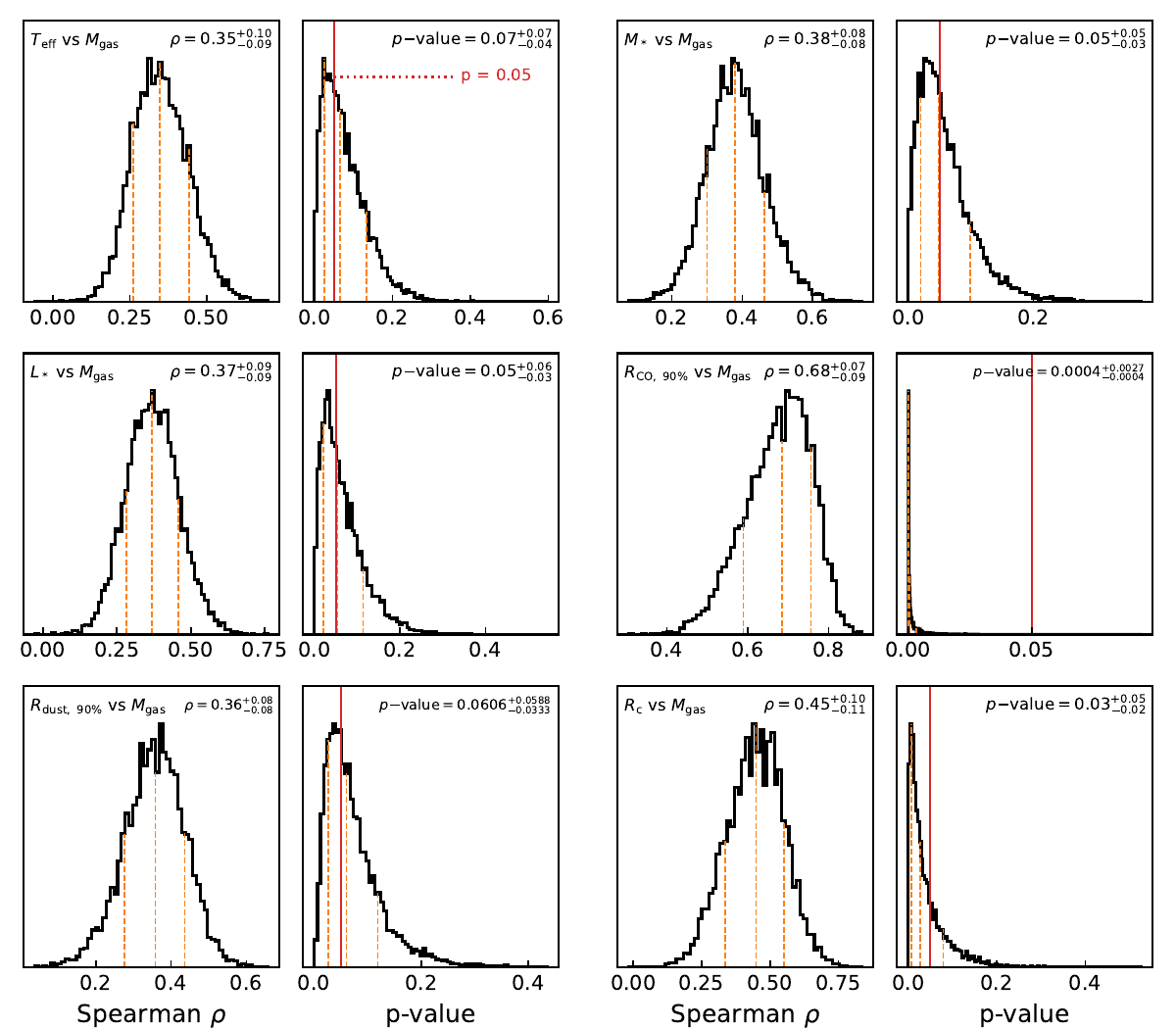}
    \caption{\label{fig: Mgas MC correlation} Results of the Monte Carlo Spearman rank tests, quantified by $\rho$, between \mgas\ and various stellar and disk parameters. Left panels show the distribution of $\rho$ and right panels show the associated $p$-value. Orange vertical lines in each panel show the $16^{\rm th}$, $50^{\rm th}$, and $84^{\rm th}$ quantile of the distribution. The red vertical line denotes $p=0.05$ to provide a guide for how statistically significant the correlation is.}
\end{figure*}

\begin{figure*}[ht]
    \centering
    \includegraphics[width=\textwidth]{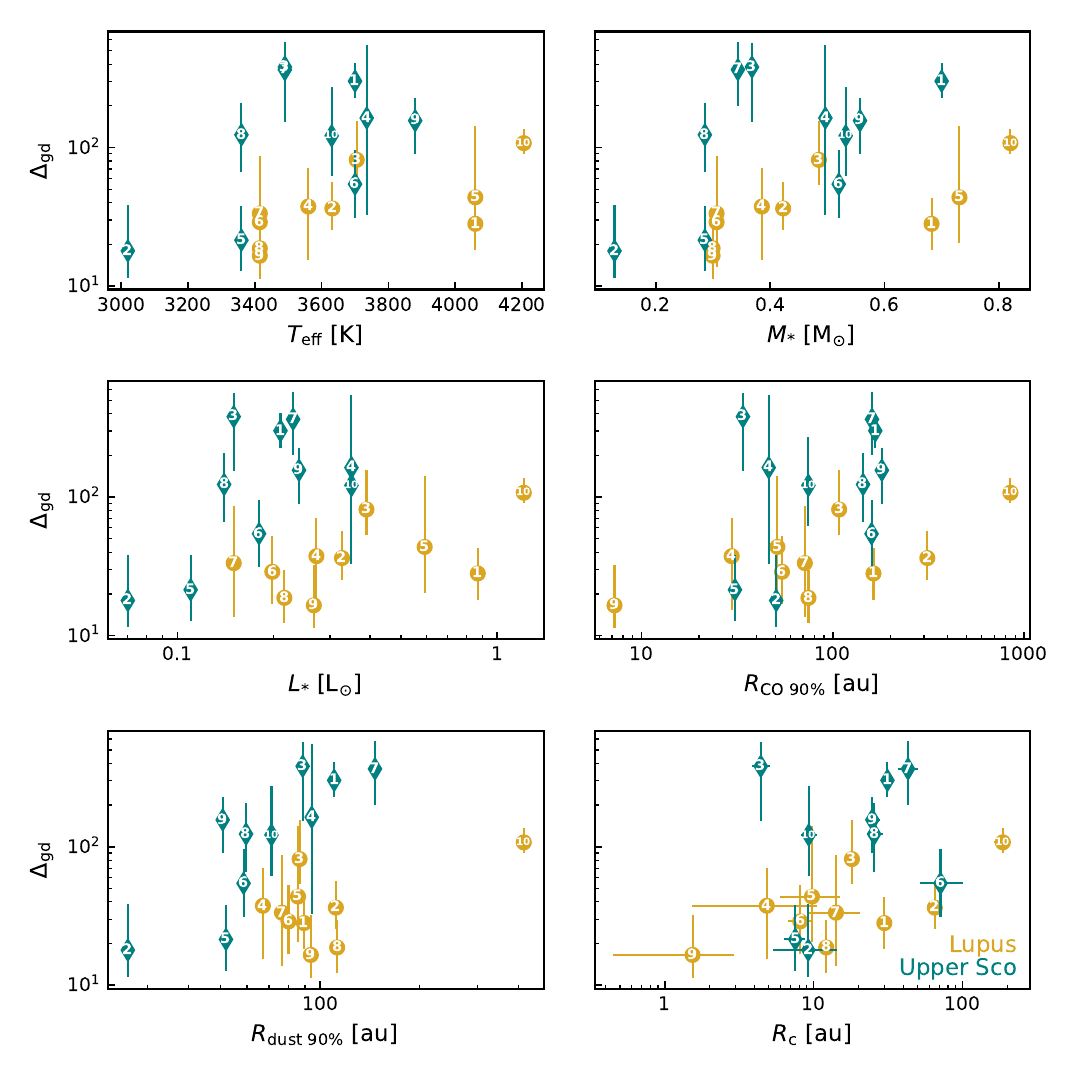}
    \caption{\label{fig: gdratio correlations} As Figure \ref{fig: correlations}, but comparing $\Delta_{\rm gd}$ to other disk and stellar parameters. }
\end{figure*}

\begin{figure*}[ht]
    \centering
    \includegraphics[width=\textwidth]{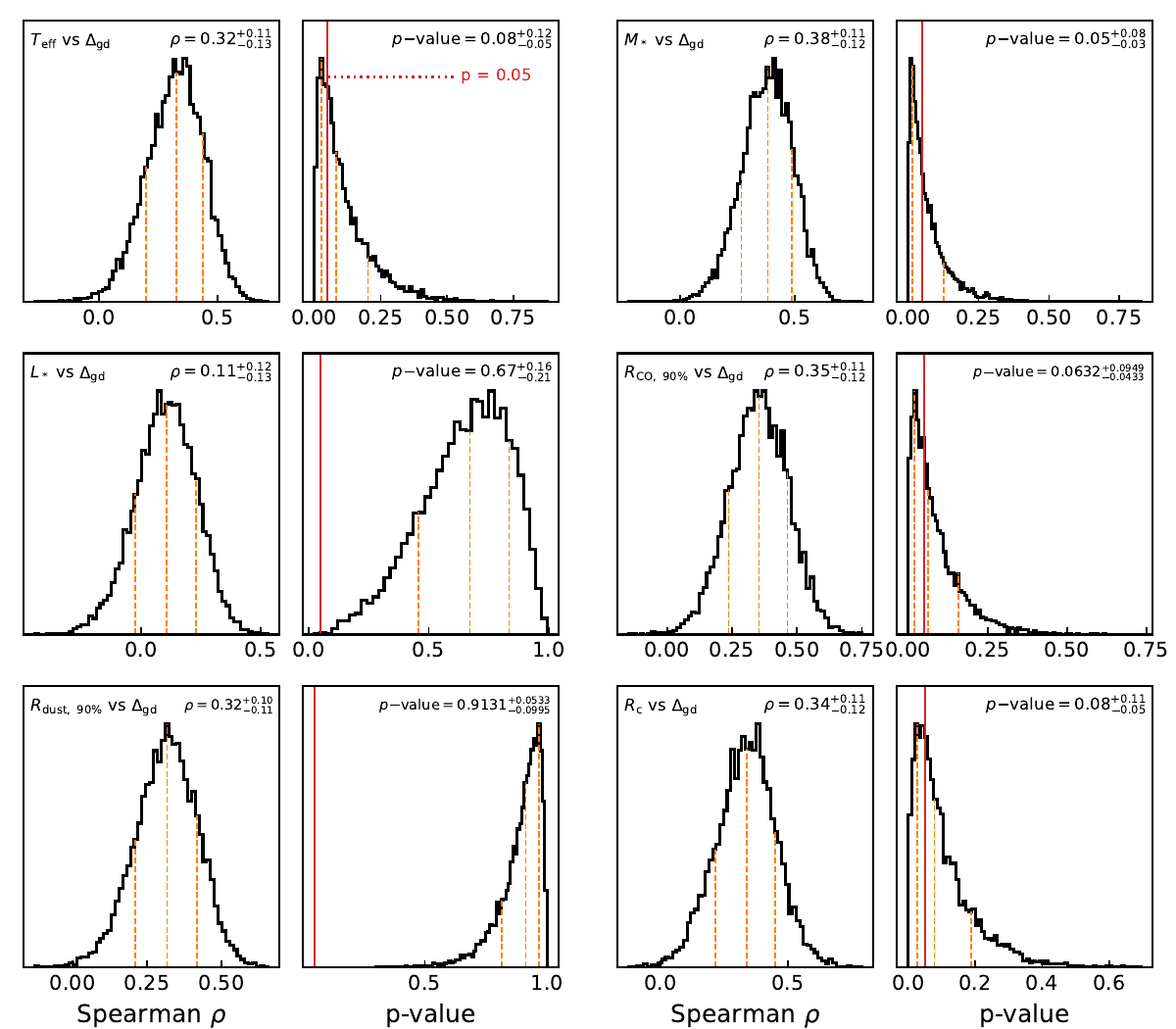}
    \caption{\label{fig: gdratio MC correlation} As Figure \ref{fig: Mgas MC correlation}, but for correlations with $\Delta_{\rm gd}$.  }
\end{figure*}

\begin{figure*}[ht]
    \centering
    \includegraphics[width=\textwidth]{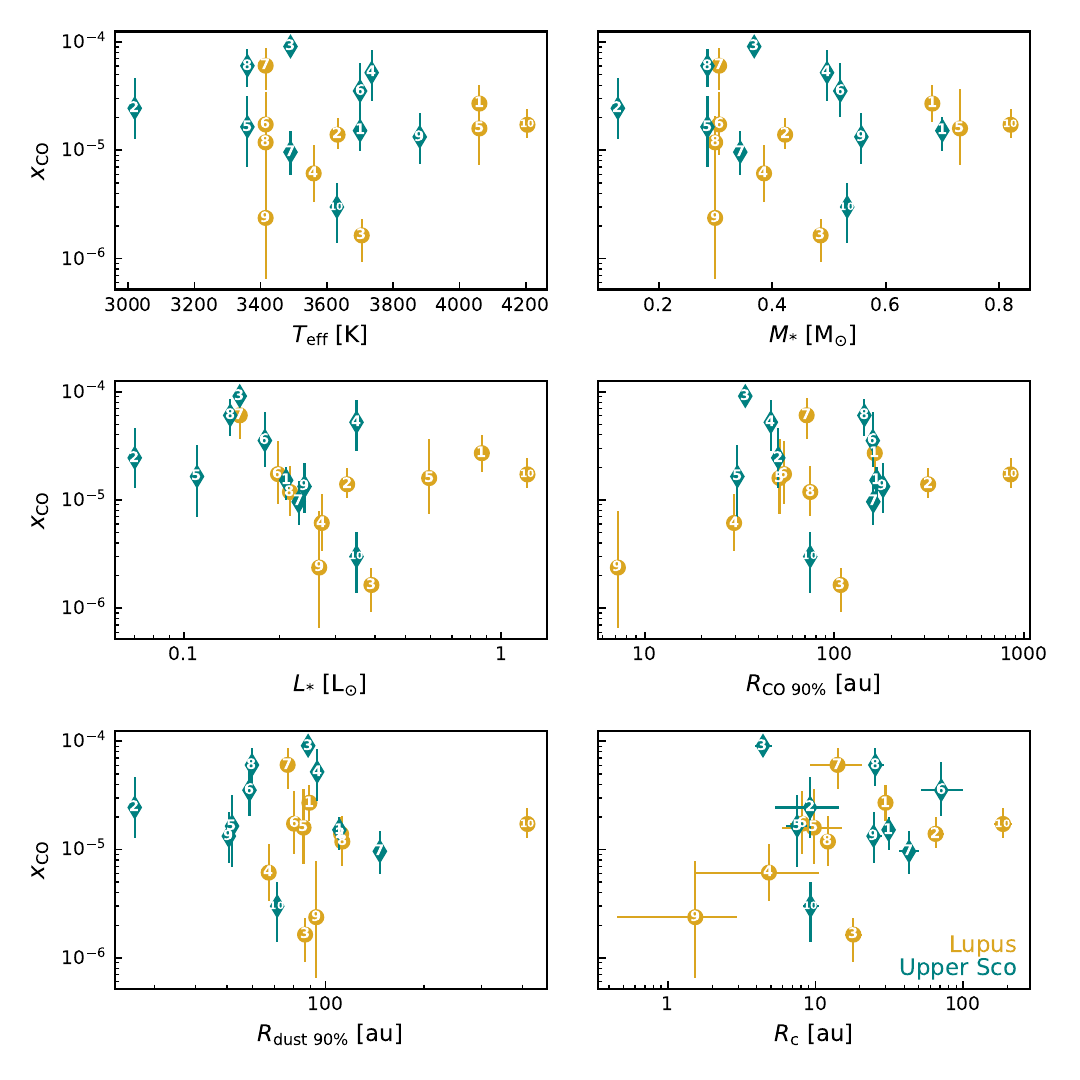}
    \caption{\label{fig: xco correlations} As Figure \ref{fig: correlations}, but comparing \abu\ to other disk and stellar parameters. }
\end{figure*}

\begin{figure*}[ht]
    \centering
    \includegraphics[width=\textwidth]{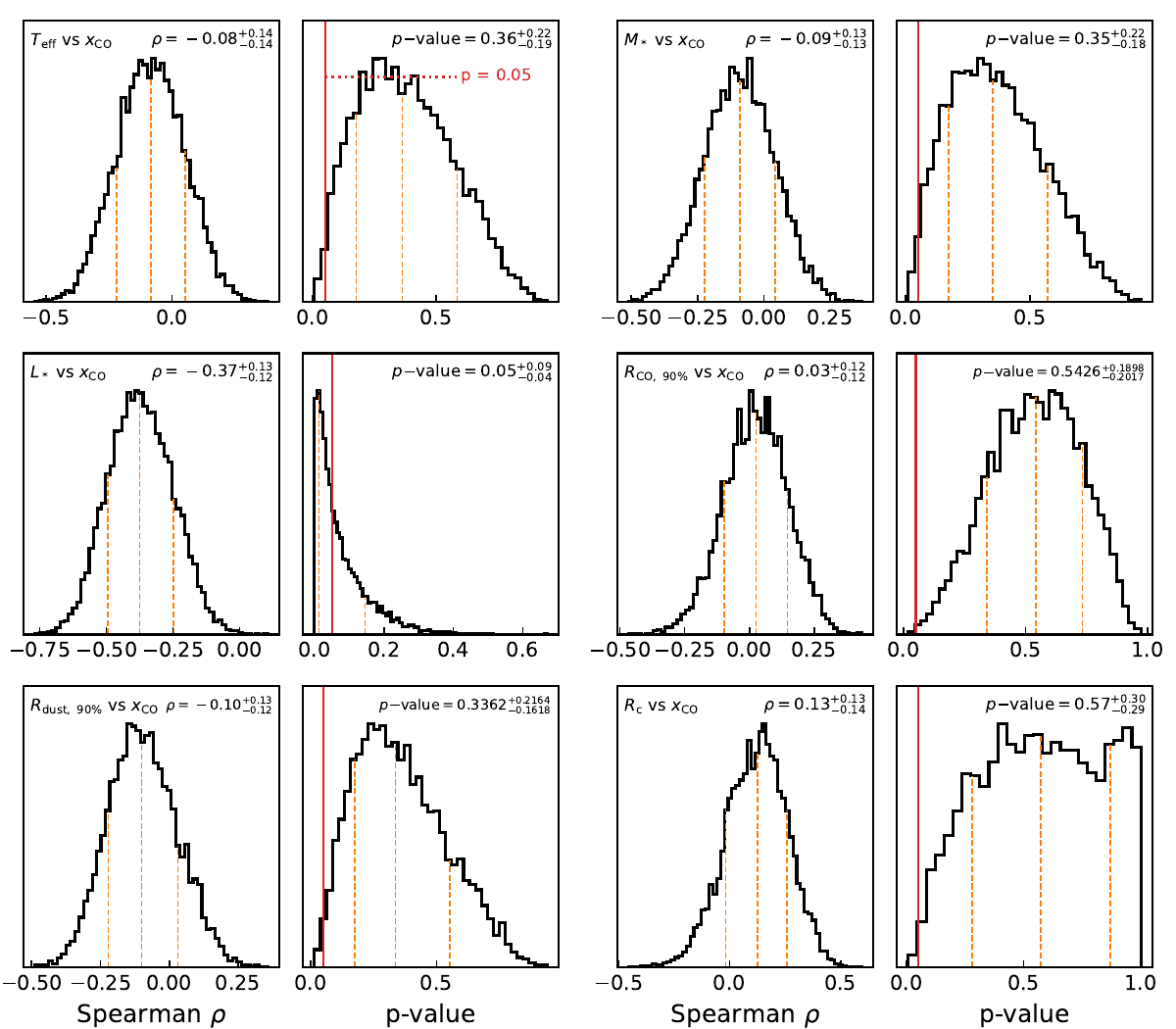}
    \caption{\label{fig: xco MC correlation} As Figure \ref{fig: Mgas MC correlation}, but for correlations with \abu. }
\end{figure*}

\section{Corner plots of the MCMC}
\label{app: MCMC corner plots }

\figsetstart
\figsetnum{26}
\figsettitle{Corner plots for the gas mass MCMC}

\figsetgrpstart
\figsetgrpnum{26.1}
\figsetgrptitle{Corner plot of Lupus 1 (Sz 65)
}
\figsetplot{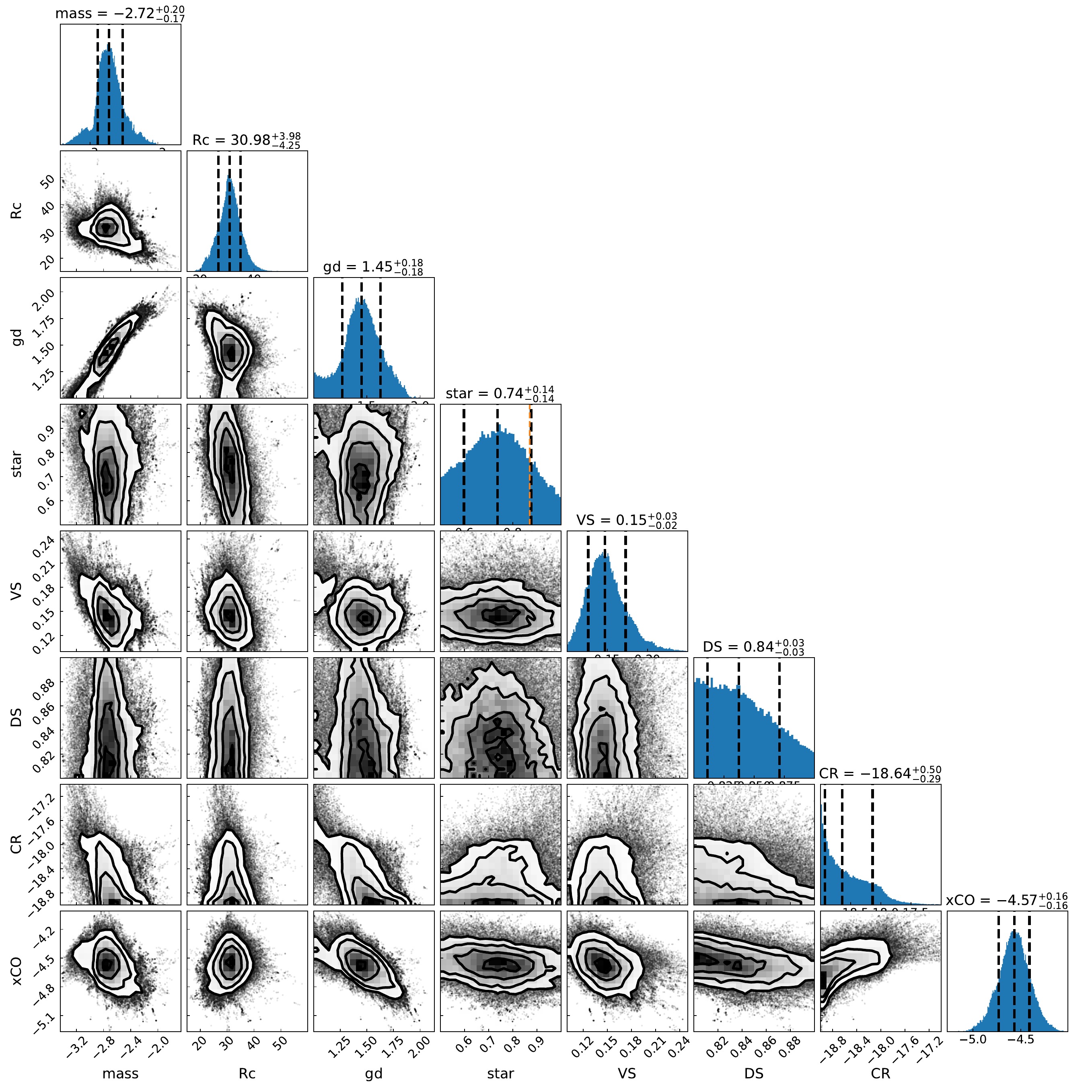}
\figsetgrpnote{Posterior probability distributions of the model parameters obtained from fitting the 13CO, C18O, N2H+, and 1.3 mm continuum fluxes and the gas disk size for the Lupus and Upper Sco disks in the AGEPRO ALMA large program.}
\figsetgrpend

\figsetend

\begin{figure}
\figurenum{26}
\plotone{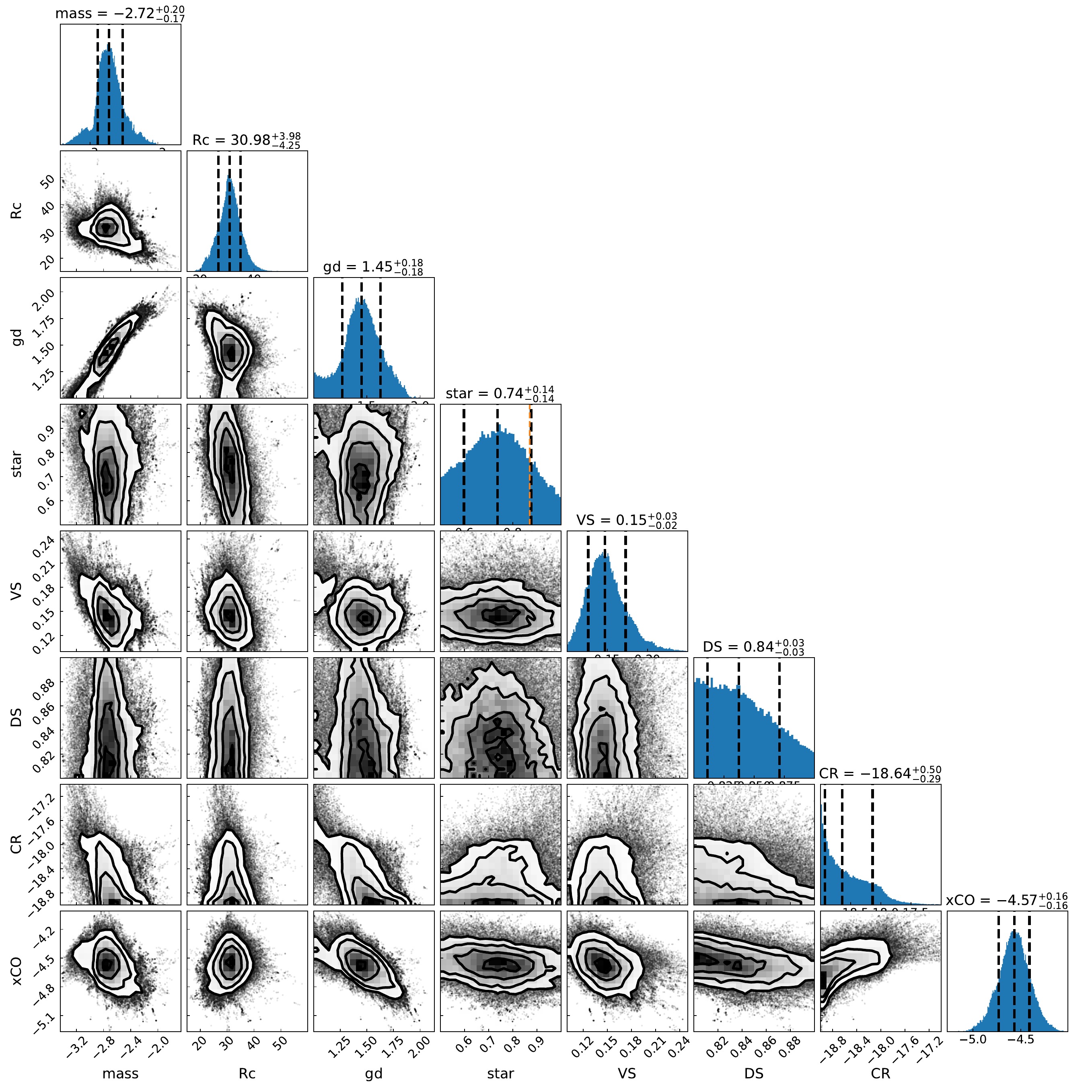}
\caption{\label{fig: corner Lupus 1} Posterior probability distributions of the model parameters obtained from fitting the observed integrated \xco\ 2-1, \cyo\ 2-1, \nhp 3-2, and 1300$\mu$m continuum fluxes and the gas disk size (\rgas) of Lupus 1 (Sz 65) using MCMC. Vertical dashed lines show the 16$^{\rm th}$, 50$^{\rm th}$, and 84$^{\rm th}$ quantile of the distribution.}
\end{figure}

\end{appendix}

\end{document}